# Two polynomial algorithms for special maximum matching constructing in trees

R.R. Kamalian

Institute for Informatics and Automation Problems of National Academy of Sciences of Republic of Armenia, 0014, Republic of Armenia

Ijevan Branch of Yerevan State University

Armenian-Russian State University

e-mail: rrkamalian@yahoo.com

and

V. V. Mkrtchyan

Department of Informatics and Applied Mathematics, Yerevan State University, Yerevan, 0025, Republic of Armenia

Institute for Informatics and Automation Problems of National Academy of Sciences of Republic of Armenia, 0014, Republic of Armenia

e-mail: vahanmkrtchyan2002@{ysu.am, ipia.sci.am, yahoo.com}

## Abstract

For an arbitrary tree we investigate the problems of constructing a maximum matching which minimizes or maximizes the cardinality of a maximum matching of the graph obtained from original one by its removal and present corresponding polynomial algorithms.

**Keywords:** Maximum matching, tree, polynomial algorithm,

## Introduction

Problem of existence and construction of matchings in various classes of graphs is one of the well-known and important directions of research carried out in Discrete Mathematics. The subject stems from the prominent "Marriage Problem" which is a problem of existence and construction of a system of distinct representatives [16], for which necessary and sufficient condition is found [17].

The interest towards problems which are related to matchings is explained not only by their theoretical importance, but also by the essential role that matchings play in the various, sometimes unexpected applications of Combinatorial Mathematics in development of methods for solving various practical problems. Many problems of existence and construction of educational time-tables, the classical "Assignment problem", the covering problems, Christofides's algorithm for solving "Traveling Salesman Problem" with the inequality of a triangle do not exhaust all cases when matchings are useful during formulation or solution of problems of Applied Combinatorial Mathematics.

A well-investigated problem related to matchings is the problem of construction of a maximum matching in a graph. The criterion for maximality of a matching is found in [3, 31].

Many polynomial algorithms are presented for solving mentioned problem in various classes of graphs. In [20] an algorithm with runtime $O(|V|^{\frac{1}{2}} \cdot |E|)$ is presented for constructing a maximum matching in bipartite graphs, where $|V|$ and $|E|$ are the number of vertices and edges of a graph, respectively. On the basis of methods and algorithms of [7, 20], in [1] an algorithm is presented for constructing a maximum matching in bipartite graphs with complexity

$$O(\min\{|V|^{\frac{1}{2}} \cdot |E|, |V|^{\frac{3}{2}} \cdot (\frac{|E|}{\log|V|})^{\frac{1}{2}}\}),$$

which is at worst

$$O(\frac{|V|^{\frac{5}{2}}}{(\log|V|)^{1/2}}).$$

Note that the problem of construction of a maximum matching is also considered in some subclasses of bipartite graphs.

A bipartite graph $G = (V_1, V_2, E)$ is referred to be convex on $V_1$, if the vertices of $V_1$ can be labeled with numbers $1, 2, ..., |V_1|$ such that vertices, adjacent to an arbitrary vertex from the set $V_2$, have consecutive numbers. For these graphs in [13] an algorithm is presented which constructs a maximum matching in time $O(|V_1| \cdot |V_2|)$. It is known [4] that there is an $O(|V| + |E|)$ algorithm which checks convexity of a given bipartite graph on $V_1$, and in case of positive answer constructs the desired enumeration of vertices of $V_1$. Let us also note that in [27] an algorithm for constructing a maximum matching of a bipartite graph which is convex on $V_1$, is described which runs in time $O(|V_1| + |V_2| A(|V_2|))$, where $A(n)$ is a slowly growing function related to the functional inverse of Ackermann's function [34].

A bipartite graph $G = (V_1, V_2, E)$ is referred to be doubly convex if it is convex on $V_1$ and $V_2$. For these graphs in [27] an $O(|V|)$ algorithm is presented which constructs a maximum matching.

Historically, the first algorithm constructing a maximum matching for arbitrary graphs was presented in [8] and had a $O(|V|^4)$ runtime. Thereafter, more effective algorithms were found which ran in time $O(|V|^{\frac{1}{2}} \cdot |E|)$ [29] and $O(|V|^{2.5})$ [11].

The problems, which require to construct not just a maximum matching but one which satisfies some additional properties, are also of remarkable importance. The classical "Assignment problem" is such a one [9, 10, 23, 24, 25]. As another example, it can be noted that in a stage of Christofides's algorithm for solving "Traveling Salesman Problem" with the inequality of a triangle, construction of a minimum weighted perfect matching [9] in a complete graph on an even number of vertices is needed. For mentioned problems polynomial algorithms are found [2, 9, 10, 23-26].

Other problems which are related with the construction of special matchings are also considered in [5, 6, 12, 14, 15, 21, 30, 32, 33].

Nevertheless, the influence of a maximum matching of the given graph on the graph obtained by its removal is not sufficiently investigated. It is clear that graphs obtained from a graph by removing its two, different maximum matchings, may have different values of parameters such as maximum degree, chromatic index, cardinality of a minimum vertex covering of a graph. This approach leads to the problem of constructing a maximum matching of a graph which maximizes or minimizes the value of a parameter that interests us for the graph obtained from original one by removing that matching.

In this paper the cardinality of a maximum matching of a graph is taken as such a parameter. This kind of problem may have interest in games which can be modeled by the language of Graph Theory and in which one step of a player represents a removal of a maximum matching.

If a player knows that after her/his step, the right of the next step passes to her/his opponent, she/he will try to remove not just a maximum matching but one which removal leads to a graph with the smallest possible maximum matching. On the other hand, if the player knows that after her/his step the right of the next step passes to her/his partner, then, of course, she/he will try to find a maximum matching which removal leads to a graph with the largest possible maximum matching.

Present paper consists of four paragraphs. In §1 we give main notations and definitions, and formulate the main problems that will be investigated in the following paragraphs. In §2 we carry out the necessary theoretical investigation related to solution of the main problems for the class of trees. In §3 we consider a subclass of trees, in which the number of maximum matchings does not exceed the number of its vertices, which, of course, implies the polynomial solvability of the main problems in the subclass mentioned above. In §4 we present two polynomial algorithms which for a given tree construct a maximum matching which minimizes or maximizes the cardinality of a maximum matching of a graph obtained from the original one by its removal. Let us note that these algorithms are based on possibility of decomposition of an arbitrary tree into the trees described in §3. In the end of the paragraph, we prove the correctness of these algorithms and estimate their runtimes.

## §1. Main notations and definitions

The cardinality of a given set $A$ will be denoted by $|A|$. The sets of natural numbers and integers will be denoted by $N$ and $Z$, respectively. Set: $Z^+ \equiv N \cup \{0\}$.

We will consider finite undirected graphs without multiple edges or loops. The sets of vertices and edges of a graph $G$ will be denoted by $V(G)$ and $E(G)$, respectively. For a vertex $v \in V(G)$ let $\psi(v)$ be the set of edges incident with $v$, and let: $d_G(v) \equiv |\psi(v)|$. Assume:

$$\Delta(G) \equiv \max_{v \in V(G)} d_G(v).$$

In a connected graph $G$ the length of the shortest $u-v$ path will be denoted by $\rho(u,v)$, where $u,v \in V(G)$. For $w \in V(G)$ and $U \subseteq V(G)$ assume:

$$\rho(w,U) \equiv \min_{u \in U} \rho(w,u).$$

Let $G$ be a graph and let $v \in V(G)$, $V' \subseteq V(G)$, $e \in E(G)$ and $E' \subseteq E(G)$. Then the graphs obtained from $G$ by deleting the vertex $v$, the subset $V'$, the edge $e$ and the subset $E'$ will be denoted by $G-v$, $G \setminus V'$, $G-e$ and $G \setminus E'$, respectively.

Let $\tau(G)$ be the set of all subgraphs of the graph $G$.

Suppose $V' \subseteq V(G)$. Then $<V'>_G$ will be the subgraph of the graph $G$ induced by the subset $V' \subseteq V(G)$.

For a graph $G$ consider the following subset of $E(G)$:

$$\theta(G) \equiv \{(u,v) \in E(G) / d_G(u) \geq 2, d_G(v) \geq 2\},$$

and assume: $\overline{\theta}(G) \equiv E(G) \setminus \theta(G)$.

Denote:

$$V_{border}(G) \equiv \{v \in V(G) / \psi(v) \cap \theta(G) \neq \emptyset, \psi(v) \cap \overline{\theta}(G) \neq \emptyset\},$$

$$E_{deep}(G) \equiv \theta(G) \setminus \bigcup_{v \in V_{border}(G)} \psi(v).$$

The set of all matchings of a graph $G$ will be denoted by $B(G)$.

Set:

$$\beta(G) \equiv \max\{|H| / H \in B(G)\},$$

$$M(G,q) \equiv \{H \in B(G) / |H| = q\}, \text{ where } q \in Z^+, 0 \leq q \leq \beta(G),$$
$$M(G) \equiv M(G, \beta(G)).$$

For $F', F'' \in M(G)$ denote: $\rho(F', F'') \equiv |F' \setminus F''|$. If $F \in M(G)$ and $M \subseteq M(G)$, then assume:
$$\rho(F, M) \equiv \min_{F' \in M} \rho(F, F').$$

For $F, F' \in M(G)$ we will write $F \prec_1 F'$, if

(a) $\rho(F, F') = 1$,

(b) there is a simple path $u_0, (u_0, u_1), u_1, (u_1, u_2), u_2, (u_2, u_3), u_3, (u_3, u_4), u_4$ of the graph $G$ such that $(u_0, u_1) \in F$, $(u_2, u_3) \in F \setminus F'$, $(u_3, u_4) \in F' \setminus F$.

For $F, F' \in M(G)$ and $k \in N$ we will write $F \prec_k F'$, if there are $F_1, ..., F_{k-1} \in M(G)$, such that
$$F \prec_1 F_1, \; F_1 \prec_1 F_2, \; ..., \; F_{k-2} \prec_1 F_{k-1}, \; F_{k-1} \prec_1 F'.$$

And finally, for $F, F' \in M(G)$ we will write $F \prec F'$, if there is $k \in N$, $k \geq \rho(F, F')$ such that $F \prec_k F'$.

For an edge $e \in E(G)$ set:
$$M(e) \equiv \{F \in M(G) / e \in F\}, \text{ and suppose:}$$
$$\pi(G) \equiv \{e \in E(G) / M(e) \neq \varnothing\},$$
$$\chi(G) \equiv E_{deep}(G) \cap \pi(G).$$

Assume:
$$P(G) \equiv \{H \in B(G) / H \cap \theta(G) = \varnothing\}, \text{ and let}$$
$$\eta(G) \equiv \max\{|H| / H \in P(G)\},$$
$$P'(G) \equiv \{H \in P(G) / |H| = \eta(G)\},$$
$$M'(G) \equiv \{F \in M(G) / |F \cap \overline{\theta}(G)| = \eta(G)\}.$$

For a graph $G$ let $\omega(G)$ be a subset of $\tau(G)$ defined as:
$$\omega(G) \equiv \{G \setminus F / F \in M(G)\}, \text{ and let:}$$
$$l(G) \equiv \min\{\beta(H) / H \in \omega(G)\},$$
$$L(G) \equiv \max\{\beta(H) / H \in \omega(G)\}.$$

For a graph $G$ and $q \in Z^+$ ($l(G) \leq q \leq L(G)$) set:
$$S(G, q) \equiv \{F \in M(G) / \beta(G \setminus F) = q\}, \text{ and assume:}$$
$$S'(G, l(G)) \equiv S(G, l(G)) \cap M'(G).$$

For a subset $E_0 \subseteq E(G)$ of the set of edges of a graph $G$ consider the family of subgraphs of $G$ defined as:
$$\chi(G, E_0) \equiv \{H \in \tau(G) / E(H) = E_0\}.$$

Let $\psi(G, E_0)$ be the subgraph of the graph $G$ which belongs to $\chi(G, E_0)$ and contains minimum number of vertices.

For a bridge $e$ of the connected graph $G$ and $i = 1, 2$ let $G(i, e)$ be $\psi(G, E(G_i) \cup \{e\})$, where $G_1, G_2$ are the connected components of the graph $G - e$, and in this case we set:
$$\lambda(e) \equiv l(G(1, e)) + l(G(2, e)),$$
$$\Lambda(e) \equiv L(G(1, e)) + L(G(2, e)).$$

If $G$ is a tree, then denote:
$$V_G(0) \equiv \{v \in V(G) / d_G(v) = 1\},$$
and for $i \geq 1$ let
$$V_G(i) \equiv \{v \in V(G) / d_{G(i)}(v) = 1\}, \text{ where } G(i) \equiv G \setminus \bigcup_{j=0}^{i-1} V_G(j);$$
and let $k_G : V(G) \to Z^+$ be the mapping, which satisfies the condition:
$$v \in V_G(k_G(v)) \text{ for every vertex } v \in V(G).$$
We will consider the following problems:

**Problem 1.1**
**Condition:** Given a graph $G$ and a positive integer $k \in N$.
**Question:** Does there exist a matching $F_0(G) \in M(G)$, such that $\beta(G \setminus F_0(G)) \leq k$?

**Problem 1.2**
**Condition:** Given a graph $G$ and a positive integer $k \in N$.
**Question:** Does there exist a matching $F_0(G) \in M(G)$, such that $\beta(G \setminus F_0(G)) \geq k$?

In §4 two polynomial algorithms are described which solve optimization versions of **Problems 1.1** and **1.2** provided that the input graph is a tree.

Note that NP-completeness of **Problem 1.2** follows from [19], since in every cubic graph $G$ $L(G) = \dfrac{|V(G)|}{2}$ if and only if its chromatic index equals to three. It turns out that: **Problems 1.1** and **1.2** are NP-complete even for connected bipartite graphs in which maximum degree is three [22].

Non-defined terms and conceptions can be found in [18, 28, 35].

## §2. Theoretical investigation

**Lemma 2.1.** Let $G$ be a tree, $e \in \theta(G)$, $\Gamma \in P'(G)$ and $\beta(G) = \beta(G-e)$. If $F \in M'(G-e)$ and $\Gamma \subseteq F$, then $F \in M'(G)$.
**Proof** is evident.
**Lemma 2.2.** Let $G$ be a tree and $e \in \chi(G)$. If $F_i \in M'(G(i,e))$ $i=1,2$, then $(F_1 \cup F_2) \in M'(G)$.
**Proof** is evident.
**Lemma 2.3.** Let $G$ be a graph, $F \in M(G)$, $e' \in F$, $e \in \overline{\theta}(G)$, $e \notin F$ and $e'$ is adjacent to the edge $e$. Then the following inequality holds:
$$\beta(G \setminus ((F \setminus \{e'\}) \cup \{e\})) \leq \beta(G \setminus F).$$
**Proof.** Let $H_0 \in M(G \setminus ((F \setminus \{e'\}) \cup \{e\}))$. Define the set $H_1$ as follows:
$$H_1 \equiv \begin{cases} (H_0 \setminus \{e'\}) \cup \{e\} & \text{if } e' \in H_0, \\ H_0 & \text{if } e' \notin H_0. \end{cases}$$
It is clear that $|H_1| = |H_0|$ and $H_1 \in B(G \setminus F)$, therefore
$$\beta(G \setminus ((F \setminus \{e'\}) \cup \{e\})) = |H_0| = |H_1| \leq \beta(G \setminus F).$$
Proof of **Lemma 2.3** is complete.

**Corollary 2.1.** Let $G$ be a graph and $e \in \overline{\theta}(G)$. There is $F \in S(G,l(G))$, such that $e \in F$.

**Corollary 2.2.** Let $G$ be a graph, $\Gamma \in P'(G)$. There is $F \in S(G,l(G))$, such that $\Gamma \subseteq F$.

**Corollary 2.3.** For an arbitrary graph $G$ $S'(G,l(G)) \neq \emptyset$.

**Corollary 2.4.** If in an arbitrary graph $G$ $F \in M(G) \cap P'(G)$, then $F \in S'(G,l(G))$.

**Corollary 2.5.** Let $G$ be a graph, $\Gamma \in P'(G)$ and $e \in \chi(G)$. If there is $F \in S(G,l(G))$ with $e \in F$, then there is $F' \in S'(G,l(G))$, such that $e \in F'$ and $\Gamma \subseteq F'$.

**Corollary 2.6.** Let $G$ be a graph, $\Gamma \in P'(G)$ and $e \in \theta(G)$. If there is $F \in S(G,l(G))$ with $e \notin F$, then there is $F' \in S'(G,l(G))$, such that $e \notin F'$ and $\Gamma \subseteq F'$.

**Lemma 2.4.** Let $G$ be a tree, $F \in M(G)$, $F \notin S(G,l(G))$, $F' \in S(G,l(G))$ and $\rho(F,F') = \rho(F,S(G,l(G)))$. There is a simple path $u_0, (u_0,u_1), u_1, (u_1,u_2), u_2, (u_2,u_3), u_3, (u_3,u_4), u_4$ of $G$ satisfying the following two conditions:
  1) $(u_0,u_1) \in F$, $(u_2,u_3) \in F$, $(u_3,u_4) \in F'$;
  2) $\psi(u_4) \cap F = \emptyset$.

**Proof.** Since $F \notin S(G,l(G))$ and $F' \in S(G,l(G))$, there is $f_1 = (v_1,w_1) \in E(G)$, such that $f_1 \in F \setminus F'$. Therefore, there is $f_1' \in (\psi(w_1) \cup \psi(v_1))$, such that $f_1' \in F' \setminus F$. Without loss of generality we assume that $f_1' = (w_1,v_2)$.

**Case 1.** $\psi(v_2) \cap F = \emptyset$.

**Lemma 2.3** implies $d_G(v_1) \geq 2$, therefore there is $(w_0,v_1) \in E(G)$, such that $w_0 \neq w_1$. Since $F \in M(G)$, there is $(v_0,w_0) \in E(G)$, such that $(v_0,w_0) \in F$. It is not hard to see that the simple path

$$v_0, (v_0,w_0), w_0, (w_0,v_1), v_1, (v_1,w_1), w_1, (w_1,v_2), v_2$$

satisfies the statement of **Lemma**.

**Case 2.** $\psi(v_2) \cap F \neq \emptyset$.

There is $f_2 = (v_2,w_2) \in E(G)$, such that $f_2 \in F$. Consider a maximal simple path

$$x_0, (x_0,x_1), x_1, (x_1,x_2), x_2, ..., x_{2r-1}, (x_{2r-1},x_{2r}), x_{2r}$$

passing through vertex $v_1$ and satisfying conditions:

$$(x_{2i-2},x_{2i-1}) \in F, (x_{2i-1},x_{2i}) \in F' \text{ for } i=1,...,r.$$

It is not hard to see that the simple path

$$x_{2r-4}, (x_{2r-4},x_{2r-3}), x_{2r-3}, (x_{2r-3},x_{2r-2}), x_{2r-2}, (x_{2r-2},x_{2r-1}), x_{2r-1}, (x_{2r-1},x_{2r}), x_{2r}$$

satisfies the statement of **Lemma**. Proof of **Lemma 2.4** is complete.

**Lemma 2.5.** Let $G$ be a tree, $F \in M(G)$, $F \notin S(G,l(G))$, $F' \in S(G,l(G))$ and $\rho(F,F') = \rho(F,S(G,l(G)))$. Then $F \underset{\rho(F,F')}{\prec} F'$.

**Proof.** Suppose $\rho(F,F') = 1$. **Lemma 2.4** implies the existence of simple path

$$u_0, (u_0,u_1), u_1, (u_1,u_2), u_2, (u_2,u_3), u_3, (u_3,u_4), u_4$$

with $(u_0,u_1) \in F$, $(u_2,u_3) \in F$, $(u_3,u_4) \in F'$. As $\rho(F,F') = 1$, we get $(u_0,u_1) \in F'$, therefore $F \underset{1}{\prec} F'$.

Now assume the statement of **Lemma 2.5** to be true for the case $\rho(F,F') \leq t-1$ and let $\rho(F,F') = t \geq 2$. **Lemma 2.4** implies the existence of simple path

$$u_0, (u_0, u_1), u_1, (u_1, u_2), u_2, (u_2, u_3), u_3, (u_3, u_4), u_4$$

with $(u_0, u_1) \in F$, $(u_2, u_3) \in F$, $(u_3, u_4) \in F'$, $\psi(u_4) \cap F = \emptyset$. Define:
$$F_1 \equiv (F \setminus \{(u_2, u_3)\}) \cup \{(u_3, u_4)\}.$$

Clearly, $F_1 \in M(G)$, $F \underset{1}{\prec} F_1$ and $\rho(F_1, F') = \rho(F, F') - 1 \geq 1$. Let us show that
$$\rho(F_1, F') = \rho(F_1, S(G, l(G))).$$

It is clear that $\rho(F_1, S(G, l(G))) \leq \rho(F_1, F')$. On the other hand, for any $F'' \in S(G, l(G))$ we have
$$\rho(F, F'') \leq \rho(F, F_1) + \rho(F_1, F'') = 1 + \rho(F_1, F''),$$

and therefore
$$\rho(F_1, F'') \geq \rho(F, F'') - 1 \geq \rho(F, F') - 1 = \rho(F_1, F'),$$

thus
$$\rho(F_1, S(G, l(G))) \geq \rho(F_1, F') \text{ or } \rho(F_1, S(G, l(G))) = \rho(F_1, F').$$

This, particularly, implies that
$$\rho(F_1, F') = \rho(F_1, S(G, l(G))) = \rho(F, F') - 1 = t - 1 \geq 1 \text{ and } F_1 \notin S(G, l(G)).$$

By the hypothesis of induction, we have
$$F_1 \underset{t-1}{\prec} F', \text{ therefore } F \underset{t}{\prec} F'.$$

Proof of **Lemma 2.5** is complete.

**Theorem 2.1.** Let $G$ be a tree and $F \in M(G)$. $F \in S(G, l(G))$ if and only if for every $F' \in M(G)$ satisfying the condition $F \prec F'$, the inequality $\beta(G \setminus F') \geq \beta(G \setminus F)$ holds.

**Proof** follows from **Lemma 2.5**.

**Lemma 2.6.** Let $G$ be a graph and $F \in M(G)$, $F' \in M(G)$. The following inequality is true:
$$|\beta(G \setminus F) - \beta(G \setminus F')| \leq \rho(F, F').$$

**Proof.** Let $H' \in M(G \setminus F')$. Then
$$\beta(G \setminus F') = |H'| = |H' \cap F| + |H' \setminus F| \leq |F \setminus F'| + \beta(G \setminus F) = \rho(F, F') + \beta(G \setminus F).$$

Similarly, the inequality $\beta(G \setminus F) \leq \rho(F, F') + \beta(G \setminus F')$ can be proved. Proof of **Lemma 2.6** is complete.

**Corollary 2.7.** In an arbitrary graph $G$ the inequality $L(G) \leq 2l(G)$ holds.

**Example 2.1.** The following example shows that there are graphs $G$ for which the equality $L(G) = 2l(G)$ holds.

For any $k \in N$, consider the tree $G_k^{(2.1)}$ shown in the figure below:

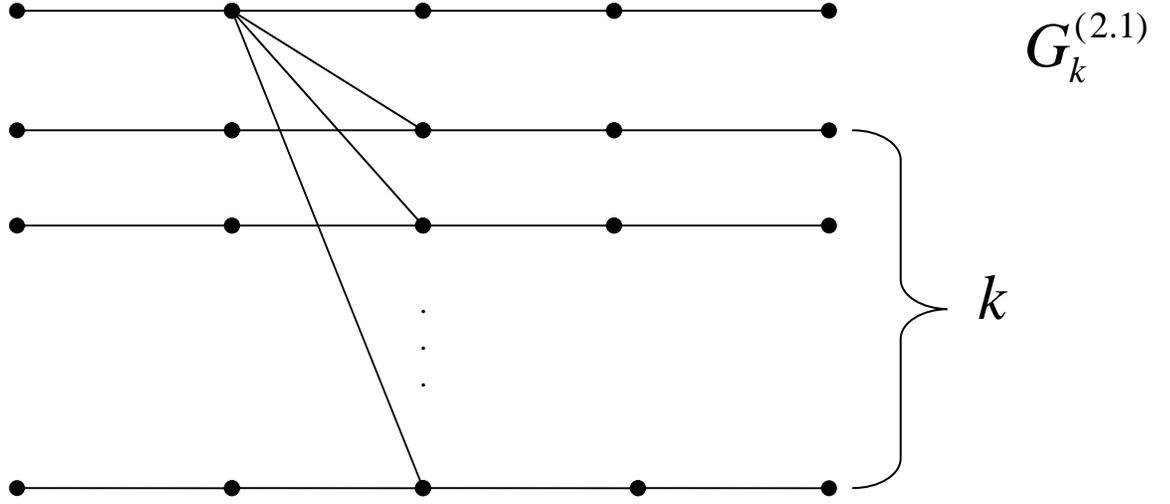

Figure 1

Note that $l(G_k^{(2.1)}) = k+1$, $L(G_k^{(2.1)}) = \beta(G_k^{(2.1)}) = 2k+2$.

**Lemma 2.7.** Let $G$ be a tree, $k \in N$, $l(G) \leq k \leq L(G)$. Then $S(G,k) \neq \varnothing$.

**Proof.** Clearly $S(G,l(G)) \neq \varnothing$, $S(G,L(G)) \neq \varnothing$. Let us show that for each $k \in N$, $l(G) < k < L(G)$,
$$S(G,k) \neq \varnothing.$$

Suppose $F_0 \in S(G,L(G))$, $\rho(F_0, S(G,l(G))) \equiv \rho_0$. Let $F_{\rho_0}$ be a matching of $G$ satisfying
$$F_{\rho_0} \in S(G,l(G)), \ \rho(F_0, F_{\rho_0}) = \rho_0.$$

**Lemma 2.5** implies $F_0 \prec_{\rho_0} F_{\rho_0}$. Therefore, there are $F_1 \in M(G), ..., F_{\rho_0-1} \in M(G)$ such that
$$F_0 \prec_1 F_1, \ F_1 \prec_1 F_2, \ ..., \ F_{\rho_0-2} \prec_1 F_{\rho_0-1} \ F_{\rho_0-1} \prec_1 F_{\rho_0}.$$

From **Lemma 2.6** we conclude that for $i = 0, ..., \rho_0 - 1$
$$|\beta(G \setminus F_i) - \beta(G \setminus F_{i+1})| \leq 1.$$

As $\beta(G \setminus F_{\rho_0}) < k < \beta(G \setminus F_0)$, there is $F \in \{F_1, ..., F_{\rho_0-1}\}$ such that $\beta(G \setminus F) = k$. Proof of **Lemma 2.7** is complete.

**Example 2.2.** It turns out that there are graphs $G$, for which $S(G,k) = \varnothing$ for every $k \in N$, $l(G) < k < L(G)$.

For $n \in N$, $n \geq 2$ consider the graph $G_n^{(2.2)}$ shown in the figure below:

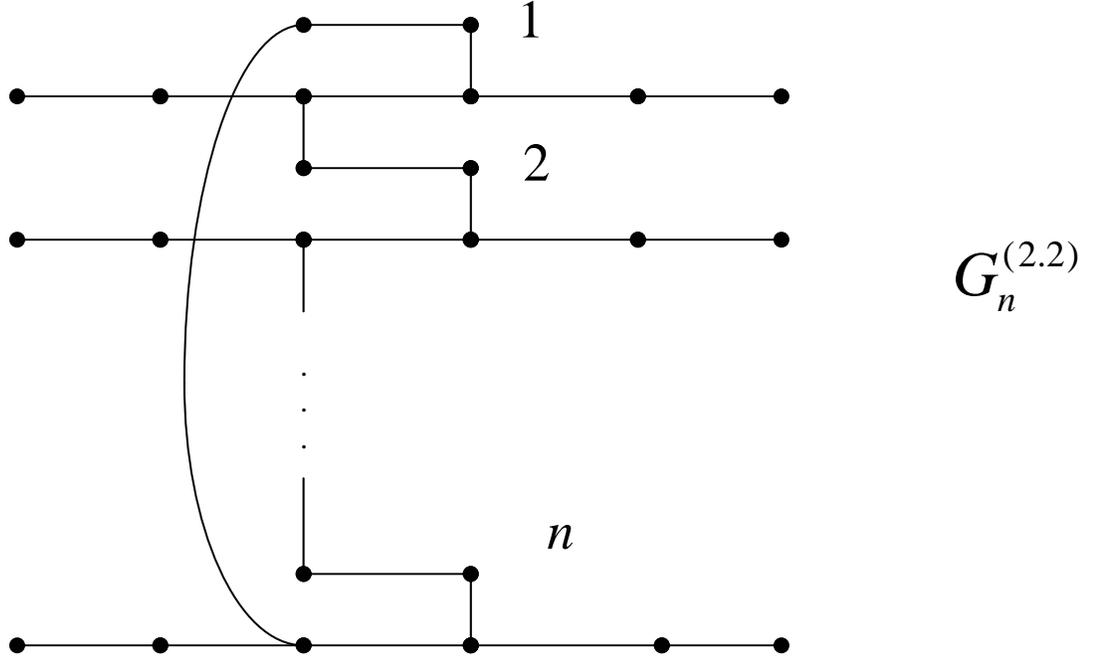

Figure 2

Note that $G_n^{(2.2)}$ contains exactly two perfect matchings $F_1^{(n)}$ and $F_2^{(n)}$, moreover, $l(G_n^{(2.2)}) = \beta(G_n^{(2.2)} \setminus F_1^{(n)}) = 2n$, $L(G_n^{(2.2)}) = \beta(G_n^{(2.2)} \setminus F_2^{(n)}) = 3n$.

**Lemma 2.8.** Let $G$ be a graph, $k \in N$, $l(G) \leq k \leq L(G)$. If $S(G,k) \neq \varnothing$ then for each $e_0 \in \overline{\theta}(G)$ there is $F \in S(G,k)$ such that $e_0 \in F$.

**Proof.** Let $e_0 = (u_0, v_0) \in \overline{\theta}(G)$, $d_G(u_0) = 1$ and $\widetilde{F} \in S(G,k)$. If $e_0 \in \widetilde{F}$ then **Lemma 2.8** is proved, therefore, without loss of generality, we may assume $e_0 \notin \widetilde{F}$. Then, there is $H_0 \in M(G \setminus \widetilde{F})$, such that $e_0 \in H_0$. Clearly $H_0 \in B(G)$. Consider a maximal alternating path

$$u_0, e_0, v_0, f_0, u_1, e_1, v_1, \ldots, u_t, e_t, v_t, f_t, u_{t+1}$$

of $G$, satisfying conditions

$$e_i = (u_i, v_i),\ f_i = (v_i, u_{i+1})\ \text{for}\ i = 0, \ldots, t,\ \{e_0, \ldots, e_t\} \subseteq H_0,\ \{f_0, \ldots, f_t\} \subseteq \widetilde{F}.$$

For $i = 0, \ldots, t$ consider a set $F_i$ defined as

$$F_i \equiv (\widetilde{F} \setminus \{f_0, \ldots, f_i\}) \cup \{e_0, \ldots, e_i\},$$

and let

$$H \equiv (H_0 \setminus \{e_0, \ldots, e_t\}) \cup \{f_0, \ldots, f_t\}.$$

Clearly $F_i \in M(G)$ $i = 0, \ldots, t$, $H \in B(G \setminus F_t)$ and $|H| = |H_0|$. **Lemma 2.3** implies that

$$\beta(G \setminus F_0) \leq \beta(G \setminus \widetilde{F}) = k.$$

On the other hand,

$$k = \beta(G \setminus \widetilde{F}) = |H_0| = |H| \leq \beta(G \setminus F_t).$$

Since for $i = 0, \ldots, t-1$ $\rho(F_i, F_{i+1}) = 1$, we get

$$|\beta(G\setminus F_i)-\beta(G\setminus F_{i+1})|\leq 1.$$

Therefore, there is $i_0\in Z^+, 0\leq i_0\leq t$, such that $\beta(G\setminus F_{i_0})=k$. As $e_0\in F_i$ for $i=0,...,t$, the proof of **Lemma 2.8** is complete.

**Corollary 2.8.** Let $G$ be a tree, $k\in N$, $l(G)\leq k\leq L(G)$ and $e_0\in\overline{\theta}(G)$. There is $F\in S(G,k)$ such that $e_0\in F$.

**Corollary 2.9.** Let $G$ be a graph, $e_0\in\overline{\theta}(G)$, $F\in S(G,L(G))$, $H\in M(G\setminus F)$, $e_0\in H$. If
$$u_0,e_0,v_0,f_0,u_1,e_1,v_1,...,u_t,e_t,v_t,f_t,u_{t+1}$$
is a maximal alternating path of $G$, satisfying conditions
$$e_i=(u_i,v_i),\ f_i=(v_i,u_{i+1})\text{ for }i=0,...,t,\ \{e_0,...,e_t\}\subseteq H,\ \{f_0,...,f_t\}\subseteq F.$$
then
$$((F\setminus\{f_0,...,f_t\})\cup\{e_0,...,e_t\})\in S(G,L(G)).$$

**Lemma 2.9.** Let $G$ be a tree and $e\in\pi(G)$. Then $\lambda(e)\geq l(G)$.

**Proof. Corollary 2.8** implies the existence of $F^{(1)}\in S(G(1,e),l(G(1,e)))$, $F^{(2)}\in S(G(2,e),l(G(2,e)))$ with $e\in F^{(1)}$, $e\in F^{(2)}$. Consider $\widetilde{F}\in B(G)$ defined as $\widetilde{F}\equiv F^{(1)}\cup F^{(2)}$. As $e\in\pi(G)$, we get $\widetilde{F}\in M(G)$, therefore
$$\lambda(e)=l(G(1,e))+l(G(2,e))=\beta(G(1,e)\setminus F^{(1)})+\beta(G(2,e)\setminus F^{(2)})=\beta(G\setminus\widetilde{F})\geq l(G).$$
Proof of **Lemma 2.9** is complete.

**Example 2.3.** The following example shows that the statement of **Lemma 2.9** may not be true if $e\notin\pi(G)$.

For every $k\in N$ consider the tree $G_k^{(2.3)}$ and its edge $e_k$ shown in the figure below:

Figure 3

Note that $l(G_k^{(2.3)}) = 3k+3$, $\lambda(e_k) = 2k+3$ and $l(G_k^{(2.3)}) - \lambda(e_k) = k$.

**Lemma 2.10.** Let $G$ be a tree, $e \in E(G)$, $F \in S(G, l(G))$ and $e \in F$. Then
$$(F \cap E(G(i,e))) \in S(G(i,e), l(G(i,e))) \text{ for } i = 1, 2.$$

**Proof.** Denote $F_i \equiv F \cap E(G(i,e))$, $i = 1, 2$. Clearly, for $i = 1, 2$ $F_i \in M(G(i,e))$. Let us show that $F_1 \in S(G(1,e), l(G(1,e)))$. Assume, on the contrary, that $F_1 \notin S(G(1,e), l(G(1,e)))$. **Corollary 2.8** implies that there is $F_1^{'} \in S(G(1,e), \beta(G(1,e) \setminus F_1) - 1)$ such that $e \in F_1^{'}$. Define $F^{'} \in B(G)$ as:
$$F^{'} \equiv (F \setminus F_1) \cup F_1^{'}.$$

Note that $F^{'} \in M(G)$ and
$$\beta(G \setminus F^{'}) = \beta(G(1,e) \setminus F_1^{'}) + \beta(G(2,e) \setminus F_2) = \beta(G(1,e) \setminus F_1) + \beta(G(2,e) \setminus F_2) - 1 = \beta(G \setminus F) - 1$$
which contradicts $F \in S(G, l(G))$, therefore $F_1 \in S(G(1,e), l(G(1,e)))$. Similarly, it can be shown that $F_2 \in S(G(2,e), l(G(2,e)))$. Proof of **Lemma 2.10** is complete.

**Lemma 2.11.** Let $G$ be a tree and $e \in E(G)$. If there is $F \in S(G, l(G))$ with $e \in F$, then $\lambda(e) = l(G)$.

**Proof.** As $F \in S(G, l(G))$ then, using **Lemma 2.10**, we get:
$$l(G) = \beta(G \setminus F) = \beta(G(1,e) \setminus (F \cap E(G(1,e)))) + \beta(G(2,e) \setminus (F \cap E(G(2,e)))) =$$
$$= l(G(1,e)) + l(G(2,e)) = \lambda(e).$$

Proof of **Lemma 2.11** is complete.

**Corollary 2.10.** Let $G$ be a tree with $\chi(G) \neq \emptyset$. Then $\min_{e \in \chi(G)} \lambda(e) = l(G)$.

**Proof. Lemma 2.9** implies that $\min_{e \in \chi(G)} \lambda(e) \geq l(G)$. As $\chi(G) \neq \emptyset$ then there is $e \in E_{deep}(G)$ and $F \in M(G)$, for which $e \in F$. Without loss of generality we can assume that $F \in M^{'}(G)$. This implies that $\eta(G) < \beta(G)$. Consider $\widetilde{F} \in S^{'}(G, l(G))$ (**Corollary 2.3**). Since $\eta(G) < \beta(G)$ there is $\tilde{e} \in \widetilde{F}$ for which $\tilde{e} \in \chi(G)$. **Lemma 2.11** implies that $\lambda(\tilde{e}) = l(G)$. **Corollary 2.10** is proved.

**Lemma 2.12.** Let $G$ be a tree, $e_0 \in \chi(G)$ and $\lambda(e_0) = l(G)$.

1. If for $i = 1, 2$ $F_i \in S(G(i, e_0), l(G(i, e_0)))$ and $e_0 \in F_i$, then $(F_1 \cup F_2) \in S(G, l(G))$;
2. If $\Gamma \in P^{'}(G)$ and for $i = 1, 2$ $F_i \in S^{'}(G(i, e_0), l(G(i, e_0)))$, $(\Gamma \cap E(G(i, e_0))) \subset F_i$ then $(F_1 \cup F_2) \in S^{'}(G, l(G))$ and $\Gamma \subset (F_1 \cup F_2)$.

**Proof.** 1. It is not hard to see that $(F_1 \cup F_2) \in M(G)$. As
$$\beta(G \setminus (F_1 \cup F_2)) = \beta(G(1, e_0) \setminus F_1) + \beta(G(2, e_0) \setminus F_2) = l(G(1, e_0)) + l(G(2, e_0)) = \lambda(e_0) = l(G)$$
then $(F_1 \cup F_2) \in S(G, l(G))$.

2. Proof of the statement follows from **Lemma 2.2** and an easily checkable inclusion $\Gamma \subset (F_1 \cup F_2)$. Proof of **Lemma 2.12** is complete.

**Corollary 2.11.** Let $G$ be a tree, $e_0 \in \chi(G)$ and suppose that there is $F \in S(G, l(G))$ such that $e_0 \in F$.

1. If for $i = 1, 2$ $F_i \in S(G(i, e_0), l(G(i, e_0)))$ and $e_0 \in F_i$, then $(F_1 \cup F_2) \in S(G, l(G))$;
2. If $\Gamma \in P^{'}(G)$ and for $i = 1, 2$ $F_i \in S^{'}(G(i, e_0), l(G(i, e_0)))$, $(\Gamma \cap E(G(i, e_0))) \subset F_i$ then $(F_1 \cup F_2) \in S^{'}(G, l(G))$ and $\Gamma \subset (F_1 \cup F_2)$.

**Proof** follows from **Lemma 2.11** and **Lemma 2.12**.

**Lemma 2.13.** Let $G$ be a tree, $\Gamma \in P^{'}(G)$, $e, f$ be adjacent edges of $G$ satisfying the inequality $\lambda(e) \geq \lambda(f)$.

(a) If $f \in \pi(G)$, then there is $F \in S(G,l(G))$, such that $e \notin F$.

(b) If $e \in \theta(G)$, $f \in \pi(G)$, then there is $F \in S'(G,l(G))$, such that $\Gamma \subseteq F$ and $e \notin F$.

**Proof.** (a) Choose $F_1 \in S(G,l(G))$. If $e \notin F_1$ then **Lemma 2.13** is proved, therefore we may assume that $e \in F_1$. **Lemma 2.11** implies that $\lambda(e) = l(G)$, and, due to **Lemma 2.9**, we get $\lambda(f) = l(G)$, therefore, using **Lemma 2.12**, we conclude that there is $F \in S(G,l(G))$ such that $f \in F$. Clearly $e \notin F$.

(b) Proof follows from (a) and **Corollary 2.6**.
Proof of **Lemma 2.13** is complete.

**Lemma 2.14.** Let $G$ be a tree, $U = \{u_0, u_1, u_2, u_3\} \subseteq V(G)$ and $d_G(u_0) = 1$, $d_G(u_1) = 2$, $d_G(u_3) \geq 2$, $(u_{i-1}, u_i) \in E(G)$ for $i = 1,2,3$. Let $e = (u_2, u_3)$, and suppose that there is $\widetilde{F} \in S(G,l(G))$ such that $e \notin \widetilde{F}$ (figure 4). Then
(a) $l(G) = l(G-e)$;
(b) If $F \in S'(G-e, l(G-e))$, then $F \in S'(G, l(G))$.

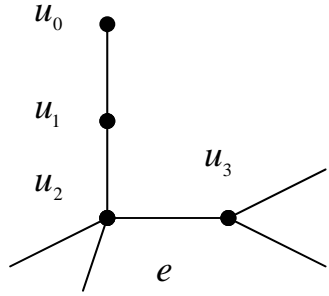

Figure 4

**Proof.** (a) First of all note that $\widetilde{F} \in M(G-e)$ and
$$l(G) = \beta(G \setminus \widetilde{F}) \geq \beta(G-e \setminus \widetilde{F}) \geq l(G-e).$$
On the other hand, for any $F_0 \in S(G-e, l(G-e))$ with $(u_0, u_1) \in F_0$ (**Corollary 2.1**), we have $F_0 \in M(G)$ and
$$l(G) \leq \beta(G \setminus F_0) = \beta(G-e \setminus F_0) = l(G-e),$$
therefore
$$l(G) = l(G-e).$$
(b) Note that $(u_0, u_1) \in F$, and as $d_G(u_3) \geq 2$, we have $F \in M'(G)$. Since
$$\beta(G \setminus F) = \beta(G-e \setminus F) = l(G-e) = l(G),$$
we imply $F \in S'(G, l(G))$. Proof of **Lemma 2.14** is complete.

**Corollary 2.12.** Let $G$ be a tree, $U = \{u_0, ..., u_4\} \subseteq V(G)$ and $d_G(u_0) = d_G(u_4) = 1$, $d_G(u_2) = d_G(u_3) = 2$, $(u_{i-1}, u_i) \in E(G)$ for $i = 1,...,4$ (figure 5). Then the equality $l(G) = l(G \setminus \{u_2, u_3, u_4\}) + 1$ holds.

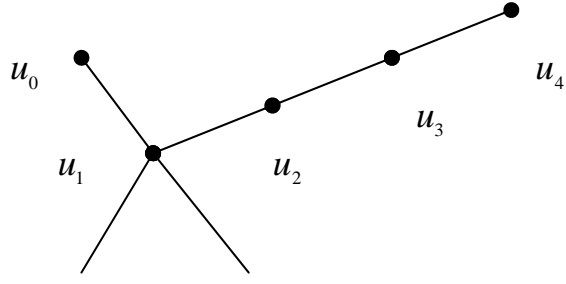

Figure 5

**Lemma 2.15.** Let $G$ be a tree, $U = \{u_0,...,u_4\} \subseteq V(G)$ and $d_G(u_0) = 1$, $d_G(u_1) = d_G(u_3) = 2$, $(u_{i-1}, u_i) \in E(G)$ for $i = 1,...,4$. Let $f = (u_2, u_3)$, $e = (u_3, u_4)$ and suppose that there is $\widetilde{F} \in S(G, l(G))$ such that $e \notin \widetilde{F}$ (figure 6). Then $f \in \pi(G)$ and $\lambda(f) = l(G)$.

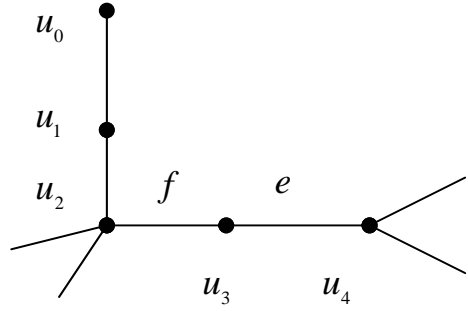

Figure 6

**Proof.** Clearly $f \in \pi(G)$. Let us show that there is $F \in S(G, l(G))$ such that $f \in F$. Suppose $f \notin \widetilde{F}$. Since $\widetilde{F} \in M(G)$ and $e \notin \widetilde{F}$, we imply that $(u_0, u_1) \in \widetilde{F}$ and there is $w \in V(G)$, $w \notin \{u_1, u_3\}$ such that $(u_2, w) \in \widetilde{F}$. Set:

$$F \equiv (\widetilde{F} \setminus \{(u_2, w)\}) \cup \{f\}.$$

Clearly, $F \in M(G)$ and $\beta(G \setminus F) = \beta(G \setminus \widetilde{F}) = l(G)$, therefore $F \in S(G, l(G))$ and, due to **Lemma 2.11**, $\lambda(f) = l(G)$. Proof of **Lemma 2.15** is complete.

**Lemma 2.16.** Let $G$ be a tree, $U = \{u_0,...,u_3\} \subseteq V(G)$ and $d_G(u_1) = d_G(u_2) = 2$, $(u_{i-1}, u_i) \in E(G)$ for $i = 1, 2, 3$. Let $g = (u_0, u_1)$, $f = (u_1, u_2)$, $e = (u_2, u_3)$ and assume that $u_3 \in V(G(2, g))$, $u_0 \in V(G(1, f))$, $u_0 \in V(G(1, e))$. Suppose that there is $\widetilde{F} \in S(G, l(G))$ such that $e \notin \widetilde{F}$ (figure 7). Then

(a) if $\beta(G(1, f)) = 1 + \beta(G(1, g))$ then $\lambda(f) = l(G)$;

(b) if $\beta(G(1, f)) = \beta(G(1, g))$ then
$$l(G(1, f)) \leq 1 + l(G(1, g)) \text{ and } l(G) = \min\{\lambda(g), \lambda(f)\};$$

(c) if $\beta(G(1, f)) = \beta(G(1, g))$ and $l(G(1, f)) = 1 + l(G(1, g))$ then $\lambda(g) = l(G)$;

(d) if $\beta(G(1, f)) = \beta(G(1, g))$ and $l(G(1, f)) \leq l(G(1, g))$ then $\lambda(f) = l(G)$.

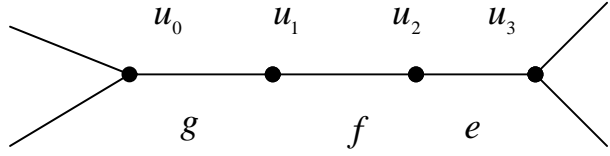

Figure 7

**Proof.** (a) Since $e \notin \widetilde{F}$ and $\beta(G(1,f)) = 1 + \beta(G(1,g))$, then $f \in \widetilde{F}$, hence, due to **Lemma 2.11**, we have $\lambda(f) = l(G)$.

(b) Choose $F \in S(G(1,g), l(G(1,g)))$ with $g \in F$ (**Corollary 2.8**). Since $\beta(G(1,f)) = \beta(G(1,g))$, we have $F \in M(G(1,f))$, and therefore
$$l(G(1,f)) \leq \beta(G(1,f) \setminus F) = 1 + \beta(G(1,g) \setminus F) = 1 + l(G(1,g)).$$
Since $\widetilde{F} \in M(G)$, we have $\widetilde{F} \cap \{g, f\} \neq \varnothing$, and therefore, due to **Lemma 2.9** and **Lemma 2.11**,
$$l(G) = \min\{\lambda(g), \lambda(f)\}.$$

(c) **Lemma 2.11** implies that it suffices to show that there is $F \in S(G, l(G))$ with $g \in F$. Suppose $g \notin \widetilde{F}$, then $f \in \widetilde{F}$. Choose $F_1 \in S(G(1,g), l(G(1,g)))$ with $g \in F_1$ (**Corollary 2.8**). Set:
$$F \equiv (\widetilde{F} \setminus (\widetilde{F} \cap E(G(1,f)))) \cup F_1.$$
As $\beta(G(1,f)) = \beta(G(1,g))$, we have $F \in M(G)$, therefore
$$\beta(G \setminus F) = l(G(1,g)) + \beta(G(2,g) \setminus (F \cap E(G(2,g)))) \leq l(G(1,f)) - 1 +$$
$$+ \beta(G(2,f) \setminus (\widetilde{F} \cap E(G(2,f)))) + 1 = l(G(1,f)) + l(G(2,f)) = l(G),$$
and $F \in S(G, l(G))$.

(d) **Lemma 2.11** implies that it suffices to show that there is $F \in S(G, l(G))$ with $f \in F$. Suppose $f \notin \widetilde{F}$, then $g \in \widetilde{F}$. Choose $F_1 \in S(G(1,f), l(G(1,f)))$ with $f \in F_1$ (**Corollary 2.8**). Set:
$$F \equiv (\widetilde{F} \setminus (\widetilde{F} \cap E(G(1,f)))) \cup F_1.$$
As $\beta(G(1,f)) = \beta(G(1,g))$, we have $F \in M(G)$, therefore
$$\beta(G \setminus F) = l(G(1,f)) + \beta(G(2,f) \setminus (F \cap E(G(2,f)))) \leq l(G(1,g)) +$$
$$+ \beta(G(2,g) \setminus (\widetilde{F} \cap E(G(2,g)))) = l(G),$$
and $F \in S(G, l(G))$. Proof of **Lemma 2.16** is complete.

**Lemma 2.17.** Let $G$ be a graph and $e \in \overline{\theta}(G)$. Then
  (a) if $\beta(G) = 1 + \beta(G - e)$ then $l(G) \geq l(G - e)$;
  (b) if $\beta(G) = \beta(G - e)$ then $l(G) \leq 1 + l(G - e)$.

**Proof.** (a) Let $F \in S(G, l(G))$. As $\beta(G) = 1 + \beta(G - e)$ then $e \in F$, therefore $(F \setminus \{e\}) \in M(G - e)$, and
$$l(G) = \beta(G \setminus F) = \beta(G - e \setminus (F \setminus \{e\})) \geq l(G - e).$$

(b) Let $F_1 \in S(G - e, l(G - e))$. As $\beta(G) = \beta(G - e)$ we get $F_1 \in M(G)$. Take $H \in M(G \setminus F_1)$ such that $e \in H$. We have
$$l(G) \leq \beta(G \setminus F_1) = |H| = 1 + |H \cap E(G - e)| \leq 1 + \beta(G - e \setminus F_1) = 1 + l(G - e).$$
Proof of **Lemma 2.17** is complete.

**Lemma 2.18.** Let $G$ be a tree, $e = (u,v) \in \overline{\theta}(G)$ and $d_G(u) = 1$ (figure 8). The following inequality holds:
$$l(G \setminus \{u,v\}) \leq l(G) \leq 1 + l(G \setminus \{u,v\}).$$

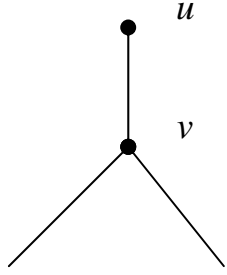

Figure 8

**Proof.** Clearly $\beta(G) = 1 + \beta(G \setminus \{u,v\})$. Choose $F \in S(G \setminus \{u,v\}, l(G \setminus \{u,v\}))$. Set $\widetilde{F} \equiv \{e\} \cup F$ and let $\widetilde{H} \in M(G \setminus \widetilde{F})$. It is not hard to see that $\widetilde{F} \in M(G)$ and
$$l(G) \leq \beta(G \setminus \widetilde{F}) = |\widetilde{H}| \leq 1 + |\widetilde{H} \cap E((G \setminus \{u,v\}) \setminus F)| \leq 1 + \beta((G \setminus \{u,v\}) \setminus F) = 1 + l(G \setminus \{u,v\}).$$

On the other hand, due to **Corollary 2.8**, there is $F' \in S(G, l(G))$ with $e \in F'$. Note that $(F' \setminus \{e\}) \in M(G \setminus \{u,v\})$, therefore
$$l(G) = \beta(G \setminus F') \geq \beta((G \setminus \{u,v\}) \setminus (F' \setminus \{e\})) \geq l(G \setminus \{u,v\}).$$

Proof of **Lemma 2.18** is complete.

**Lemma 2.19.** Let $G$ be a tree, $U = \{u_0, ..., u_3\} \subseteq V(G)$, $d_G(u_0) = 1$, $d_G(u_1) = d_G(u_2) = 2$, $(u_{i-1}, u_i) \in E(G)$ for $i = 1, 2, 3$ (figure 9). The following inequality holds:
$$1 + l(G \setminus U) \leq l(G) \leq 2 + l(G \setminus U).$$

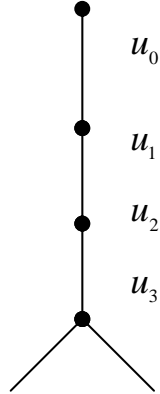

Figure 9

**Proof.** Applying **Lemma 2.18** two times, we get
$$l(G) \leq 1 + l(G \setminus \{u_0, u_1\}) \leq 2 + l(G \setminus U).$$

On the other hand, due to **Corollary 2.8**, there is $F \in S(G, l(G))$ such that $(u_0, u_1) \in F$. Consider the following two cases:

**Case 1.** $(u_2, u_3) \in F$. Note that $(F \setminus \{(u_0, u_1)\}) \in M(G \setminus \{u_0, u_1\})$, therefore, using **Lemma 2.10** and **Lemma 2.18**, we get:
$$l(G) = \beta(G \setminus F) = 1 + \beta((G \setminus \{u_0, u_1\}) \setminus (F \setminus \{(u_0, u_1)\})) = 1 + l(G \setminus \{u_0, u_1\}) \geq 1 + l(G \setminus U).$$

**Case 2.** $(u_2, u_3) \notin F$. Since $F \in M(G)$, there is $v \in V(G)$, $v \notin U$, such that $(u_3, v) \in F$. Consider the trees $G(1, (u_3, v))$ and $G(2, (u_3, v))$, where, without loss of generality, it is assumed that $u_0 \in V(G(1, (u_3, v)))$. Note that
$$\beta(G) = 2 + \beta(G \setminus U) = 2 + \beta(G(2, (u_3, v)) - u_3) + \beta(G(1, (u_3, v)) \setminus (U \cup \{v\})) \text{ and}$$
$$\beta(G) = \beta(G(1, (u_3, v))) + \beta(G(2, (u_3, v))) - 1 = 1 + \beta(G(2, (u_3, v))) + \beta(G(1, (u_3, v)) \setminus (U \cup \{v\})),$$
therefore
$$\beta(G(2, (u_3, v))) = 1 + \beta(G(2, (u_3, v)) - u_3),$$
thus, using **Lemma 2.17**, we get
$$l(G(2, (u_3, v))) \geq l(G(2, (u_3, v)) - u_3).$$

**Lemma 2.10** and **Lemma 2.18** imply that
$$l(G) = \beta(G \setminus F) = l(G(1, (u_3, v))) + l(G(2, (u_3, v))) \geq$$
$$\geq 1 + l(G(1, (u_3, v)) \setminus (U \cup \{v\})) + l(G(2, (u_3, v)) - u_3) = 1 + l(G \setminus U).$$

Proof of **Lemma 2.19** is complete.

**Lemma 2.20.** Let $G$ be a tree, $U = \{u_0, u_1, u_2\} \subseteq V(G)$ and $d_G(u_0) = d_G(u_2) = 1$, $(u_{i-1}, u_i) \in E(G)$ for $i = 1, 2$ (figure 10). Then
  (a) $l(G) = l(G \setminus U) + 1$;
  (b) If $F' \in S(G \setminus U, l(G \setminus U))$, then $F \equiv (F' \cup \{(u_0, u_1)\}) \in S(G, l(G))$;
  (c) If $\Gamma \in P'(G)$, $(u_0, u_1) \in \Gamma$, $F' \in S'(G \setminus U, l(G \setminus U))$ and $\Gamma \cap E(G \setminus U) \subseteq F'$, then $\overline{F} \equiv (F' \cup \{(u_0, u_1)\}) \in S'(G, l(G))$ and $\Gamma \subseteq \overline{F}$.

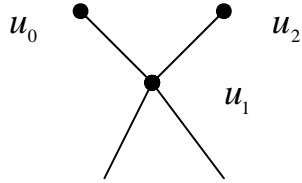

Figure 10

**Proof.** (a) **Lemma 2.18** implies that $l(G) \leq l(G \setminus U) + 1$. On the other hand, since $d_G(u_0) = 1$, then, due to **Corollary 2.8**, there is $\widetilde{F} \in S(G, l(G))$ such that $(u_0, u_1) \in \widetilde{F}$. As $d_{G \setminus \widetilde{F}}(u_2) = 1$, we imply that there is $H \in M(G \setminus \widetilde{F})$, such that $(u_1, u_2) \in H$. Note that
$$(\widetilde{F} \setminus (u_0, u_1)) \in M(G \setminus U) \text{ and } (H \setminus (u_1, u_2)) \in M((G \setminus U) \setminus (\widetilde{F} \setminus (u_0, u_1))),$$
therefore
$$l(G) = \beta(G \setminus \widetilde{F}) = |H| = 1 + |H \cap E(G \setminus U)| = 1 + \beta((G \setminus U) \setminus (\widetilde{F} \setminus \{(u_0, u_1)\})) \geq 1 + l(G \setminus U),$$
or
$$l(G) = l(G \setminus U) + 1.$$

(b) As $d_{G\setminus F}(u_2)=1$, we imply that there is $\widetilde{H}\in M(G\setminus F)$, such that $(u_1,u_2)\in\widetilde{H}$. It is not hard to see that
$$\beta(G\setminus F)=|\widetilde{H}|=1+|\widetilde{H}\cap E((G\setminus U)\setminus F')|=1+\beta((G\setminus U)\setminus F')=1+l(G\setminus U),$$
therefore, due to (a),
$$F\in S(G,l(G)).$$
(c) Clearly $\Gamma\subseteq\overline{F}$ and $\overline{F}\in M'(G)$, therefore, due to (b), $\overline{F}\in S'(G,l(G))$. Proof of **Lemma 2.20** is complete.

**Lemma 2.21.** Let $G$ be a tree, $U=\{u_0,...,u_3\}\subseteq V(G)$ and $d_G(u_0)=d_G(u_3)=1$, $d_G(u_2)=2$, $(u_{i-1},u_i)\in E(G)$ for $i=1,2,3$ (figure 11). Then
  (a) $l(G)=l(G\setminus U)+1$;
  (b) If $F'\in S(G\setminus U,l(G\setminus U))$, then $F\equiv(F'\cup\{(u_0,u_1),(u_2,u_3)\})\in S(G,l(G))$;
  (c) If $\Gamma\in P'(G)$, $(u_0,u_1)\in\Gamma$, $F'\in S'(G\setminus U,l(G\setminus U))$ and $\Gamma\cap E(G\setminus U)\subseteq F'$, then $\overline{F}\equiv(F'\cup\{(u_0,u_1),(u_2,u_3)\})\in S'(G,l(G))$ and $\Gamma\subseteq\overline{F}$.

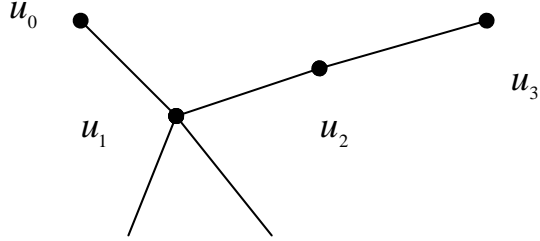

Figure 11

**Proof.** (a) Note that $(u_1,u_2)\notin\pi(G)$, therefore $\beta(G)=1+\beta(G-u_3)$. Due to **Lemma 2.17** and **Lemma 2.20**, we have
$$l(G)\geq l(G-u_3)=1+l(G\setminus U).$$
Now, let us show that $l(G)\leq 1+l(G\setminus U)$. Choose $F_0\in S(G\setminus U,l(G\setminus U))$. Since $\beta(G)=2+\beta(G\setminus U)$, then $(F_0\cup\{(u_0,u_1),(u_2,u_3)\})\in M(G)$ and
$$l(G)\leq\beta(G\setminus(F_0\cup\{(u_0,u_1),(u_2,u_3)\}))=1+\beta((G\setminus U)\setminus F_0)=1+l(G\setminus U),$$
therefore
$$l(G)=l(G\setminus U)+1.$$
(b) Since $\beta(G)=2+\beta(G\setminus U)$, then $F\in M(G)$. It is not hard to see that
$$\beta(G\setminus F)=1+\beta((G\setminus U)\setminus F')=1+l(G\setminus U)=l(G),$$
therefore $F\in S(G,l(G))$.
(c) Clearly $\Gamma\subseteq\overline{F}$ and $\overline{F}\in M'(G)$, therefore, due to (b), $\overline{F}\in S'(G,l(G))$. Proof of **Lemma 2.21** is complete.

**Lemma 2.22.** Let $G$ be a tree, $U=\{u_0,...,u_5\}\subseteq V(G)$, $d_G(u_0)=1$, $d_G(u_1)=d_G(u_2)=$ $=d_G(u_3)=d_G(u_4)=2$, $(u_{i-1},u_i)\in E(G)$ for $i=1,...,5$ (figure 12). Then the following inequality holds:
$$2+l(G\setminus U)\leq l(G)\leq 3+l(G\setminus U).$$

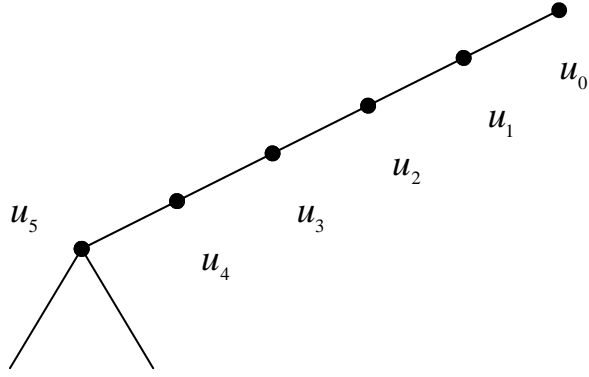

Figure 12

**Proof.** Applying **Lemma 2.18** and **Lemma 2.19**, we get
$$l(G) \leq 1 + l(G \setminus \{u_0, u_1\}) \leq 3 + l(G \setminus U).$$
On the other hand, due to **Corollary 2.8**, there is $F \in S(G, l(G))$ such that $(u_0, u_1) \in F$. Consider the following two cases:

**Case 1.** $(u_2, u_3) \in F$. Note that $(F \setminus \{(u_0, u_1)\}) \in M(G \setminus \{u_0, u_1\})$, therefore, using **Lemma 2.10** and **Lemma 2.19**, we get:
$$l(G) = \beta(G \setminus F) = 1 + \beta((G \setminus \{u_0, u_1\}) \setminus (F \setminus \{(u_0, u_1)\})) = 1 + l(G \setminus \{u_0, u_1\}) \geq 2 + l(G \setminus U).$$

**Case 2.** $(u_2, u_3) \notin F$. Since $F \in M(G)$, we imply that $(u_3, u_4) \in F$ and there is $v \in V(G)$, $v \notin U$, such that $(u_5, v) \in F$. Consider the trees $G(1, (u_5, v))$ and $G(2, (u_5, v))$, where, without loss of generality, it is assumed that $u_0 \in V(G(1, (u_5, v)))$. Note that
$$\beta(G) = 3 + \beta(G \setminus U) = 3 + \beta(G(2, (u_5, v)) - u_5) + \beta(G(1, (u_5, v)) \setminus (U \cup \{v\})) \text{ and}$$
$$\beta(G) = \beta(G(1, (u_5, v))) + \beta(G(2, (u_5, v))) - 1 = 2 + \beta(G(2, (u_5, v))) + \beta(G(1, (u_5, v)) \setminus (U \cup \{v\}))$$
therefore
$$\beta(G(2, (u_5, v))) = 1 + \beta(G(2, (u_5, v)) - u_5),$$
thus, using **Lemma 2.17**, we get
$$l(G(2, (u_5, v))) \geq l(G(2, (u_5, v)) - u_5).$$
**Lemma 2.10** and **Lemma 2.21** imply
$$l(G) = \beta(G \setminus F) = l(G(1, (u_5, v))) + l(G(2, (u_5, v))) = 2 + l(G(1, (u_5, v)) \setminus (U \cup \{v\})) +$$
$$+ l(G(2, (u_5, v))) \geq 2 + l(G(1, (u_5, v)) \setminus (U \cup \{v\})) + l(G(2, (u_5, v)) - u_5) = 2 + l(G \setminus U).$$
Proof of **Lemma 2.22** is complete.

**Lemma 2.23.** Let $G$ be a tree, $U = \{u_0, ..., u_4\} \subseteq V(G)$ and $d_G(u_0) = d_G(u_4) = 1$, $d_G(u_1) = d_G(u_3) = 2$, $(u_{i-1}, u_i) \in E(G)$ for $i = 1, ..., 4$ (figure 13). Let $\Gamma \in P'(G)$ and $G' \equiv G \setminus \{u_0, u_1\}$. If $F' \in S'(G', l(G'))$, $\Gamma \cap E(G') \subseteq F'$, then $F \in S'(G, l(G))$ and $\Gamma \subseteq F$, where $F \equiv F' \cup \{(u_0, u_1)\}$.

**Proof** follows from **Corollary 2.8** and **Lemma 2.14**.

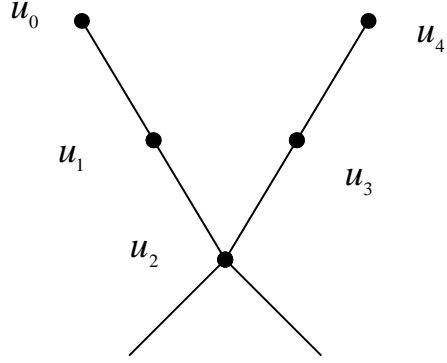

Figure 13

**Lemma 2.24.** Let $G$ be a tree, $U = \{u_0,...,u_4\} \subseteq V(G)$ and $d_G(u_0) = d_G(u_4) = 1$, $d_G(u_1) = 2$, $(u_{i-1}, u_i) \in E(G)$ for $i = 1,...,4$, $\Gamma \in P'(G)$ and $(u_3, u_4) \in \Gamma$ (figure 14). If $F \in S'(G-(u_2,u_3), l(G-(u_2,u_3)))$ and $\Gamma \subseteq F$, then $F \in S'(G, l(G))$ and $\Gamma \subseteq F$.

**Proof** follows from **Corollary 2.8** and **Lemma 2.14**.

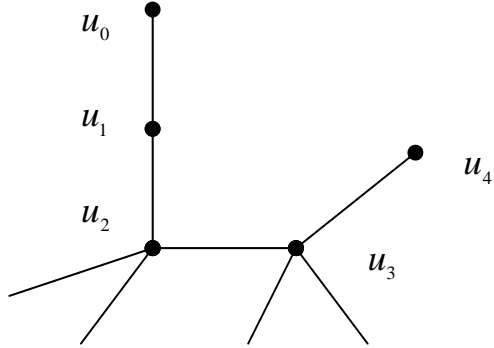

Figure 14

**Lemma 2.25.** Let $G$ be a tree, $U = \{u_0,...,u_5\} \subseteq V(G)$ and $d_G(u_0) = d_G(u_5) = 1$, $d_G(u_1) = d_G(u_3) = 2$, $(u_{i-1}, u_i) \in E(G)$ for $i = 1,...,5$, $(u_2, u_3) \in E_{deep}(G)$, $\Gamma \in P'(G)$ and $(u_4, u_5) \in \Gamma$ (figure 15). If $F_i \in S'(G(i,(u_2,u_3)), l(G(i,(u_2,u_3))))$ and $\Gamma \cap E(G(i,(u_2,u_3))) \subseteq F_i$, for $i = 1,2$, then $F_1 \cup F_2 \in S'(G, l(G))$ and $\Gamma \subseteq F_1 \cup F_2$.

**Proof** follows from **Corollary 2.8**, **Lemma 2.15** and **Lemma 2.12**.

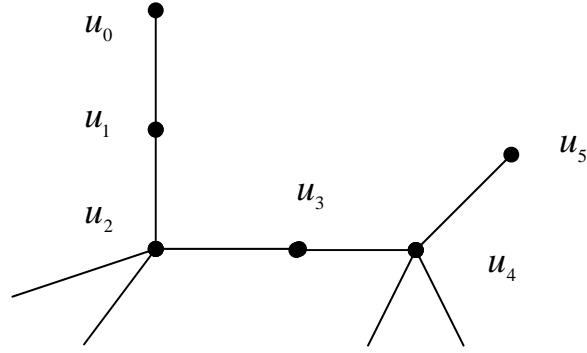

Figure 15

**Lemma 2.26.** Let $G$ be a tree, $U = \{u_0,...,u_4\} \subseteq V(G)$ and $d_G(u_4) = 1$, $d_G(u_1) = d_G(u_2) = 2$, $(u_{i-1}, u_i) \in E(G)$ for $i = 1, 2, 3, 4$ (figure 16). Suppose $\Gamma \in P'(G)$, $(u_3, u_4) \in \Gamma$ and let $g = (u_0, u_1)$, $f = (u_1, u_2)$, $e = (u_2, u_3)$, $u_4 \in V(G(2,g))$, $u_4 \in V(G(2,f))$, $u_4 \in V(G(2,e))$, $g \in E_{deep}(G)$. Then

(a) if $e \notin \pi(G)$ and $F_i \in S'(G(i,f), l(G(i,f)))$, $\Gamma \cap E(G(i,f)) \subseteq F_i$, for $i = 1, 2$, then $F_1 \cup F_2 \in S'(G, l(G))$ and $\Gamma \subseteq F_1 \cup F_2$;

(b) if $e \in \pi(G)$ and $l(G(1,f)) = 1 + l(G(1,g))$, then if $F_i \in S'(G(i,g), l(G(i,g)))$, $\Gamma \cap E(G(i,g)) \subseteq F_i$, for $i = 1, 2$, then $F_1 \cup F_2 \in S'(G, l(G))$ and $\Gamma \subseteq F_1 \cup F_2$;

(c) if $e \in \pi(G)$ and $l(G(1,f)) \leq l(G(1,g))$, then if $F_i \in S'(G(i,f), l(G(i,f)))$, $\Gamma \cap E(G(i,f)) \subseteq F_i$, for $i = 1, 2$, then $F_1 \cup F_2 \in S'(G, l(G))$ and $\Gamma \subseteq F_1 \cup F_2$.

**Proof** follows from **Corollary 2.8**, **Lemma 2.16** and **Lemma 2.12**.

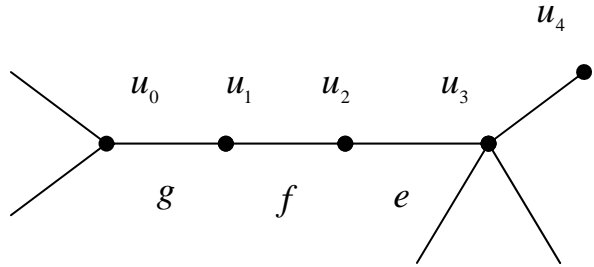

Figure 16

**Lemma 2.27.** Let $G$ be a tree, $U = \{u_0,...,u_6\} \subseteq V(G)$ and $d_G(u_0) = d_G(u_6) = 1$, $d_G(u_1) = d_G(u_5) = 2$, $(u_{i-1}, u_i) \in E(G)$ for $i = 1,...,6$, $\Gamma \in P'(G)$. Set: $e = (u_2, u_3)$, $e' = (u_3, u_4)$, and let $u_0 \in V(G(1,e))$, $u_0 \in V(G(1,e'))$ (figure 17). Then

(a) if $e \notin \pi(G)$, then if $F \in S'(G-e, l(G-e))$ and $\Gamma \subseteq F$, then $F \in S'(G, l(G))$ and $\Gamma \subseteq F$;

(b) if $e' \notin \pi(G)$, then if $F' \in S'(G-e', l(G-e'))$ and $\Gamma \subseteq F'$, then $F' \in S'(G, l(G))$ and $\Gamma \subseteq F'$;

(c) if $e \in \pi(G)$, $e' \in \pi(G)$ and $l(G(1,e)) + l(G(2,e') - u_3) \le l(G(1,e) - u_3) + l(G(2,e'))$, then if $F' \in S'(G-e', l(G-e'))$ and $\Gamma \subseteq F'$, then $F' \in S'(G, l(G))$ and $\Gamma \subseteq F'$;

(d) if $e \in \pi(G)$, $e' \in \pi(G)$ and $l(G(1,e)) + l(G(2,e') - u_3) \ge l(G(1,e) - u_3) + l(G(2,e'))$, then if $F \in S'(G-e, l(G-e))$ and $\Gamma \subseteq F$, then $F \in S'(G, l(G))$ and $\Gamma \subseteq F$.

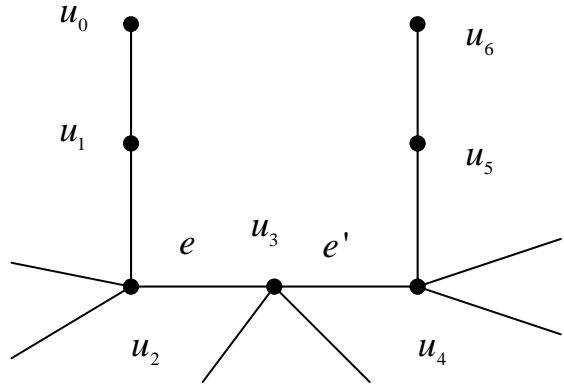

Figure 17

**Proof.** (a) and (b) follow from **Lemma 2.14**.

(c) and (d) follow from **Lemma 2.13, Lemma 2.14**, and the following two equalities:
$$\lambda(e) = l(G(1,e)) + l(G(2,e') - u_3) + l(G(2,e) \setminus (V(G(2,e')) \setminus \{u_3\})),$$
$$\lambda(e') = l(G(1,e) - u_3) + l(G(2,e')) + l(G(1,e') \setminus (V(G(1,e)) \setminus \{u_3\})).$$

**Lemma 2.28.** Let $G$ be a tree, $U = \{u_0,...,u_7\} \subseteq V(G)$ and $d_G(u_0) = d_G(u_7) = 1$, $d_G(u_1) = d_G(u_4) = d_G(u_6) = 2$, $(u_{i-1}, u_i) \in E(G)$ for $i = 1,...,7$, $\Gamma \in P'(G)$. Set: $e = (u_2, u_3)$, $e' = (u_3, u_4)$, $f' = (u_4, u_5)$, and let $u_0 \in V(G(1,e))$, $u_0 \in V(G(1,e'))$, $u_0 \in V(G(1,f'))$ (figure 18). Then

(a) if $e \notin \pi(G)$, then if $F \in S'(G-e, l(G-e))$ and $\Gamma \subseteq F$, then $F \in S'(G, l(G))$ and $\Gamma \subseteq F$;

(b) if $e' \notin \pi(G)$, then if $F_i' \in S'(G(i,f'), l(G(i,f')))$, $\Gamma \cap E(G(i,f')) \subseteq F_i'$, for $i = 1,2$, then $F_1' \cup F_2' \in S'(G, l(G))$ and $\Gamma \subseteq F_1' \cup F_2'$;

(c) if $e \in \pi(G)$, $e' \in \pi(G)$ and $l(G(1,e) - u_3) + l(G(2,e')) \le l(G(1,e)) + l(G(2,f'))$, then if $F \in S'(G-e, l(G-e))$ and $\Gamma \subseteq F$, then $F \in S'(G, l(G))$ and $\Gamma \subseteq F$;

(d) if $e \in \pi(G)$, $e' \in \pi(G)$ and $l(G(1,e) - u_3) + l(G(2,e')) \ge 1 + l(G(1,e)) + l(G(2,f'))$, then if $F_i' \in S'(G(i,f'), l(G(i,f')))$, $\Gamma \cap E(G(i,f')) \subseteq F_i'$, for $i = 1,2$, then $F_1' \cup F_2' \in S'(G, l(G))$ and $\Gamma \subseteq F_1' \cup F_2'$.

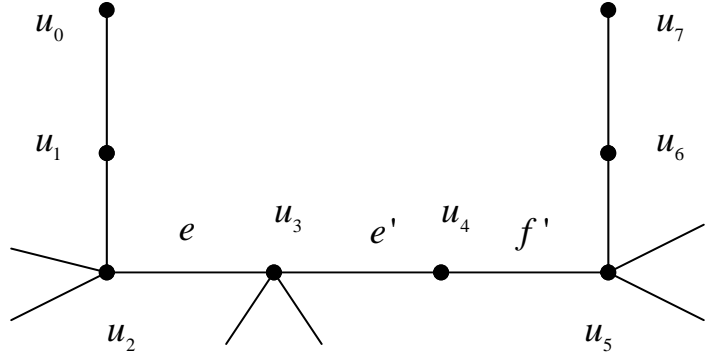

Figure 18

**Proof.** (a) follows from **Lemma 2.14**. (b) follows from **Lemma 2.15** and **Lemma 2.12**.
(c) and (d) follow from **Lemma 2.13, Lemma 2.14**, **Lemma 2.15**, **Lemma 2.12**, **Lemma 2.21** and the following two equalities:

$$\lambda(e) = 1 + l(G(1,e)) + l(G(2, f')) + l(G(2,e) \setminus (\{u_2\} \cup V(G(2,e')))),$$

$$\lambda(e') = l(G(1,e) - u_3) + l(G(2,e')) + l(G(1,e') \setminus (V(G(1,e)) \setminus \{u_3\})).$$

**Lemma 2.29.** Let $G$ be a tree, $U = \{u_0,...,u_8\} \subseteq V(G)$ and $d_G(u_0) = d_G(u_8) = 1$, $d_G(u_1) = d_G(u_3) = = d_G(u_5) = d_G(u_7) = 2$, $(u_{i-1}, u_i) \in E(G)$ for $i = 1,...,8$, $\Gamma \in P'(G)$. Set: $f = (u_2, u_3)$, $e = (u_3, u_4)$, $e' = (u_4, u_5)$, $f' = (u_5, u_6)$, and let $u_0 \in V(G(1, f))$, $u_0 \in V(G(1, e))$, $u_0 \in V(G(1, e'))$, $u_0 \in V(G(1, f'))$ (figure 19). Then

(a) if $e \notin \pi(G)$, then if $F_i \in S'(G(i, f), l(G(i, f)))$, $\Gamma \cap E(G(i, f)) \subseteq F_i$, for $i = 1, 2$, then $F_1 \cup F_2 \in S'(G, l(G))$ and $\Gamma \subseteq F_1 \cup F_2$;

(b) if $e' \notin \pi(G)$, then if $F_i' \in S'(G(i, f'), l(G(i, f')))$, $\Gamma \cap E(G(i, f')) \subseteq F_i'$, for $i = 1, 2$, then $F_1' \cup F_2' \in S'(G, l(G))$ and $\Gamma \subseteq F_1' \cup F_2'$;

(c) if $e \in \pi(G)$, $e' \in \pi(G)$ and $l(G(1,e)) + l(G(2, f')) \leq l(G(1, f)) + l(G(2, e'))$, then if $F_i' \in S'(G(i, f'), l(G(i, f')))$, $\Gamma \cap E(G(i, f')) \subseteq F_i'$, for $i = 1, 2$, then $F_1' \cup F_2' \in S'(G, l(G))$ and $\Gamma \subseteq F_1' \cup F_2'$;

(d) if $e \in \pi(G)$, $e' \in \pi(G)$ and $l(G(1,e)) + l(G(2, f')) \geq l(G(1, f)) + l(G(2, e'))$, then if $F_i \in S'(G(i, f), l(G(i, f)))$, $\Gamma \cap E(G(i, f)) \subseteq F_i$, for $i = 1, 2$, then $F_1 \cup F_2 \in S'(G, l(G))$ and $\Gamma \subseteq F_1 \cup F_2$.

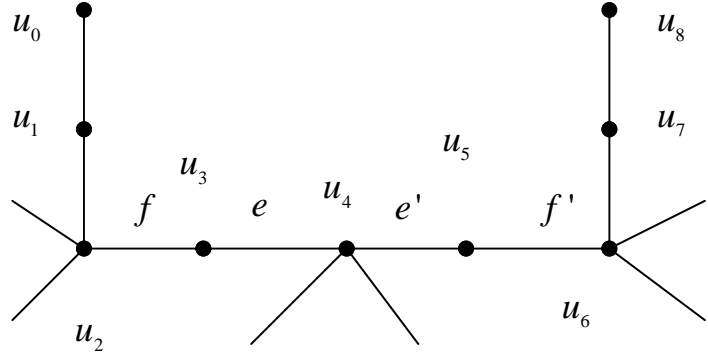

Figure 19

**Proof.** (a) and (b) follow from **Lemma 2.15** and **Lemma 2.12**.
(c) and (d) follow from **Lemma 2.13, Lemma 2.15**, **Lemma 2.12**, **Lemma 2.21** and the following two equalities:
$$\lambda(e) = 1 + l(G(1,e)) + l(G(2,f')) + l(G(2,e) \setminus (\{u_3\} \cup V(G(2,e')))),$$
$$\lambda(e') = 1 + l(G(1,f)) + l(G(2,e')) + l(G(1,e') \setminus (\{u_5\} \cup V(G(1,e)))).$$

**Lemma 2.30.** Let $G$ be a tree, $U = \{u_0,...,u_6\} \subseteq V(G)$ and $d_G(u_0) = 1$, $d_G(u_1) = d_G(u_4) = d_G(u_5) = 2$, $(u_{i-1}, u_i) \in E(G)$ for $i = 1,...,6$, $\Gamma \in P'(G)$. Set: $e = (u_2, u_3)$, $e' = (u_3, u_4)$, $f' = (u_4, u_5)$, $g' = (u_5, u_6)$ and let $u_0 \in V(G(1,e))$, $u_0 \in V(G(1,e'))$, $u_0 \in V(G(1,f'))$, $u_0 \in V(G(1,g'))$ (figure 20). Then

(a) if $e \notin \pi(G)$, then if $F \in S'(G-e, l(G-e))$ and $\Gamma \subseteq F$, then $F \in S'(G, l(G))$ and $\Gamma \subseteq F$;

(b) if $e' \notin \pi(G)$, then if $F_i' \in S'(G(i,f'), l(G(i,f')))$, $\Gamma \cap E(G(i,f')) \subseteq F_i'$, for $i = 1, 2$, then $F_1' \cup F_2' \in S'(G, l(G))$ and $\Gamma \subseteq F_1' \cup F_2'$;

(c) if $e \in \pi(G)$, $e' \in \pi(G)$ then the following equality holds:
$$\beta(G(1,e) - u_3) + \beta(G(2,e')) = \beta(G(1,e)) + \beta(G(2,f'));$$

(d) if $e \in \pi(G)$, $e' \in \pi(G)$, $\beta(G(2,f')) = 1 + \beta(G(2,g'))$ and $l(G(1,e) - u_3) + l(G(2,e')) \geq l(G(1,e)) + l(G(2,f')) + 1$, then if $F_i' \in S'(G(i,f'), l(G(i,f')))$, $\Gamma \cap E(G(i,f')) \subseteq F_i'$, for $i = 1, 2$, then $F_1' \cup F_2' \in S'(G, l(G))$ and $\Gamma \subseteq F_1' \cup F_2'$;

(e) if $e \in \pi(G)$, $e' \in \pi(G)$, $\beta(G(2,f')) = 1 + \beta(G(2,g'))$ and $l(G(1,e) - u_3) + l(G(2,e')) \leq l(G(1,e)) + l(G(2,f'))$, then if $F \in S'(G-e, l(G-e))$ and $\Gamma \subseteq F$, then $F \in S'(G, l(G))$ and $\Gamma \subseteq F$;

(f) if $e \in \pi(G)$, $e' \in \pi(G)$, $\beta(G(2,f')) = \beta(G(2,g'))$ and $l(G(2,f')) = 1 + l(G(2,g'))$, $l(G(1,e) - u_3) + l(G(2,e')) \leq l(G(1,e)) + l(G(2,g'))$ then if $F \in S'(G-e, l(G-e))$ and $\Gamma \subseteq F$, then $F \in S'(G, l(G))$ and $\Gamma \subseteq F$;

(g) if $e \in \pi(G)$, $e' \in \pi(G)$, $\beta(G(2,f')) = \beta(G(2,g'))$ and $l(G(2,f')) = 1 + l(G(2,g'))$, $l(G(1,e) - u_3) + l(G(2,e')) \geq 1 + l(G(1,e)) + l(G(2,g'))$ then if $F_i' \in S'(G(i,f'), l(G(i,f')))$, $\Gamma \cap E(G(i,f')) \subseteq F_i'$, for $i = 1, 2$, then $F_1' \cup F_2' \in S'(G, l(G))$ and $\Gamma \subseteq F_1' \cup F_2'$;

(h) if $e \in \pi(G)$, $e' \in \pi(G)$, $\beta(G(2, f')) = \beta(G(2, g'))$, $l(G(2, f')) \leq l(G(2, g'))$ and $l(G(1,e) - u_3) + l(G(2,e')) \geq l(G(1,e)) + l(G(2, f')) + 1$, then if $F_i' \in S'(G(i, f'), l(G(i, f')))$, $\Gamma \cap E(G(i, f')) \subseteq F_i'$, for $i = 1, 2$, then $F_1' \cup F_2' \in S'(G, l(G))$ and $\Gamma \subseteq F_1' \cup F_2'$;

(i) if $e \in \pi(G)$, $e' \in \pi(G)$, $\beta(G(2, f')) = \beta(G(2, g'))$, $l(G(2, f')) \leq l(G(2, g'))$ and $l(G(1,e) - u_3) + l(G(2,e')) \leq l(G(1,e)) + l(G(2, f'))$, then if $F \in S'(G-e, l(G-e))$ and $\Gamma \subseteq F$, then $F \in S'(G, l(G))$ and $\Gamma \subseteq F$.

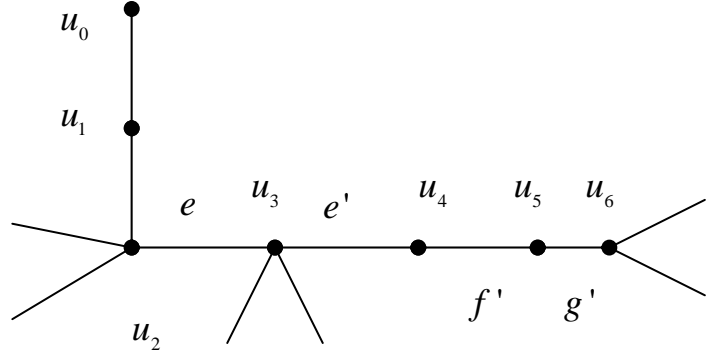

Figure 20

**Proof.** (a) follows from **Lemma 2.14**. (b) follows from **Lemma 2.16** and **Lemma 2.12.**
(c) Since $e \in \pi(G)$, $e' \in \pi(G)$, we have
$$\beta(G) = \beta(G(1,e)) + \beta(G(2, f')) + \beta(G(2,e) \setminus (\{u_2\} \cup V(G(2,e')))),$$
$$\beta(G) = \beta(G(1,e) - u_3) + \beta(G(2,e')) + \beta(G(1,e') \setminus (\{u_4\} \cup V(G(1,e)))),$$
therefore, the equality $\beta(G(1,e) - u_3) + \beta(G(2,e')) = \beta(G(1,e)) + \beta(G(2, f'))$ holds.

(d) Note that **Lemma 2.16** and **Lemma 2.12** imply that it suffices to show that there is $\widetilde{F} \in S(G, l(G))$ such that $e' \notin \widetilde{F}$. Using **Lemma 2.16**, **Lemma 2.14**, **Lemma 2.21**, **Lemma 2.18**, we get
$$\lambda(e') = l(G(1,e) - u_3) + l(G(2,e')) + l(G(1,e') \setminus (V(G(1,e)) \setminus \{u_3\})) \geq l(G(1,e)) +$$
$$+ l(G(2, f')) + 1 + l(G(2,e) \setminus (\{u_2\} \cup V(G(2,e')))) = \lambda(e),$$
therefore, due to **Lemma 2.13**, there is $\widetilde{F} \in S(G, l(G))$ such that $e' \notin \widetilde{F}$.

(e) Note that **Lemma 2.14** implies that it suffices to show that there is $\widetilde{F} \in S(G, l(G))$ such that $e \notin \widetilde{F}$. Using **Lemma 2.16**, **Lemma 2.14**, **Lemma 2.21**, **Lemma 2.18**, we get
$$\lambda(e') = l(G(1,e) - u_3) + l(G(2,e')) + l(G(1,e') \setminus (V(G(1,e)) \setminus \{u_3\})) \leq l(G(1,e)) +$$
$$+ l(G(2, f')) + 1 + l(G(2,e) \setminus (\{u_2\} \cup V(G(2,e')))) = \lambda(e),$$
therefore, due to **Lemma 2.13**, there is $\widetilde{F} \in S(G, l(G))$ such that $e \notin \widetilde{F}$.

(f) Note that **Lemma 2.14** implies that it suffices to show that there is $\widetilde{F} \in S(G, l(G))$ such that $e \notin \widetilde{F}$. Using **Lemma 2.16**, **Lemma 2.14**, **Lemma 2.21**, and **Corollary 2.12**, we get
$$\lambda(e) = l(G(1,e)) + 1 + l(G(2, g')) + l(G(2,e) \setminus V(G(2, f'))) > l(G(1,e) - u_3) +$$
$$+ l(G(2,e')) + l(G(1,e') \setminus (V(G(1,e)) \setminus \{u_3\})) = \lambda(e'),$$
therefore, due to **Lemma 2.13**, there is $\widetilde{F} \in S(G, l(G))$ such that $e \notin \widetilde{F}$.

(g) Note that **Lemma 2.16** implies that it suffices to show that there is $\widetilde{F} \in S(G, l(G))$ such that $e' \notin \widetilde{F}$. Using **Lemma 2.16**, **Lemma 2.14**, **Lemma 2.21**, and **Corollary 2.12**, we get
$$\lambda(e') = l(G(1,e) - u_3) + l(G(2,e')) + l(G(1,e') \setminus (V(G(1,e)) \setminus \{u_3\})) \geq$$
$$\geq l(G(1,e)) + 1 + l(G(2,g')) + l(G(2,e) \setminus V(G(2,f'))) = \lambda(e),$$
therefore, due to **Lemma 2.13**, there is $\widetilde{F} \in S(G, l(G))$ such that $e' \notin \widetilde{F}$.

(h) Note that **Lemma 2.16** and **Lemma 2.12** imply that it suffices to show that there is $\widetilde{F} \in S(G, l(G))$ such that $e' \notin \widetilde{F}$. Using **Lemma 2.14**, **Lemma 2.16**, **Lemma 2.18**, **Lemma 2.21**, we get
$$\lambda(e') = l(G(1,e) - u_3) + l(G(2,e')) + l(G(1,e') \setminus (V(G(1,e)) \setminus \{u_3\})) \geq l(G(1,e)) +$$
$$+ l(G(2,f')) + 1 + l(G(1,e') \setminus (\{u_4\} \cup V(G(1,e)))) = \lambda(e),$$
therefore, due to **Lemma 2.13**, there is $\widetilde{F} \in S(G, l(G))$ such that $e' \notin \widetilde{F}$.

(i) Note that **Lemma 2.14** implies that it suffices to show that there is $\widetilde{F} \in S(G, l(G))$ such that $e \notin \widetilde{F}$. Using **Lemma 2.14**, **Lemma 2.16**, **Lemma 2.18**, **Lemma 2.21**, we get
$$\lambda(e) = l(G(1,e)) + l(G(2,f')) + 1 + l(G(2,e) \setminus (\{u_2\} \cup V(G(2,e')))) \geq l(G(1,e) - u_3) +$$
$$+ l(G(2,e')) + l(G(1,e') \setminus (V(G(1,e)) \setminus \{u_3\})) = \lambda(e'),$$
therefore, due to **Lemma 2.13**, there is $\widetilde{F} \in S(G, l(G))$ such that $e \notin \widetilde{F}$. Proof of **Lemma 2.30** is complete.

**Lemma 2.31.** Let $G$ be a tree, $U = \{u_0, ..., u_7\} \subseteq V(G)$ and $d_G(u_0) = 1$, $d_G(u_1) = d_G(u_3) = d_G(u_5) = d_G(u_6) = 2$, $(u_{i-1}, u_i) \in E(G)$ for $i = 1, ..., 7$, $\Gamma \in P'(G)$. Set: $f = (u_2, u_3)$, $e = (u_3, u_4)$, $e' = (u_4, u_5)$, $f' = (u_5, u_6)$, $g' = (u_6, u_7)$ and let $u_0 \in V(G(1,f))$, $u_0 \in V(G(1,e))$, $u_0 \in V(G(1,e'))$, $u_0 \in V(G(1,f'))$, $u_0 \in V(G(1,g'))$ (figure 21). Then

(a) if $e \notin \pi(G)$, then if $F_i \in S'(G(i,f), l(G(i,f)))$, $\Gamma \cap E(G(i,f)) \subseteq F_i$, for $i = 1, 2$, then $F_1 \cup F_2 \in S'(G, l(G))$ and $\Gamma \subseteq F_1 \cup F_2$;

(b) if $e' \notin \pi(G)$, then if $F_i' \in S'(G(i,f'), l(G(i,f')))$, $\Gamma \cap E(G(i,f')) \subseteq F_i'$, for $i = 1, 2$, then $F_1' \cup F_2' \in S'(G, l(G))$ and $\Gamma \subseteq F_1' \cup F_2'$;

(c) if $e \in \pi(G)$, $e' \in \pi(G)$ then the following equality holds:
$$\beta(G(1,f)) + \beta(G(2,e')) = \beta(G(1,e)) + \beta(G(2,f'));$$

(d) if $e \in \pi(G)$, $e' \in \pi(G)$, $\beta(G(2,f')) = 1 + \beta(G(2,g'))$ and $l(G(1,f)) + l(G(2,e')) \geq l(G(1,e)) + l(G(2,f'))$, then if $F_i' \in S'(G(i,f'), l(G(i,f')))$, $\Gamma \cap E(G(i,f')) \subseteq F_i'$, for $i = 1, 2$, then $F_1' \cup F_2' \in S'(G, l(G))$ and $\Gamma \subseteq F_1' \cup F_2'$;

(e) if $e \in \pi(G)$, $e' \in \pi(G)$, $\beta(G(2,f')) = 1 + \beta(G(2,g'))$ and $l(G(1,f)) + l(G(2,e')) \leq l(G(1,e)) + l(G(2,f'))$, then if $F_i \in S'(G(i,f), l(G(i,f)))$, $\Gamma \cap E(G(i,f)) \subseteq F_i$, for $i = 1, 2$, then $F_1 \cup F_2 \in S'(G, l(G))$ and $\Gamma \subseteq F_1 \cup F_2$;

(f) if $e \in \pi(G)$, $e' \in \pi(G)$, $\beta(G(2,f')) = \beta(G(2,g'))$ and $l(G(2,f')) = 1 + l(G(2,g'))$, $l(G(1,f)) + l(G(2,e')) \leq l(G(1,e)) + l(G(2,g'))$, then if $F_i \in S'(G(i,f), l(G(i,f)))$, $\Gamma \cap E(G(i,f)) \subseteq F_i$, for $i = 1, 2$, then $F_1 \cup F_2 \in S'(G, l(G))$ and $\Gamma \subseteq F_1 \cup F_2$;

(g) if $e \in \pi(G)$, $e' \in \pi(G)$, $\beta(G(2,f')) = \beta(G(2,g'))$ and $l(G(2,f')) = 1 + l(G(2,g'))$, $l(G(1,f)) + l(G(2,e')) \geq 1 + l(G(1,e)) + l(G(2,g'))$ then if $F_i' \in S'(G(i,f'), l(G(i,f')))$, $\Gamma \cap E(G(i,f')) \subseteq F_i'$, for $i = 1, 2$, then $F_1' \cup F_2' \in S'(G, l(G))$ and $\Gamma \subseteq F_1' \cup F_2'$;

(h) if $e \in \pi(G)$, $e' \in \pi(G)$, $\beta(G(2,f')) = \beta(G(2,g'))$, $l(G(2,f')) \leq l(G(2,g'))$ and $l(G(1,f)) + l(G(2,e')) \geq l(G(1,e)) + l(G(2,f'))$, then if $F_i' \in S'(G(i,f'), l(G(i,f')))$, $\Gamma \cap E(G(i,f')) \subseteq F_i'$, for $i = 1, 2$, then $F_1' \cup F_2' \in S'(G, l(G))$ and $\Gamma \subseteq F_1' \cup F_2'$;

(i) if $e \in \pi(G)$, $e' \in \pi(G)$, $\beta(G(2,f')) = \beta(G(2,g'))$, $l(G(2,f')) \leq l(G(2,g'))$ and $l(G(1,f)) + l(G(2,e')) \leq l(G(1,e)) + l(G(2,f'))$, then if $F_i \in S'(G(i,f), l(G(i,f)))$, $\Gamma \cap E(G(i,f)) \subseteq F_i$, for $i = 1, 2$, then $F_1 \cup F_2 \in S'(G, l(G))$ and $\Gamma \subseteq F_1 \cup F_2$.

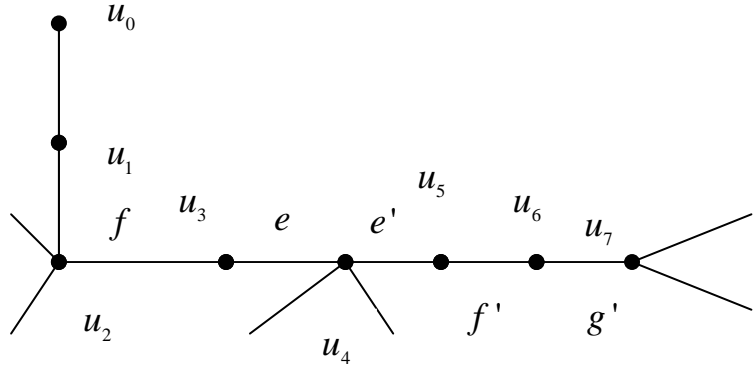

Figure 21

**Proof.** (a) follows from **Lemma 2.15**. (b) follows from **Lemma 2.16** and **Lemma 2.12.**

(c) Since $e \in \pi(G)$, $e' \in \pi(G)$, we have
$$\beta(G) = \beta(G(1,e)) + \beta(G(2,f')) + \beta(G(2,e) \setminus (\{u_3\} \cup V(G(2,e')))),$$
$$\beta(G) = \beta(G(1,f)) + \beta(G(2,e')) + \beta(G(1,e') \setminus (\{u_5\} \cup V(G(1,e)))),$$
therefore, the equality $\beta(G(1,f)) + \beta(G(2,e')) = \beta(G(1,e)) + \beta(G(2,f'))$ holds.

(d) Note that **Lemma 2.16** and **Lemma 2.12** imply that it suffices to show that there is $\widetilde{F} \in S(G, l(G))$ such that $e' \notin \widetilde{F}$. Using **Lemma 2.16**, **Lemma 2.15**, **Lemma 2.21**, **Lemma 2.18**, we get
$$\lambda(e') = l(G(1,f)) + l(G(2,e')) + 1 + l(G(1,e') \setminus (V(G(1,e)) \cup \{u_5\})) \geq l(G(1,e)) +$$
$$+ l(G(2,f')) + 1 + l(G(2,e) \setminus (\{u_3\} \cup V(G(2,e')))) = \lambda(e),$$
therefore, due to **Lemma 2.13**, there is $\widetilde{F} \in S(G, l(G))$ such that $e' \notin \widetilde{F}$.

(e) Note that **Lemma 2.15** implies that it suffices to show that there is $\widetilde{F} \in S(G, l(G))$ such that $e \notin \widetilde{F}$. Using **Lemma 2.16**, **Lemma 2.15**, **Lemma 2.21**, **Lemma 2.18**, we get
$$\lambda(e') = l(G(1,f)) + l(G(2,e')) + 1 + l(G(1,e') \setminus (V(G(1,e)) \cup \{u_5\})) \leq l(G(1,e)) +$$
$$+ l(G(2,f')) + 1 + l(G(2,e) \setminus (\{u_3\} \cup V(G(2,e')))) = \lambda(e),$$
therefore, due to **Lemma 2.13**, there is $\widetilde{F} \in S(G, l(G))$ such that $e \notin \widetilde{F}$.

(f) Note that **Lemma 2.15** implies that it suffices to show that there is $\widetilde{F} \in S(G, l(G))$ such that $e \notin \widetilde{F}$. Using **Lemma 2.16**, **Lemma 2.15**, **Lemma 2.21**, and **Corollary 2.12**, **Lemma 2.18**, we get

$$\lambda(e) = l(G(1,e)) + 1 + l(G(2,g')) + l(G(2,e) \setminus V(G(2,f'))) \geq l(G(1,f)) +$$
$$+ l(G(2,e')) + 1 + l(G(1,e') \setminus (V(G(1,e)) \cup \{u_5\})) = \lambda(e'),$$

therefore, due to **Lemma 2.13**, there is $\widetilde{F} \in S(G, l(G))$ such that $e \notin \widetilde{F}$.

(g) Note that **Lemma 2.16** implies that it suffices to show that there is $\widetilde{F} \in S(G, l(G))$ such that $e' \notin \widetilde{F}$.
Using **Lemma 2.16**, **Lemma 2.15**, **Lemma 2.21**, and **Corollary 2.12**, **Lemma 2.18**, we get
$$\lambda(e) = l(G(1,e)) + 1 + l(G(2,g')) + l(G(2,e) \setminus V(G(2,f'))) \leq l(G(1,f)) +$$
$$+ l(G(2,e')) + 1 + l(G(1,e') \setminus (V(G(1,e)) \cup \{u_5\})) = \lambda(e'),$$

therefore, due to **Lemma 2.13**, there is $\widetilde{F} \in S(G, l(G))$ such that $e' \notin \widetilde{F}$.

(h) Note that **Lemma 2.16** and **Lemma 2.12** imply that it suffices to show that there is $\widetilde{F} \in S(G, l(G))$ such that $e' \notin \widetilde{F}$. Using **Lemma 2.15**, **Lemma 2.16**, **Lemma 2.21**, we get
$$\lambda(e') = l(G(1,f)) + l(G(2,e')) + 1 + l(G(1,e') \setminus (V(G(1,e)) \cup \{u_5\})) \geq l(G(1,e)) +$$
$$+ l(G(2,f')) + 1 + l(G(2,e) \setminus (\{u_3\} \cup V(G(2,e')))) = \lambda(e),$$

therefore, due to **Lemma 2.13**, there is $\widetilde{F} \in S(G, l(G))$ such that $e' \notin \widetilde{F}$.

(i) Note that **Lemma 2.15** implies that it suffices to show that there is $\widetilde{F} \in S(G, l(G))$ such that $e \notin \widetilde{F}$.
Using **Lemma 2.15**, **Lemma 2.16**, **Lemma 2.21**, we get
$$\lambda(e') = l(G(1,f)) + l(G(2,e')) + 1 + l(G(1,e') \setminus (V(G(1,e)) \cup \{u_5\})) \leq l(G(1,e)) +$$
$$+ l(G(2,f')) + 1 + l(G(2,e) \setminus (\{u_3\} \cup V(G(2,e')))) = \lambda(e),$$

therefore, due to **Lemma 2.13**, there is $\widetilde{F} \in S(G, l(G))$ such that $e \notin \widetilde{F}$. Proof of **Lemma 2.31** is complete.

**Lemma 2.32.** Let $G$ be a tree, $U = \{u_0, ..., u_6\} \subseteq V(G)$ and $d_G(u_1) = d_G(u_2) = d_G(u_4) = d_G(u_5) = 2$, $(u_{i-1}, u_i) \in E(G)$ for $i = 1, ..., 6$, $\Gamma \in P'(G)$. Set: $g = (u_0, u_1)$, $f = (u_1, u_2)$, $e = (u_2, u_3)$, $e' = (u_3, u_4)$, $f' = (u_4, u_5)$, $g' = (u_5, u_6)$ and let $u_6 \in V(G(2,g))$, $u_0 \in V(G(1,f))$, $u_0 \in V(G(1,e))$, $u_0 \in V(G(1,e'))$, $u_0 \in V(G(1,f'))$, $u_0 \in V(G(1,g'))$ (figure 22). Then

(a) if $e \notin \pi(G)$, then if $F_i \in S'(G(i,f), l(G(i,f)))$, $\Gamma \cap E(G(i,f)) \subseteq F_i$, for $i = 1, 2$, then $F_1 \cup F_2 \in S'(G, l(G))$ and $\Gamma \subseteq F_1 \cup F_2$;

(b) if $e' \notin \pi(G)$, then if $F_i' \in S'(G(i,f'), l(G(i,f')))$, $\Gamma \cap E(G(i,f')) \subseteq F_i'$, for $i = 1, 2$, then $F_1' \cup F_2' \in S'(G, l(G))$ and $\Gamma \subseteq F_1' \cup F_2'$;

(c) if $e \in \pi(G)$, $e' \in \pi(G)$ then the following equality holds:
$$\beta(G(1,f)) + \beta(G(2,e')) = \beta(G(1,e)) + \beta(G(2,f'));$$

(d) if $e \in \pi(G)$, $e' \in \pi(G)$, $\beta(G(2,f')) = 1 + \beta(G(2,g'))$ and $l(G(1,e)) + l(G(2,f')) \geq l(G(1,f)) + l(G(2,e'))$, then if $F_i \in S'(G(i,f), l(G(i,f)))$, $\Gamma \cap E(G(i,f)) \subseteq F_i$, for $i = 1, 2$, then $F_1 \cup F_2 \in S'(G, l(G))$ and $\Gamma \subseteq F_1 \cup F_2$;

(e) if $e \in \pi(G)$, $e' \in \pi(G)$, $\beta(G(2,f')) = 1 + \beta(G(2,g'))$ and $l(G(1,e)) + l(G(2,f')) \leq l(G(1,f)) + l(G(2,e'))$, then if $F_i' \in S'(G(i,f'), l(G(i,f')))$, $\Gamma \cap E(G(i,f')) \subseteq F_i'$, for $i = 1, 2$, then $F_1' \cup F_2' \in S'(G, l(G))$ and $\Gamma \subseteq F_1' \cup F_2'$;

(f) if $e \in \pi(G)$, $e' \in \pi(G)$, $\beta(G(2,f')) = \beta(G(2,g'))$, $l(G(2,f')) \leq l(G(2,g'))$, $l(G(1,f)) \leq l(G(1,g))$ and $l(G(1,e)) + l(G(2,f')) \geq l(G(1,f)) + l(G(2,e'))$, then if $F_i \in S'(G(i,f), l(G(i,f)))$, $\Gamma \cap E(G(i,f)) \subseteq F_i$, for $i=1,2$, then $F_1 \cup F_2 \in S'(G, l(G))$ and $\Gamma \subseteq F_1 \cup F_2$;

(g) if $e \in \pi(G)$, $e' \in \pi(G)$, $\beta(G(2,f')) = \beta(G(2,g'))$, $l(G(2,f')) \leq l(G(2,g'))$, $l(G(1,f)) \leq l(G(1,g))$ and $l(G(1,e)) + l(G(2,f')) \leq l(G(1,f)) + l(G(2,e'))$, then if $F_i' \in S'(G(i,f'), l(G(i,f')))$, $\Gamma \cap E(G(i,f')) \subseteq F_i'$, for $i=1,2$, then $F_1' \cup F_2' \in S'(G, l(G))$ and $\Gamma \subseteq F_1' \cup F_2'$;

(h) if $e \in \pi(G)$, $e' \in \pi(G)$, $\beta(G(2,f')) = \beta(G(2,g'))$, $l(G(2,f')) = 1 + l(G(2,g'))$ and $l(G(1,f)) = 1 + l(G(1,g))$, then if $F_i' \in S'(G(i,g'), l(G(i,g')))$, $\Gamma \cap E(G(i,g')) \subseteq F_i'$, for $i=1,2$, then $F_1' \cup F_2' \in S'(G, l(G))$ and $\Gamma \subseteq F_1' \cup F_2'$;

(i) if $e \in \pi(G)$, $e' \in \pi(G)$, $\beta(G(2,f')) = \beta(G(2,g'))$, $l(G(2,f')) = 1 + l(G(2,g'))$ and $l(G(1,f)) \leq l(G(1,g))$, then if $F_i \in S'(G(i,f), l(G(i,f)))$, $\Gamma \cap E(G(i,f)) \subseteq F_i$, for $i=1,2$, then $F_1 \cup F_2 \in S'(G, l(G))$ and $\Gamma \subseteq F_1 \cup F_2$;

(j) if $e \in \pi(G)$, $e' \in \pi(G)$, $\beta(G(2,f')) = \beta(G(2,g'))$, $l(G(2,f')) \leq l(G(2,g'))$ and $l(G(1,f)) = 1 + l(G(1,g))$, then if $F_i' \in S'(G(i,f'), l(G(i,f')))$, $\Gamma \cap E(G(i,f')) \subseteq F_i'$, for $i=1,2$, then $F_1' \cup F_2' \in S'(G, l(G))$ and $\Gamma \subseteq F_1' \cup F_2'$.

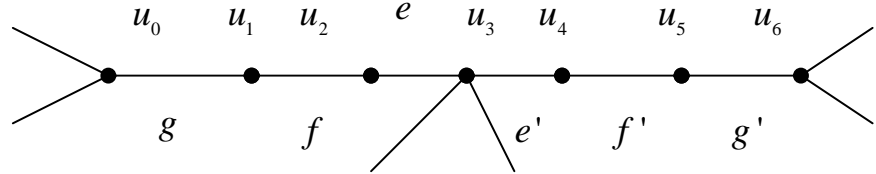

Figure 22

**Proof.** (a) and (b) follow from **Lemma 2.16** and **Lemma 2.12**.

(c) Since $e \in \pi(G)$, $e' \in \pi(G)$, we have
$$\beta(G) = \beta(G(1,e)) + \beta(G(2,f')) + \beta(G(2,e) \setminus (\{u_2\} \cup V(G(2,e')))),$$
$$\beta(G) = \beta(G(1,f)) + \beta(G(2,e')) + \beta(G(1,e') \setminus (\{u_4\} \cup V(G(1,e)))),$$
therefore, the equality $\beta(G(1,e)) + \beta(G(2,f')) = \beta(G(1,f)) + \beta(G(2,e'))$ holds.

(d) Note that **Lemma 2.16** and **Lemma 2.12** imply that it suffices to show that there is $\widetilde{F} \in S(G, l(G))$ such that $e \notin \widetilde{F}$. Using **Lemma 2.16** and **Lemma 2.21**, we get
$$\lambda(e) = l(G(1,e)) + 1 + l(G(2,f')) + l(G(2,e) \setminus (\{u_2\} \cup V(G(2,e')))) \geq$$
$$\geq l(G(1,f)) + 1 + l(G(2,e')) + l(G(1,e') \setminus (\{u_4\} \cup V(G(1,e)))) = \lambda(e'),$$
therefore, due to **Lemma 2.13**, there is $\widetilde{F} \in S(G, l(G))$ such that $e \notin \widetilde{F}$.

(e) Note that **Lemma 2.16** and **Lemma 2.12** imply that it suffices to show that there is $\widetilde{F} \in S(G,l(G))$ such that $e' \notin \widetilde{F}$. Using **Lemma 2.16** and **Lemma 2.21**, we get
$$\lambda(e') = l(G(1,f)) + 1 + l(G(2,e')) + l(G(1,e') \setminus (\{u_4\} \cup V(G(1,e)))) \geq$$
$$\geq l(G(1,e)) + 1 + l(G(2,f')) + l(G(2,e) \setminus (\{u_2\} \cup V(G(2,e')))) = \lambda(e),$$
therefore, due to **Lemma 2.13**, there is $\widetilde{F} \in S(G,l(G))$ such that $e' \notin \widetilde{F}$.

(f) Note that **Lemma 2.16** and **Lemma 2.12** imply that it suffices to show that there is $\widetilde{F} \in S(G,l(G))$ such that $e \notin \widetilde{F}$. Using **Lemma 2.16**, **Lemma 2.21**, we get
$$\lambda(e) = l(G(1,e)) + 1 + l(G(2,f')) + l(G(2,e) \setminus (\{u_2\} \cup V(G(2,e')))) \geq$$
$$\geq l(G(1,f)) + 1 + l(G(2,e')) + l(G(1,e') \setminus (\{u_4\} \cup V(G(1,e)))) = \lambda(e'),$$
therefore, due to **Lemma 2.13**, there is $\widetilde{F} \in S(G,l(G))$ such that $e \notin \widetilde{F}$.

(g) Note that **Lemma 2.16** and **Lemma 2.12** imply that it suffices to show that there is $\widetilde{F} \in S(G,l(G))$ such that $e' \notin \widetilde{F}$. Using **Lemma 2.16**, **Lemma 2.21**, we get
$$\lambda(e) = l(G(1,e)) + 1 + l(G(2,f')) + l(G(2,e) \setminus (\{u_2\} \cup V(G(2,e')))) \leq$$
$$\leq l(G(1,f)) + 1 + l(G(2,e')) + l(G(1,e') \setminus (\{u_4\} \cup V(G(1,e)))) = \lambda(e'),$$
therefore, due to **Lemma 2.13**, there is $\widetilde{F} \in S(G,l(G))$ such that $e' \notin \widetilde{F}$.

(h) Note that **Lemma 2.17** and **Lemma 2.12** imply that it suffices to show that there is $\widetilde{F} \in S(G,l(G))$ such that $e' \notin \widetilde{F}$. Using **Lemma 2.17** and **Corollary 2.12**, we get
$$\lambda(e') = l(G(1,g)) + 1 + l(G(2,e')) + l(G(1,e') \setminus V(G(1,f))) =$$
$$= l(G(1,g)) + 2 + l(G(2,g')) + l(G(1,e') \setminus V(G(1,f))) =$$
$$= l(G(1,e)) + 1 + l(G(2,g')) + l(G(2,e) \setminus V(G(2,f'))) = \lambda(e),$$
therefore, due to **Lemma 2.13**, there is $\widetilde{F} \in S(G,l(G))$ such that $e' \notin \widetilde{F}$.

(i) Note that **Lemma 2.16** and **Lemma 2.12** imply that it suffices to show that there is $\widetilde{F} \in S(G,l(G))$ such that $e \notin \widetilde{F}$. Using **Lemma 2.16**, **Lemma 2.21**, **Lemma 2.18**, we get
$$\lambda(e) = l(G(1,e)) + 1 + l(G(2,g')) + l(G(2,e) \setminus V(G(2,f'))) =$$
$$= l(G(1,g)) + 1 + l(G(2,e')) + l(G(1,e') \setminus V(G(1,f))) \geq$$
$$\geq l(G(1,f)) + 1 + l(G(2,e')) + l(G(1,e') \setminus (\{u_4\} \cup V(G(1,e)))) = \lambda(e'),$$
therefore, due to **Lemma 2.13**, there is $\widetilde{F} \in S(G,l(G))$ such that $e \notin \widetilde{F}$.

(j) Note that **Lemma 2.16** and **Lemma 2.12** imply that it suffices to show that there is $\widetilde{F} \in S(G,l(G))$ such that $e' \notin \widetilde{F}$. Using **Lemma 2.16**, **Lemma 2.21**, **Lemma 2.18**, we get
$$\lambda(e') = l(G(1,g)) + 1 + l(G(2,e')) + l(G(1,e') \setminus V(G(1,f))) =$$
$$= l(G(1,e)) + 1 + l(G(2,g')) + l(G(2,e) \setminus V(G(2,f'))) \geq$$
$$\geq l(G(1,e)) + 1 + l(G(2,f')) + l(G(2,e) \setminus (\{u_2\} \cup V(G(2,e')))) = \lambda(e),$$
therefore, due to **Lemma 2.13**, there is $\widetilde{F} \in S(G,l(G))$ such that $e' \notin \widetilde{F}$. Proof of **Lemma 2.32** is complete.

**Lemma 2.33.** Let $G$ be a tree and $e \in \pi(G)$. Then $\Lambda(e) \leq L(G)$.

**Proof. Corollary 2.8** implies the existence of $F^{(1)} \in S(G(1,e), L(G(1,e)))$, $F^{(2)} \in S(G(2,e), L(G(2,e)))$ with $e \in F^{(1)}$, $e \in F^{(2)}$. Consider $\widetilde{F} \in B(G)$ defined as $\widetilde{F} \equiv F^{(1)} \cup F^{(2)}$. Since $e \in \pi(G)$, we have $\widetilde{F} \in M(G)$ and

$$\Lambda(e) = L(G(1,e)) + L(G(2,e)) = \beta(G(1,e) \setminus F^{(1)}) + \beta(G(2,e) \setminus F^{(2)}) = \beta(G \setminus \widetilde{F}) \le L(G).$$

Proof of **Lemma 2.33** is complete.

**Example 2.4.** Simply modifying the tree from **Example 2.3** we are able to show that the statement of **Lemma 2.33** may not be true if $e \notin \pi(G)$.

For every $k \in N$ consider the tree $G_k^{(2.4)}$ and its edge $e_k$ shown in the figure below:

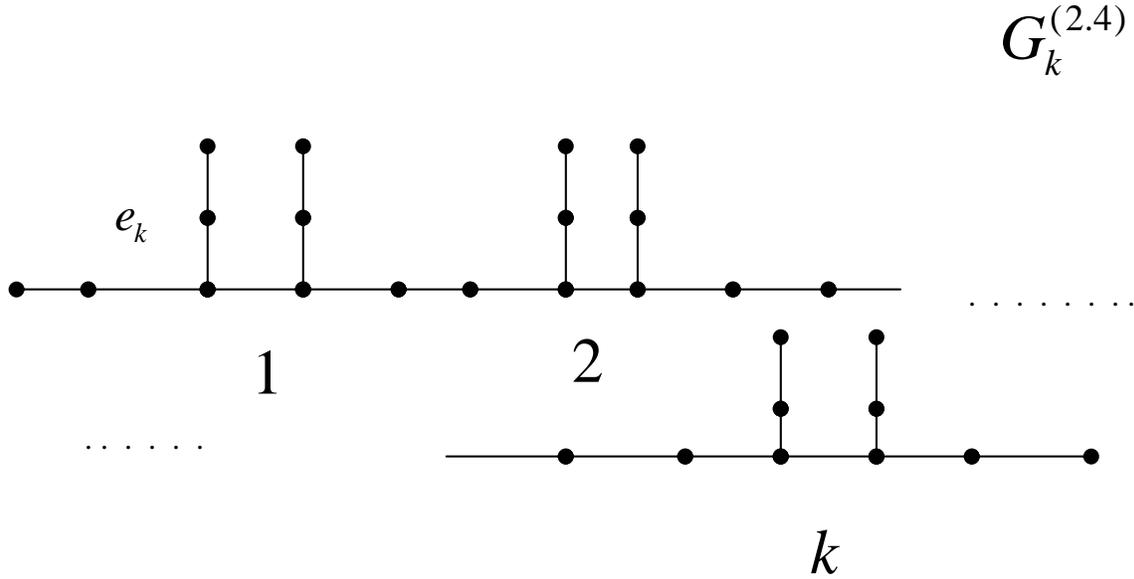

Figure 23

Note that $L(G_k^{(2.4)}) = 2k$, $\Lambda(e_k) = 3k+1$ and $\Lambda(e_k) - L(G_k^{(2.4)}) = k+1$.

**Lemma 2.34.** Let $G$ be a tree, $e \in E(G)$, $F \in S(G, L(G))$ and $e \in F$. Then for $i = 1, 2$

$$(F \cap E(G(i,e))) \in S(G(i,e), L(G(i,e))).$$

**Proof.** For $i = 1, 2$ denote $F_i \equiv F \cap E(G(i,e))$. Clearly, for $i = 1, 2$ $F_i \in M(G(i,e))$. Let us show that $F_1 \in S(G(1,e)), L(G(1,e))))$. Assume that $F_1 \notin S(G(1,e), L(G(1,e)))$. **Corollary 2.8** implies that there is $F_1^{'} \in S(G(1,e), \beta(G(1,e) \setminus F_1)+1)$ such that $e \in F_1^{'}$. Consider the matching $F^{'}$ defined as $F^{'} \equiv (F \setminus F_1) \cup F_1^{'}$. Note that $F^{'} \in M(G)$ and

$$\beta(G \setminus F^{'}) = \beta(G(1,e) \setminus F_1^{'}) + \beta(G(2,e) \setminus F_2) = \beta(G(1,e) \setminus F_1) + \beta(G(2,e) \setminus F_2) + 1 = \beta(G \setminus F) + 1$$

which contradicts the condition $F \in S(G, L(G))$, therefore $F_1 \in S(G(1,e)), L(G(1,e))))$. Similarly, it can be shown that $F_2 \in S(G(2,e), L(G(2,e)))$. Proof of **Lemma 2.34** is complete.

**Lemma 2.35.** Let $G$ be a tree and $e \in E(G)$. If there is $F \in S(G, L(G))$ such that $e \in F$, then

$$\Lambda(e) = L(G).$$

**Proof.** Since $F \in S(G, L(G))$, then, using **Lemma 2.34**, we get:

$$L(G) = \beta(G \setminus F) = \beta(G(1,e) \setminus (F \cap E(G(1,e)))) + \beta(G(2,e) \setminus (F \cap E(G(2,e)))) =$$
$$= L(G(1,e)) + L(G(2,e)) = \Lambda(e).$$

Proof of **Lemma 2.35** is complete.

**Lemma 2.36.** Let $G$ be a graph and suppose that there is $F \in M(G)$ such that $F \cap \theta(G) \neq \varnothing$. Then there is $\widetilde{F} \in S(G, L(G))$ such that $\widetilde{F} \cap \theta(G) \neq \varnothing$.

**Proof.** Assume that for each $\widetilde{F} \in S(G, L(G))$ $\widetilde{F} \cap \theta(G) = \varnothing$. Then, due to **Corollary 2.4**, $l(G) = L(G)$, thus $F \in S(G, L(G))$ and $F \cap \theta(G) = \varnothing$, which contradicts the condition of **Lemma**. Proof of **Lemma 2.36** is complete.

**Corollary 2.13.** If in a tree $G$ $\pi(G) \cap \theta(G) \neq \varnothing$, then $\max\limits_{e \in \pi(G) \cap \theta(G)} \Lambda(e) = L(G)$.

**Proof. Lemma 2.33** implies that $\max\limits_{e \in \pi(G) \cap \theta(G)} \Lambda(e) \leq L(G)$. Since $\pi(G) \cap \theta(G) \neq \varnothing$, there is $F \in M(G)$ such that $F \cap \theta(G) \neq \varnothing$, therefore due to **Lemma 2.36**, there is $\widetilde{F} \in S(G, L(G))$ such that $\widetilde{F} \cap \theta(G) \neq \varnothing$. Consider an edge $e_0 \in \widetilde{F} \cap \theta(G)$. **Lemma 2.35** implies that $\Lambda(e_0) = L(G)$. Since $\widetilde{F} \cap \theta(G) \subseteq \pi(G) \cap \theta(G)$, **Corollary 2.13** is proved.

**Lemma 2.37.** Let $G$ be a tree, $e_0 \in \pi(G)$ and $\Lambda(e_0) = L(G)$. If for $i = 1, 2$ $F_i \in S(G(i, e_0), L(G(i, e_0)))$ and $e_0 \in F_i$, then $(F_1 \cup F_2) \in S(G, L(G))$.

**Proof.** It is not hard to see that $(F_1 \cup F_2) \in M(G)$. Since
$$\beta(G \setminus (F_1 \cup F_2)) = \beta(G(1, e_0) \setminus F_1) + \beta(G(2, e_0) \setminus F_2) = L(G(1, e_0)) + L(G(2, e_0)) = \Lambda(e_0) = L(G)$$
then $(F_1 \cup F_2) \in S(G, L(G))$. Proof of **Lemma 2.37** is complete.

**Lemma 2.38.** Let $G$ be a tree, $e \in \pi(G)$, $f \in \pi(G)$, $e$ and $f$ are adjacent. If $\Lambda(e) \leq \Lambda(f)$, then there is $\widetilde{F} \in S(G, L(G))$, such that $e \notin \widetilde{F}$.

**Proof.** Let $F \in S(G, L(G))$. If $e \notin F$ then **Lemma 2.38** is proved, therefore we may assume that $e \in F$. **Lemma 2.35** implies that $\Lambda(e) = L(G)$, and, due to **Lemma 2.33**, we get $\Lambda(f) = L(G)$, therefore, using **Lemma 2.37**, we conclude that there is $\widetilde{F} \in S(G, L(G))$ such that $f \in \widetilde{F}$. Clearly $e \notin \widetilde{F}$. Proof of **Lemma 2.38** is complete.

**Lemma 2.39.** Let $G$ be a graph and $e \in \overline{\theta}(G)$. Then
     (a) if $\beta(G) = 1 + \beta(G - e)$ then $L(G) \leq L(G - e)$;
     (b) if $\beta(G) = \beta(G - e)$ then $L(G) \geq L(G - e)$.

**Proof.** (a) Let $F \in S(G, L(G))$. As $\beta(G) = 1 + \beta(G - e)$ then $e \in F$, therefore $(F \setminus \{e\}) \in M(G - e)$, and
$$L(G) = \beta(G \setminus F) = \beta(G - e \setminus (F \setminus \{e\})) \leq L(G - e).$$
(b) Let $F \in S(G - e, L(G - e))$. Since $\beta(G) = \beta(G - e)$, we have $F \in M(G)$, therefore
$$L(G) \geq \beta(G \setminus F) \geq \beta(G - e \setminus F) = L(G - e).$$
Proof of **Lemma 2.39** is complete.

**Lemma 2.40.** Let $G$ be a tree, $e = (u, v) \in \overline{\theta}(G)$ and $d_G(u) = 1$ (figure 8). Then the following inequality holds:
$$L(G \setminus \{u, v\}) \leq L(G) \leq 1 + L(G \setminus \{u, v\}).$$

**Proof.** Clearly, $\beta(G) = 1 + \beta(G \setminus \{u, v\})$. Let $F \in S(G \setminus \{u, v\}, L(G \setminus \{u, v\}))$. Set: $\widetilde{F} \equiv \{e\} \cup F$. Clearly, $\widetilde{F} \in M(G)$ and

$$L(G) \geq \beta(G \setminus \widetilde{F}) \geq \beta((G \setminus \{u,v\}) \setminus F) = L(G \setminus \{u,v\}).$$

On the other hand, due to **Corollary 2.8**, there is $F' \in S(G, L(G))$ such that $e \in F'$. Note that $(F' \setminus \{e\}) \in M(G \setminus \{u,v\})$, therefore

$$L(G) = \beta(G \setminus F') \leq 1 + \beta((G \setminus \{u,v\}) \setminus (F' \setminus \{e\})) \leq 1 + L(G \setminus \{u,v\}).$$

Proof of **Lemma 2.40** is complete.

**Lemma 2.41.** Let $G$ be a tree, $U = \{u_0, u_1, u_2\} \subseteq V(G)$, $d_G(u_0) = d_G(u_2) = 1$, $(u_{i-1}, u_i) \in E(G)$ for $i = 1, 2$ (figure 10). Then

(a) $L(G) = L(G \setminus U) + 1$;

(b) If $F' \in S(G \setminus U, L(G \setminus U))$, then $F \equiv (F' \cup \{(u_0, u_1)\}) \in S(G, L(G))$.

**Proof.** (a) **Lemma 2.40** implies that $L(G) \leq 1 + L(G \setminus U)$. Choose any $F_0 \in S(G \setminus U, L(G \setminus U))$ and consider $F_1 \in M(G)$ defined as:

$$F_1 \equiv F_0 \cup \{(u_0, u_1)\}.$$

As $d_{G \setminus F_1}(u_2) = 1$, we imply that there is $\widetilde{H} \in M(G \setminus F_1)$, such that $(u_1, u_2) \in \widetilde{H}$. It is not hard to see that

$$L(G) \geq \beta(G \setminus F_1) = |\widetilde{H}| = 1 + |\widetilde{H} \cap E((G \setminus U) \setminus F_0)| = 1 + \beta((G \setminus U) \setminus F_0) = 1 + L(G \setminus U),$$

or

$$L(G) = L(G \setminus U) + 1.$$

(b) As $d_{G \setminus F}(u_2) = 1$, we imply that there is $\widetilde{H} \in M(G \setminus F)$, such that $(u_1, u_2) \in \widetilde{H}$. It is not hard to see that

$$\beta(G \setminus F) = |\widetilde{H}| = 1 + |\widetilde{H} \cap E((G \setminus U) \setminus F')| = 1 + \beta((G \setminus U) \setminus F') = 1 + L(G \setminus U),$$

therefore, due to (a),

$$F \in S(G, L(G)).$$

Proof of **Lemma 2.41** is complete.

**Lemma 2.42.** Let $G$ be a tree, $U = \{u_0, \ldots, u_3\} \subseteq V(G)$, $d_G(u_0) = d_G(u_3) = 1$, $d_G(u_2) = 2$, $(u_{i-1}, u_i) \in E(G)$ for $i = 1, 2, 3$ (figure 11). Then

(a) $L(G) = L(G \setminus U) + 1$;

(b) If $F' \in S(G \setminus U, L(G \setminus U))$, then $F \equiv (F' \cup \{(u_0, u_1), (u_2, u_3)\}) \in S(G, L(G))$.

**Proof.** (a) Note that $\beta(G) = 1 + \beta(G - u_3)$, therefore, due to **Lemma 2.39** and **Lemma 2.41**, we get:

$$L(G) \leq L(G - u_3) = 1 + L(G \setminus U).$$

Choose any $F_0 \in S(G \setminus U, L(G \setminus U))$ and consider $F_1 \in M(G)$ defined as:

$$F_1 \equiv F_0 \cup \{(u_0, u_1), (u_2, u_3)\}.$$

As $d_{G \setminus F_1}(u_2) = 1$, we imply that there is $\widetilde{H} \in M(G \setminus F_1)$, such that $(u_1, u_2) \in \widetilde{H}$. It is not hard to see that

$$L(G) \geq \beta(G \setminus F_1) = |\widetilde{H}| = 1 + |\widetilde{H} \cap E((G \setminus U) \setminus F_0)| = 1 + \beta((G \setminus U) \setminus F_0) = 1 + L(G \setminus U),$$

or

$$L(G) = L(G \setminus U) + 1.$$

(b) As $d_{G \setminus F}(u_2) = 1$, we imply that there is $\widetilde{H} \in M(G \setminus F)$, such that $(u_1, u_2) \in \widetilde{H}$. It is not hard to see that

$$\beta(G \setminus F) = |\widetilde{H}| = 1 + |\widetilde{H} \cap E((G \setminus U) \setminus F')| = 1 + \beta((G \setminus U) \setminus F') = 1 + L(G \setminus U),$$

therefore, due to (a),
$$F \in S(G, L(G)).$$
Proof of **Lemma 2.42** is complete.

**Lemma 2.43.** Let $G$ be a tree, $U = \{u_0,...,u_5\} \subseteq V(G)$, $d_G(u_0) = d_G(u_4) = 1$, $d_G(u_1) = d_G(u_3) = 2$, $(u_{i-1}, u_i) \in E(G)$ for $i = 1,...,4$, $(u_2, u_5) \in E(G)$. Set $f = (u_2, u_5)$, $e = (u_1, u_2)$, $e' = (u_2, u_3)$ and let $u_0 \in V(G(2, f))$ (figure 24). Then

(a) if there is $\widetilde{F} \in S(G, L(G))$ such that $f \notin \widetilde{F}$, then if $F \in S(G-f, L(G-f))$ then $F \in S(G, L(G))$;

(b) if for each $F_0 \in S(G, L(G))$ $e \notin F_0$, then if $F' \in S(G', L(G'))$ and $(u_0, u_1) \in F'$, then $F \equiv (F' \cup \{(u_3, u_4)\}) \in S(G, L(G))$, where $G' \equiv G \setminus \{u_3, u_4\}$;

(c) if $e \in \pi(G)$, $f \in \pi(G)$, then $L(G(1, f)) \geq L(G(1, f) - u_2)$;

(d) if $e \in \pi(G)$, $f \in \pi(G)$ and $L(G(1, f)) \leq 1 + L(G(1, f) - u_2)$, then there is $\widetilde{F} \in S(G, L(G))$ such that $f \notin \widetilde{F}$;

(e) if $e \in \pi(G)$, $f \in \pi(G)$ and $L(G(1, f)) \geq 2 + L(G(1, f) - u_2)$, then $e \notin F_0$ for every $F_0 \in S(G, L(G))$.

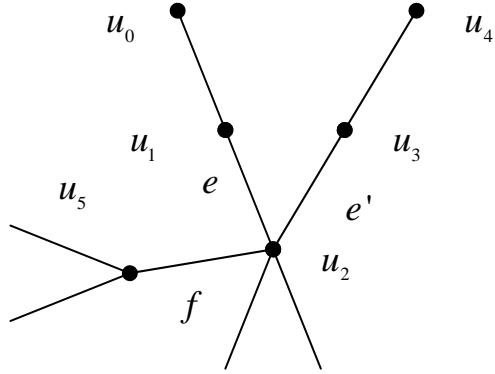

Figure 24

**Proof.** (a) Since $\widetilde{F} \in M(G)$, we get $\widetilde{F} \cap \{(u_0, u_1), (u_3, u_4)\} \neq \emptyset$, therefore
$$L(G) = \beta(G \setminus \widetilde{F}) = \beta(G - f \setminus \widetilde{F}) \leq L(G - f).$$
On the other hand, since
$$\beta(G \setminus F) \geq \beta(G - f \setminus F) = L(G - f),$$
we have
$$L(G) = L(G - f) \text{ and } F \in S(G, L(G)).$$

(b) Let $F_0 \in S(G, L(G))$ and $e \notin F_0$. Clearly $e' \notin F_0$. Therefore $\{(u_0, u_1), (u_3, u_4)\} \subseteq F_0$ and
$$L(G) = \beta(G \setminus F_0) = \beta(G' \setminus (F_0 \cap E(G'))) \leq L(G').$$
On the other hand, $F \in M(G)$ and, since $(u_0, u_1) \in F'$, we get:
$$\beta(G \setminus F) = \beta(G' \setminus F') = L(G'),$$
therefore
$$L(G) = L(G') \text{ and } F \in S(G, L(G)).$$

(c) Since $e \in \pi(G)$, $f \in \pi(G)$, we have:
$$\beta(G) = \beta(G(1,f) - u_2) + 2 + \beta(G(2,f) \setminus U),$$
$$\beta(G) = \beta(G(1,f)) + 2 + \beta(G(2,f) \setminus U),$$
therefore $\beta(G(1,f)) = \beta(G(1,f) - u_2)$ and, due to **Lemma 2.39**, the inequality $L(G(1,f)) \geq$
$\geq L(G(1,f) - u_2)$ is true.

(d) Using **Lemma 2.42**, we get
$$\Lambda(f) = 1 + L(G(1,f)) + L(G(2,f) \setminus U) \leq 2 + L(G(1,f) - u_2) + L(G(2,f) \setminus U) = \Lambda(e),$$
hence, due to **Lemma 2.38**, there is there is $\widetilde{F} \in S(G, L(G))$ such that $f \notin \widetilde{F}$;

(e) Using **Lemma 2.42** and **Lemma 2.33**, we get
$$L(G) \geq \Lambda(f) = 1 + L(G(1,f)) + L(G(2,f) \setminus U) \geq 3 + L(G(1,f) - u_2) + L(G(2,f) \setminus U) = 1 + \Lambda(e) > \Lambda(e),$$
hence, due to **Lemma 2.35**, $e \notin F_0$ for every $F_0 \in S(G, L(G))$. Proof of **Lemma 2.43** is complete.

**Lemma 2.44.** Let $G$ be a tree, $U = \{u_0, u_1, u_2\} \subseteq V(G)$, $d_G(u_1) = 2$, $(u_{i-1}, u_i) \in E(G)$ for $i = 1, 2$. Set: $f = (u_0, u_1)$, $e = (u_1, u_2)$, and let $u_2 \in V(G(2,f))$, $u_0 \in V(G(1,e))$ (figure 25). Suppose that there is $\widetilde{F} \in S(G, L(G))$ such that $e \notin \widetilde{F}$. Then $f \in \pi(G)$ and $\Lambda(f) = L(G)$.

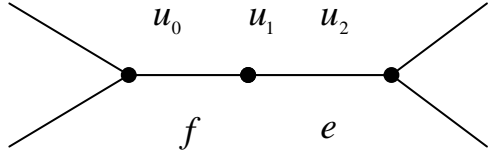

Figure 25

**Proof.** It is clear that $f \in \pi(G)$. If $f \in \widetilde{F}$, then the proof of **Lemma 2.44** follows from **Lemma 2.35**, therefore we can assume that $f \notin \widetilde{F}$. This implies that there is $w \in V(G)$, $w \neq u_1$ such that $(u_0, w) \in \widetilde{F}$.

**Case 1:** There is $H \in M(G \setminus \widetilde{F})$ with $e \in H$. Define:
$$F \equiv (\widetilde{F} \setminus \{(u_0, w)\}) \cup \{f\}.$$
Note that $F \in M(G)$ and
$$\beta(G \setminus F) \geq |H| = \beta(G \setminus \widetilde{F}) = L(G),$$
therefore $F \in S(G, L(G))$ and $\Lambda(f) = L(G)$ (**Lemma 2.35**).

**Case 2:** For each $H \in M(G \setminus \widetilde{F})$ $e \notin H$. It is clear that
$$L(G) = \beta(G \setminus \widetilde{F}) = \beta(G - e \setminus \widetilde{F}) = L(G(1,f)) + L(G(2,e) - u_1).$$
Choose $F_1 \in S(G(1,f), L(G(1,f)))$ with $f \in F_1$ (**Corollary 2.8**), and define:
$$F \equiv (\widetilde{F} \setminus (\widetilde{F} \cap E(G(1,f)))) \cup F_1.$$
Note that $F \in M(G)$ and
$$\beta(G \setminus F) \geq \beta(G - e \setminus F) = L(G(1,f)) + L(G(2,e) - u_1) = L(G)$$
therefore $F \in S(G, L(G))$ and $\Lambda(f) = L(G)$ (**Lemma 2.35**). Proof of **Lemma 2.44** is complete.

**Corollary 2.14.** Let $G$ be a tree, $U = \{u_0,...,u_3\} \subseteq V(G)$, $d_G(u_0) = 1$, $d_G(u_1) = d_G(u_2) = 2$, $(u_{i-1}, u_i) \in E(G)$ for $i = 1, 2, 3$ (figure 26). Set: $G' \equiv G \setminus \{u_0, u_1\}$. If $F' \in S(G, L(G))$ and $(u_2, u_3) \in F'$, then $F \equiv F' \cup \{(u_0, u_1)\} \in S(G, L(G))$.

**Proof** follows from **Corollary 2.8**, **Lemma 2.44** and **Lemma 2.37**.

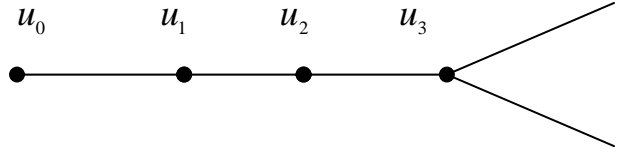

Figure 26

**Lemma 2.45.** Let $G$ be a tree, $U = \{u_0,...,u_5\} \subseteq V(G)$, $d_G(u_4) = 1$, $d_G(u_1) = d_G(u_3) = 2$, $d_G(u_2) = 3$, $(u_{i-1}, u_i) \in E(G)$ for $i = 1,...,4$, $(u_2, u_5) \in E(G)$. Set $g = (u_0, u_1)$, $f = (u_1, u_2)$, $e = (u_2, u_5)$ and let $u_4 \in V(G(2,g))$, $u_4 \in V(G(2,f))$, $u_4 \in V(G(1,e))$ (figure 27). Suppose that there is $\widetilde{F} \in S(G, L(G))$ such that $e \notin \widetilde{F}$. Then $f \in \pi(G)$ and $\Lambda(f) = L(G)$.

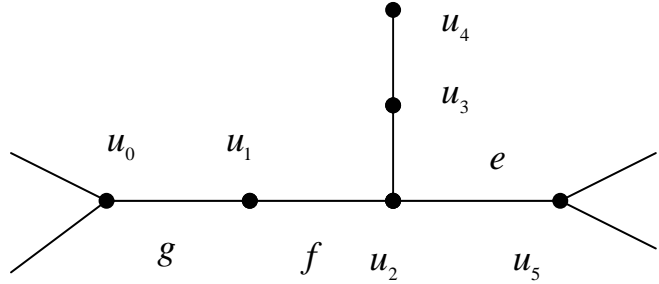

Figure 27

**Proof.** It is clear that $f \in \pi(G)$. **Lemma 2.35** implies that it suffices to show that there is $F \in S(G, L(G))$ such that $f \in F$.

Consider $\widetilde{F} \in S(G, L(G))$ and suppose that $f \notin \widetilde{F}$. Then $g \in \widetilde{F}$. **Lemma 2.3** implies that, without loss of generality, we may assume that $(u_2, u_3) \in \widetilde{F}$. Using **Lemma 2.42**, we get:

$$L(G) = \beta(G \setminus \widetilde{F}) = L(G(1, g)) + 2 + \beta((G(2, e) - u_2) \setminus (\widetilde{F} \cap E(G(2, e) - u_2))).$$

Note that conditions $e \notin \widetilde{F}$ and $g \in \widetilde{F}$ imply that $L(G(1, f)) \geq 1 + L(G(1, g))$. Since $d_{G(1,f)}(u_2) = 1$, due to **Corollary 2.8**, there is $F_0 \in S(G(1, f), L(G(1, f)))$ with $f \in F_0$. Assume:

$$F \equiv (\widetilde{F} \setminus ((\widetilde{F} \cap E(G(1, f))) \cup \{(u_2, u_3)\})) \cup F_0 \cup \{(u_3, u_4)\}.$$

Clearly, $F \in M(G)$, $f \in F$. Using **Lemma 2.42**, we get:

$$\beta(G \setminus F) = L(G(1, f)) + 1 + \beta((G(2, e) - u_2) \setminus (\widetilde{F} \cap E(G(2, e) - u_2))) \geq 2 + L(G(1, g)) +$$
$$+ \beta((G(2, e) - u_2) \setminus (\widetilde{F} \cap E(G(2, e) - u_2))) = L(G),$$

therefore $F \in S(G, L(G))$. Proof of **Lemma 2.45** is complete.

**Lemma 2.46.** Let $G$ be a tree, $U = \{u_0,...,u_7\} \subseteq V(G)$, $d_G(u_0) = d_G(u_5) = 1$, $d_G(u_1) = d_G(u_4) = 2$, $d_G(u_2) = d_G(u_3) = 3$, $(u_{i-1}, u_i) \in E(G)$ for $i = 1,...,5$, $(u_2, u_6) \in E(G)$, $(u_3, u_7) \in E(G)$. Set $g = (u_2, u_6)$, $f = (u_2, u_3)$, $e = (u_3, u_7)$ and let $u_0 \in V(G(2,g))$, $u_0 \in V(G(1,f))$, $u_0 \in V(G(1,e))$ (figure 28). Suppose that there is $\widetilde{F} \in S(G, L(G))$ such that $e \notin \widetilde{F}$. Then
   (a) if $\beta(G(1,g)) = \beta(G(1,g) - u_2)$ then $\Lambda(f) = L(G)$;
   (b) if $\beta(G(1,g)) = 1 + \beta(G(1,g) - u_2)$ then
$$L(G) = \max\{\Lambda(g), \Lambda(f)\} \text{ and } L(G(1,g)) \le L(G(1,g) - u_2);$$
   (c) if $\beta(G(1,g)) = 1 + \beta(G(1,g) - u_2)$ and $L(G(1,g)) = L(G(1,g) - u_2)$ then $\Lambda(g) = L(G)$;
   (d) if $\beta(G(1,g)) = 1 + \beta(G(1,g) - u_2)$ and $L(G(1,g)) \le L(G(1,g) - u_2) - 1$ then $\Lambda(f) = L(G)$.

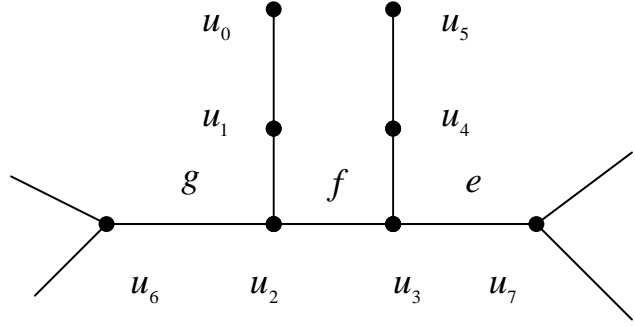

Figure 28

**Proof.** (a) Since $e \notin \widetilde{F}$ and $\beta(G(1,g)) = \beta(G(1,g) - u_2)$, then, clearly, $f \in \widetilde{F}$, therefore, due to **Lemma 2.35**, the equality $\Lambda(f) = L(G)$ is true.

(b) Note that **Lemma 2.39** immediately implies that $L(G(1,g)) \le L(G(1,g) - u_2)$. On the other hand, since $e \notin \widetilde{F}$ and $\widetilde{F} \in M(G)$, then clearly, $\widetilde{F} \cap \{g, f\} \ne \varnothing$, therefore, using **Lemma 2.33** and **Lemma 2.35**, we get $L(G) = \max\{\Lambda(g), \Lambda(f)\}$.

(c) Note that if $g \in \widetilde{F}$, then **Lemma 2.35** immediately implies $\Lambda(g) = L(G)$, therefore we may assume that $g \notin \widetilde{F}$, hence $f \in \widetilde{F}$ and, using **Lemma 2.42** we get:
$$L(G) = \beta(G \setminus \widetilde{F}) = 2 + L(G(1,g) - u_2) + L(G(2,e) - u_3).$$
Since $\beta(G(1,g)) = 1 + \beta(G(1,g) - u_2)$ and $e \notin \widetilde{F}$, then $\{g, f\} \subseteq \pi(G)$, therefore
$$\beta(G) = 2 + \beta(G(1,g)) + \beta(G(2,e)),$$
$$\beta(G) = 3 + \beta(G(1,g) - u_2) + \beta(G(2,e) - u_3),$$
hence $\beta(G(2,e)) = \beta(G(2,e) - u_3)$ and, due to **Lemma 2.39**, $L(G(2,e)) \ge L(G(2,e) - u_3)$. Using **Corollary 2.14** and **Lemma 2.42**, we get
$$\Lambda(g) = L(G(1,g)) + 2 + L(G(2,e)) \ge L(G(1,g) - u_2) + 2 + L(G(2,e) - u_3) = L(G),$$
thus due to **Lemma 2.33**, $\Lambda(g) = L(G)$.

(d) Note that if $f \in \widetilde{F}$, then **Lemma 2.35** immediately implies $\Lambda(f) = L(G)$, therefore we may assume that $f \notin \widetilde{F}$, hence $g \in \widetilde{F}$. **Lemma 2.3** implies that, without loss of generality, we may assume that $(u_3, u_4) \in \widetilde{F}$ and, using **Lemma 2.42**, we get:

$$L(G) = \beta(G \setminus \widetilde{F}) = 2 + L(G(1, g)) + L(G(2, f) \setminus \{u_2, u_5\}).$$

**Lemma 2.40** and **Lemma 2.42** imply

$$\Lambda(f) = L(G(1, g) - u_2) + 2 + L(G(2, e) - u_3) \geq$$
$$\geq L(G(1, g)) + 3 + L(G(2, e) - u_3) \geq L(G(1, g)) + 2 + L(G(2, f) \setminus \{u_2, u_5\}) = L(G),$$

therefore, due to **Lemma 2.33**, we get $\Lambda(f) = L(G)$. Proof of **Lemma 2.46** is complete.

**Lemma 2.47.** Let $G$ be a tree, $U = \{u_0, ..., u_3\} \subseteq V(G)$, $d_G(u_1) = 2$, $d_G(u_3) = 1$, $(u_{i-1}, u_i) \in E(G)$ for $i = 1, ..., 3$ (figure 29). If $F_i \in S(G(i, (u_0, u_1)), L(G(i, (u_0, u_1))))$ and $(u_0, u_1) \in F_i$ for $i = 1, 2$, then $F_1 \cup F_2 \in S(G, L(G))$.

**Proof** follows from **Lemma 2.37**, **Lemma 2.44** and **Corollary 2.8**.

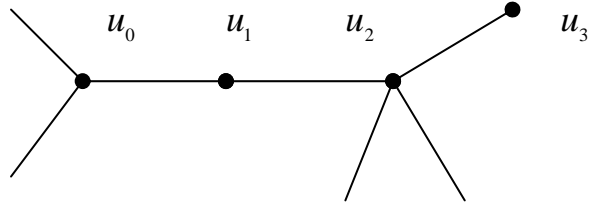

Figure 29

**Lemma 2.48.** Let $G$ be a tree, $U = \{u_0, ..., u_6\} \subseteq V(G)$, $d_G(u_4) = d_G(u_6) = 1$, $d_G(u_1) = d_G(u_5) = 2$, $d_G(u_2) = 3$, $(u_{i-1}, u_i) \in E(G)$ for $i = 1, 2, 3, 4, 6$, $(u_2, u_5) \in E(G)$ (figure 30). If $F_i \in S(G(i, (u_1, u_2)), L(G(i, (u_1, u_2))))$ and $(u_1, u_2) \in F_i$ for $i = 1, 2$, then $F_1 \cup F_2 \in S(G, L(G))$.

**Proof** follows from **Lemma 2.37**, **Lemma 2.45** and **Corollary 2.8**.

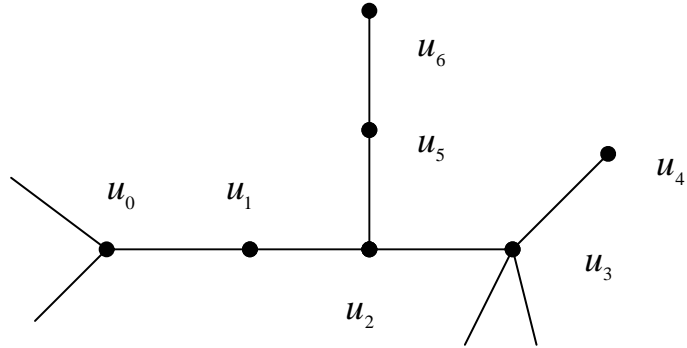

Figure 30

**Lemma 2.49.** Let $G$ be a tree, $U = \{u_0, ..., u_8\} \subseteq V(G)$, $d_G(u_0) = d_G(u_5) = d_G(u_7) = 1$, $d_G(u_1) = d_G(u_4) = 2$, $d_G(u_2) = d_G(u_3) = 3$, $(u_{i-1}, u_i) \in E(G)$ for $i = 1, 2, 3, 4, 5, 7$, $(u_2, u_8) \in E(G)$, $(u_3, u_6) \in E(G)$. Set: $g = (u_2, u_8)$, $f = (u_2, u_3)$, and let $u_7 \in V(G(2, g))$, $u_7 \in V(G(2, f))$ (figure 31). Then

(a) if $\beta(G(1, g)) = \beta(G(1, g) - u_2)$ then if $F_i \in S(G(i, f), L(G(i, f)))$ and $f \in F_i$, for $i = 1, 2$, then $F_1 \cup F_2 \in S(G, L(G))$;

(b) if $\beta(G(1, g)) = 1 + \beta(G(1, g) - u_2)$ and $L(G(1, g)) = L(G(1, g) - u_2)$, then if $F_i \in S(G(i, g), L(G(i, g)))$ and $g \in F_i$, for $i = 1, 2$, then $F_1 \cup F_2 \in S(G, L(G))$;

(c) if $\beta(G(1,g)) = 1 + \beta(G(1,g) - u_2)$ and $L(G(1,g)) \le L(G(1,g) - u_2) - 1$ then if $F_i \in S(G(i,f), L(G(i,f)))$ and $f \in F_i$, for $i = 1, 2$, then $F_1 \cup F_2 \in S(G, L(G))$.

**Proof** follows from **Lemma 2.37**, **Lemma 2.46** and **Corollary 2.8**.

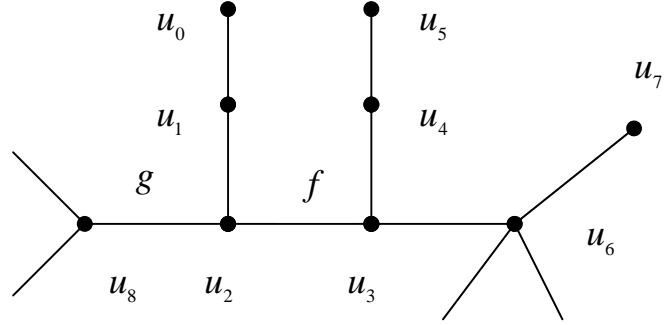

Figure 31

**Lemma 2.50.** Let $G$ be a tree, $U = \{u_0, ..., u_4\} \subseteq V(G)$, $d_G(u_1) = d_G(u_3) = 2$, $(u_{i-1}, u_i) \in E(G)$ for $i = 1, ..., 4$. Set: $f = (u_0, u_1)$, $e = (u_1, u_2)$, $e' = (u_2, u_3)$, $f' = (u_3, u_4)$, and let $u_4 \in V(G(2, f))$, $u_0 \in V(G(1, e))$, $u_0 \in V(G(1, e'))$, $u_0 \in V(G(1, f'))$ (figure 32). Then

(a) if $e \notin \pi(G)$, then if $F_i \in S(G(i,f), L(G(i,f)))$ and $f \in F_i$, for $i = 1, 2$, then $F_1 \cup F_2 \in S(G, L(G))$;

(b) if $e' \notin \pi(G)$, then if $F_i' \in S(G(i,f'), L(G(i,f')))$ and $f' \in F_i'$, for $i = 1, 2$, then $F_1' \cup F_2' \in S(G, L(G))$;

(c) if $e \in \pi(G)$, $e' \in \pi(G)$, and $L(G(1,e)) + L(G(2,f')) \ge L(G(1,f)) + L(G(2,e'))$, then if $F_i' \in S(G(i,f'), L(G(i,f')))$ and $f' \in F_i'$, for $i = 1, 2$, then $F_1' \cup F_2' \in S(G, L(G))$;

(d) if $e \in \pi(G)$, $e' \in \pi(G)$, and $L(G(1,e)) + L(G(2,f')) \le L(G(1,f)) + L(G(2,e'))$, then if $F_i \in S(G(i,f), L(G(i,f)))$ and $f \in F_i$, for $i = 1, 2$, then $F_1 \cup F_2 \in S(G, L(G))$.

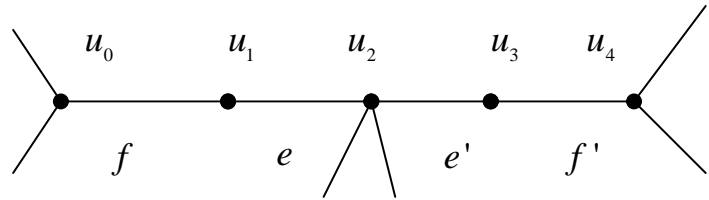

Figure 32

**Proof** follows from **Lemma 2.44**, **Lemma 2.37**, **Lemma 2.38**, **Lemma 2.42**, **Corollary 2.8**, and the following two equalities:
$$\Lambda(e) = L(G(1,e)) + L(G(2,f')) + 1 + L(G(2,e) \setminus (\{u_1\} \cup V(G(2,e')))),$$
$$\Lambda(e') = L(G(1,f)) + L(G(2,e')) + 1 + L(G(1,e') \setminus (\{u_3\} \cup V(G(1,e)))).$$

**Lemma 2.51.** Let $G$ be a tree, $U = \{u_0,...,u_7\} \subseteq V(G)$, $d_G(u_1) = d_G(u_4) = d_G(u_6) = 2$, $d_G(u_3) = 3$, $d_G(u_7) = 1$, $(u_{i-1}, u_i) \in E(G)$ for $i = 1,...,5,7$, $(u_3, u_6) \in E(G)$. Set: $f = (u_0, u_1)$, $e = (u_1, u_2)$, $e' = (u_2, u_3)$, $f' = (u_3, u_4)$, and let $u_7 \in V(G(2, f))$, $u_0 \in V(G(1, e))$, $u_0 \in V(G(1, e'))$, $u_0 \in V(G(1, f'))$ (figure 33). Then

(a) if $e \notin \pi(G)$, then if $F_i \in S(G(i, f), L(G(i, f)))$ and $f \in F_i$, for $i = 1, 2$, then $F_1 \cup F_2 \in S(G, L(G))$;

(b) if $e' \notin \pi(G)$, then if $F_i' \in S(G(i, f'), L(G(i, f')))$ and $f' \in F_i'$, for $i = 1, 2$, then $F_1' \cup F_2' \in S(G, L(G))$;

(c) if $e \in \pi(G)$, $e' \in \pi(G)$, and $L(G(1,e)) + L(G(2, f')) \geq L(G(1, f)) + L(G(2, e'))$, then if $F_i' \in S(G(i, f'), L(G(i, f')))$ and $f' \in F_i'$, for $i = 1, 2$, then $F_1' \cup F_2' \in S(G, L(G))$;

(d) if $e \in \pi(G)$, $e' \in \pi(G)$, and $L(G(1,e)) + L(G(2, f')) \leq L(G(1, f)) + L(G(2, e')) - 1$, then if $F_i \in S(G(i, f), L(G(i, f)))$ and $f \in F_i$, for $i = 1, 2$, then $F_1 \cup F_2 \in S(G, L(G))$.

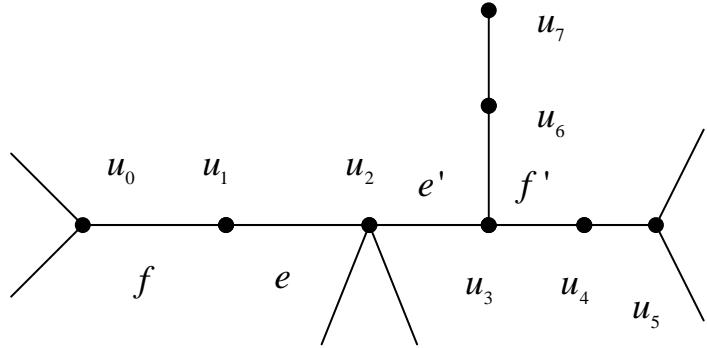

Figure 33

**Proof** follows from **Lemma 2.44**, **Lemma 2.45**, **Lemma 2.37**, **Lemma 2.38**, **Lemma 2.40**, **Lemma 2.42**, **Corollary 2.8**, and the following two equalities:
$$\Lambda(e) = L(G(1, e)) + L(G(2, f')) + 1 + L(G(2, e) \setminus (\{u_6, u_7\} \cup V(G(2, f')))),$$
$$\Lambda(e') = L(G(1, f)) + L(G(2, e')) + 1 + L(G(1, e') \setminus (\{u_3\} \cup V(G(1, e)))).$$

**Lemma 2.52.** Let $G$ be a tree, $U = \{u_0,...,u_{10}\} \subseteq V(G)$, $d_G(u_1) = d_G(u_5) = d_G(u_7) = d_G(u_9) = 2$, $d_G(u_2) = d_G(u_4) = 3$, $d_G(u_8) = d_G(u_{10}) = 1$, $(u_{i-1}, u_i) \in E(G)$ for $i = 1,...,6,8,10$, $(u_2, u_7) \in E(G)$, $(u_4, u_9) \in E(G)$. Set: $f = (u_1, u_2)$, $e = (u_2, u_3)$, $e' = (u_3, u_4)$, $f' = (u_4, u_5)$, and let $u_0 \in V(G(1, f))$, $u_0 \in V(G(1, e))$, $u_0 \in V(G(1, e'))$, $u_0 \in V(G(1, f'))$ (figure 34). Then

(a) if $e \notin \pi(G)$, then if $F_i \in S(G(i, f), L(G(i, f)))$ and $f \in F_i$, for $i = 1, 2$, then $F_1 \cup F_2 \in S(G, L(G))$;

(b) if $e' \notin \pi(G)$, then if $F_i' \in S(G(i, f'), L(G(i, f')))$ and $f' \in F_i'$, for $i = 1, 2$, then $F_1' \cup F_2' \in S(G, L(G))$;

(c) if $e \in \pi(G)$, $e' \in \pi(G)$, and $L(G(1,e)) + L(G(2, f')) \geq L(G(1, f)) + L(G(2, e'))$, then if $F_i' \in S(G(i, f'), L(G(i, f')))$ and $f' \in F_i'$, for $i = 1, 2$, then $F_1' \cup F_2' \in S(G, L(G))$;

(d) if $e \in \pi(G)$, $e' \in \pi(G)$, and $L(G(1,e)) + L(G(2,f')) \leq L(G(1,f)) + L(G(2,e'))$, then if $F_i \in S(G(i,f), L(G(i,f)))$ and $f \in F_i$, for $i = 1, 2$, then $F_1 \cup F_2 \in S(G, L(G))$.

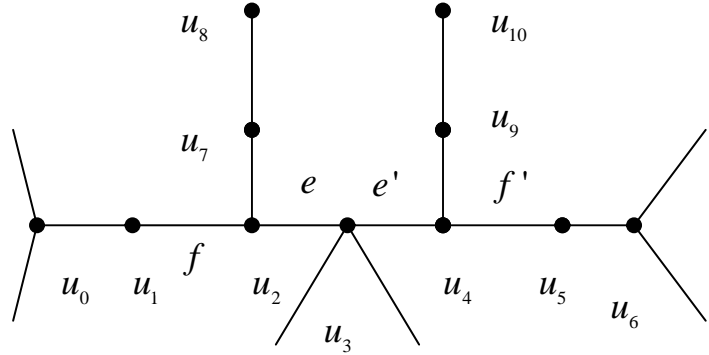

Figure 34

**Proof** follows from **Lemma 2.45**, **Lemma 2.37**, **Lemma 2.38**, **Lemma 2.42**, **Corollary 2.8**, and the following two equalities:
$$\Lambda(e) = L(G(1,e)) + L(G(2,f')) + 1 + L(G(2,e) \setminus (\{u_9, u_{10}\} \cup V(G(2,f')))),$$
$$\Lambda(e') = L(G(1,f)) + L(G(2,e')) + 1 + L(G(1,e) \setminus (\{u_7, u_8\} \cup V(G(1,f)))).$$

**Lemma 2.53.** Let $G$ be a tree, $U = \{u_0, ..., u_9\} \subseteq V(G)$, $d_G(u_0) = d_G(u_7) = 1$, $d_G(u_1) = d_G(u_8) = d_G(u_5) = 2$, $d_G(u_2) = d_G(u_3) = 3$, $(u_{i-1}, u_i) \in E(G)$ for $i = 1, ..., 6, 8$, $(u_2, u_9) \in E(G)$, $(u_3, u_8) \in E(G)$. Set: $g = (u_9, u_2)$, $f = (u_2, u_3)$, $e = (u_3, u_4)$, $e' = (u_4, u_5)$, $f' = (u_5, u_6)$, and let $u_0 \in V(G(2,g))$, $u_0 \in V(G(1,f))$, $u_0 \in V(G(1,e))$, $u_0 \in V(G(1,e'))$, $u_0 \in V(G(1,f'))$ (figure 35). Then

(a) if $e \notin \pi(G)$, then if $F_i \in S(G(i,f), L(G(i,f)))$ and $f \in F_i$, for $i = 1, 2$, then $F_1 \cup F_2 \in S(G, L(G))$;

(b) if $e' \notin \pi(G)$, then if $F_i' \in S(G(i,f'), L(G(i,f')))$ and $f' \in F_i'$, for $i = 1, 2$, then $F_1' \cup F_2' \in S(G, L(G))$;

(c) if $e \in \pi(G)$, $e' \in \pi(G)$ then
$$\beta(G(1,g)) + \beta(G(2,f')) = \beta(G(1,g) - u_2) + \beta(G(2,e'));$$

(d) if $e \in \pi(G)$, $e' \in \pi(G)$, $\beta(G(1,g)) = \beta(G(1,g) - u_2)$ and $L(G(1,g) - u_2) + L(G(2,e')) \leq L(G(1,g)) + L(G(2,f'))$, then if $F_i' \in S(G(i,f'), L(G(i,f')))$ and $f' \in F_i'$, for $i = 1, 2$, then $F_1' \cup F_2' \in S(G, L(G))$;

(e) if $e \in \pi(G)$, $e' \in \pi(G)$, $\beta(G(1,g)) = \beta(G(1,g) - u_2)$ and $L(G(1,g) - u_2) + L(G(2,e')) \geq L(G(1,g)) + L(G(2,f')) + 1$, then if $F_i \in S(G(i,f), L(G(i,f)))$ and $f \in F_i$, for $i = 1, 2$, then $F_1 \cup F_2 \in S(G, L(G))$;

(f) if $e \in \pi(G)$, $e' \in \pi(G)$, $\beta(G(1,g)) = \beta(G(1,g) - u_2) + 1$, $L(G(1,g)) = L(G(1,g) - u_2)$, then if $F_i' \in S(G(i,f'), L(G(i,f')))$ and $f' \in F_i'$, for $i = 1, 2$, then $F_1' \cup F_2' \in S(G, L(G))$;

(g) if $e \in \pi(G)$, $e' \in \pi(G)$, $\beta(G(1,g)) = \beta(G(1,g)-u_2)+1$, $L(G(1,g)) \leq L(G(1,g)-u_2)-1$, $L(G(1,g)-u_2)+L(G(2,e')) \leq L(G(1,g))+L(G(2,f'))$, then if $F_i' \in S(G(i,f'),L(G(i,f')))$ and $f' \in F_i'$, for $i=1,2$, then $F_1' \cup F_2' \in S(G,L(G))$;

(h) if $e \in \pi(G)$, $e' \in \pi(G)$, $\beta(G(1,g)) = \beta(G(1,g)-u_2)+1$, $L(G(1,g)) \leq L(G(1,g)-u_2)-1$, $L(G(1,g)-u_2)+L(G(2,e')) \geq L(G(1,g))+L(G(2,f'))+1$, then if $F_i \in S(G(i,f),L(G(i,f)))$ and $f \in F_i$, for $i=1,2$, then $F_1 \cup F_2 \in S(G,L(G))$.

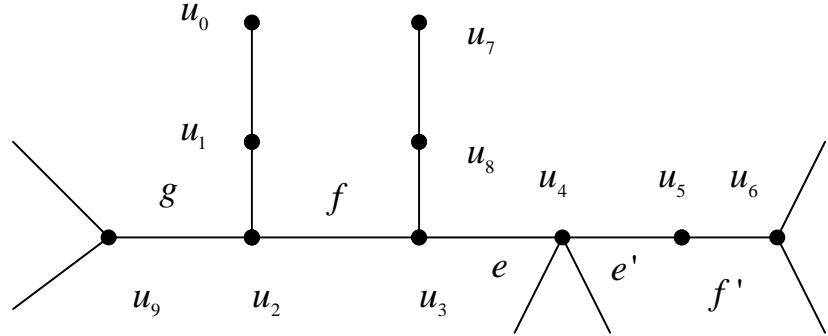

Figure 35

**Proof.** (a) follows from **Lemma 2.46** and **Lemma 2.37**. (b) follows from **Lemma 2.44** and **Lemma 2.37**.

(c) Since $e \in \pi(G), e' \in \pi(G)$, we have
$$\beta(G) = 3 + \beta(G(1,g)) + \beta(G(2,f')) + \beta(G(2,e) \setminus (\{u_3\} \cup V(G(2,e')))),$$
$$\beta(G) = 3 + \beta(G(1,g)-u_2) + \beta(G(2,e')) + \beta(G(1,e') \setminus (\{u_5\} \cup V(G(1,e)))),$$
therefore
$$\beta(G(1,g)) + \beta(G(2,f')) = \beta(G(1,g)-u_2) + \beta(G(2,e')).$$

(d) **Lemma 2.44** and **Lemma 2.37** imply that it suffices to show that there is $F \in S(G,L(G))$ such that $e' \notin F$.

Using **Lemma 2.47**, **Lemma 2.42, Corollary 2.14, Lemma 2.49**, we get
$$\Lambda(e) = 3 + L(G(1,g)) + L(G(2,f')) + L(G(2,e) \setminus (\{u_3\} \cup V(G(2,e')))) \geq$$
$$\geq L(G(1,g)-u_2) + 2 + L(G(2,e')) + L(G(1,e') \setminus (\{u_7,u_8\} \cup V(G(1,f)))) = \Lambda(e'),$$
therefore, due to **Lemma 2.38**, there is $F \in S(G,L(G))$ such that $e' \notin F$.

(e) **Lemma 2.46** and **Lemma 2.37** imply that it suffices to show that there is $F \in S(G,L(G))$ such that $e \notin F$.

Using **Lemma 2.47**, **Lemma 2.42, Corollary 2.14, Lemma 2.49**, we get
$$\Lambda(e') = L(G(1,g)-u_2) + 2 + L(G(2,e')) + L(G(1,e') \setminus (\{u_7,u_8\} \cup V(G(1,f)))) \geq$$
$$\geq 3 + L(G(1,g)) + L(G(2,f')) + L(G(2,e) \setminus (\{u_3\} \cup V(G(2,e')))) = \Lambda(e),$$
therefore, due to **Lemma 2.38**, there is $F \in S(G,L(G))$ such that $e \notin F$.

(f) **Lemma 2.44** and **Lemma 2.37** imply that it suffices to show that there is $F \in S(G,L(G))$ such that $e' \notin F$.

Using **Lemma 2.49**, **Lemma 2.42, Lemma 2.40**, **Corollary 2.14, Lemma 2.47** and the inequality $L(G(2, f')) \geq L(G(2, e'))$, we get

$$\Lambda(e) = 3 + L(G(1, g)) + L(G(2, f')) + L(G(2,e) \setminus (\{u_3\} \cup V(G(2,e')))) \geq$$
$$\geq L(G(1, g)) + 3 + L(G(2, e')) + L(G(1, e') \setminus (\{u_5\} \cup V(G(1,e)))) = \Lambda(e'),$$

therefore, due to **Lemma 2.38**, there is $F \in S(G, L(G))$ such that $e' \notin F$.

(g) **Lemma 2.45** and **Lemma 2.37** imply that it suffices to show that there is $F \in S(G, L(G))$ such that $e' \notin F$.

Using **Lemma 2.47**, **Lemma 2.42, Corollary 2.14, Lemma 2.49**, we get

$$\Lambda(e) = 3 + L(G(1, g)) + L(G(2, f')) + L(G(2,e) \setminus (\{u_3\} \cup V(G(2,e')))) \geq$$
$$\geq L(G(1, g) - u_2) + 2 + L(G(2, e')) + L(G(1, e') \setminus (\{u_7, u_8\} \cup V(G(1, f')))) = \Lambda(e'),$$

therefore, due to **Lemma 2.38**, there is $F \in S(G, L(G))$ such that $e' \notin F$.

(h) **Lemma 2.46** and **Lemma 2.37** imply that it suffices to show that there is $F \in S(G, L(G))$ such that $e \notin F$.

Using **Lemma 2.49**, **Lemma 2.42, Corollary 2.14, Lemma 2.47**, we get

$$\Lambda(e') = L(G(1, g) - u_2) + 2 + L(G(2, e')) + L(G(1, e') \setminus (\{u_7, u_8\} \cup V(G(1, f')))) \geq$$
$$\geq 3 + L(G(1, g)) + L(G(2, f')) + L(G(2,e) \setminus (\{u_3\} \cup V(G(2, e')))) = \Lambda(e),$$

therefore, due to **Lemma 2.38**, there is $F \in S(G, L(G))$ such that $e \notin F$. Proof of **Lemma 2.53** is complete.

**Lemma 2.54.** Let $G$ be a tree, $U = \{u_0, ..., u_{12}\} \subseteq V(G)$, $d_G(u_0) = d_G(u_9) = d_G(u_{11}) = 1$, $d_G(u_1) = d_G(u_8) = d_G(u_{10}) = d_G(u_6) = 2$, $d_G(u_2) = d_G(u_3) = d_G(u_5) = 3$, $(u_{i-1}, u_i) \in E(G)$ for $i = 1, ..., 7, 9, 11$, $(u_2, u_{12}) \in E(G)$, $(u_3, u_8) \in E(G)$, $(u_5, u_{10}) \in E(G)$. Set: $g = (u_{12}, u_2)$, $f = (u_2, u_3)$, $e = (u_3, u_4)$, $e' = (u_4, u_5)$, $f' = (u_5, u_6)$, $g' = (u_6, u_7)$, and let $u_0 \in V(G(2, g))$, $u_0 \in V(G(1, f))$, $u_0 \in V(G(1, e))$, $u_0 \in V(G(1, e'))$, $u_0 \in V(G(1, f'))$, $u_0 \in V(G(1, g'))$ (figure 36). Then

   (a) if $e \notin \pi(G)$, then if $F_i \in S(G(i, f), L(G(i, f)))$ and $f \in F_i$, for $i = 1, 2$, then $F_1 \cup F_2 \in S(G, L(G))$;

   (b) if $e' \notin \pi(G)$, then if $F_i' \in S(G(i, f'), L(G(i, f')))$ and $f' \in F_i'$, for $i = 1, 2$, then $F_1' \cup F_2' \in S(G, L(G))$;

   (c) if $e \in \pi(G)$, $e' \in \pi(G)$ then
   $$\beta(G(1, g)) + \beta(G(2, f')) = \beta(G(1, g) - u_2) + \beta(G(2, g')) + 1;$$

   (d) if $e \in \pi(G)$, $e' \in \pi(G)$, $\beta(G(1, g)) = \beta(G(1, g) - u_2)$ and $L(G(1, g) - u_2) + L(G(2, g')) \leq L(G(1, g)) + L(G(2, f'))$, then if $F_i' \in S(G(i, f'), L(G(i, f')))$ and $f' \in F_i'$, for $i = 1, 2$, then $F_1' \cup F_2' \in S(G, L(G))$;

   (e) if $e \in \pi(G)$, $e' \in \pi(G)$, $\beta(G(1, g)) = \beta(G(1, g) - u_2)$ and $L(G(1, g) - u_2) + L(G(2, g')) \geq L(G(1, g)) + L(G(2, f'))$, then if $F_i \in S(G(i, f), L(G(i, f)))$ and $f \in F_i$, for $i = 1, 2$, then $F_1 \cup F_2 \in S(G, L(G))$;

   (f) if $e \in \pi(G)$, $e' \in \pi(G)$, $\beta(G(1, g)) = \beta(G(1, g) - u_2) + 1$, $L(G(1, g)) = L(G(1, g) - u_2)$, then if $F_i' \in S(G(i, f'), L(G(i, f')))$ and $f' \in F_i'$, for $i = 1, 2$, then $F_1' \cup F_2' \in S(G, L(G))$;

(g) if $e \in \pi(G)$, $e' \in \pi(G)$, $\beta(G(1,g)) = \beta(G(1,g) - u_2) + 1$, $L(G(1,g)) \leq L(G(1,g) - u_2) - 1$, $L(G(1,g) - u_2) + L(G(2,g')) \leq L(G(1,g)) + L(G(2,f'))$, then if $F_i' \in S(G(i,f'), L(G(i,f')))$ and $f' \in F_i'$, for $i = 1, 2$, then $F_1' \cup F_2' \in S(G, L(G))$;

(h) if $e \in \pi(G)$, $e' \in \pi(G)$, $\beta(G(1,g)) = \beta(G(1,g) - u_2) + 1$, $L(G(1,g)) \leq L(G(1,g) - u_2) - 1$, $L(G(1,g) - u_2) + L(G(2,g')) \geq L(G(1,g)) + L(G(2,f'))$, then if $F_i \in S(G(i,f), L(G(i,f)))$ and $f \in F_i$, for $i = 1, 2$, then $F_1 \cup F_2 \in S(G, L(G))$.

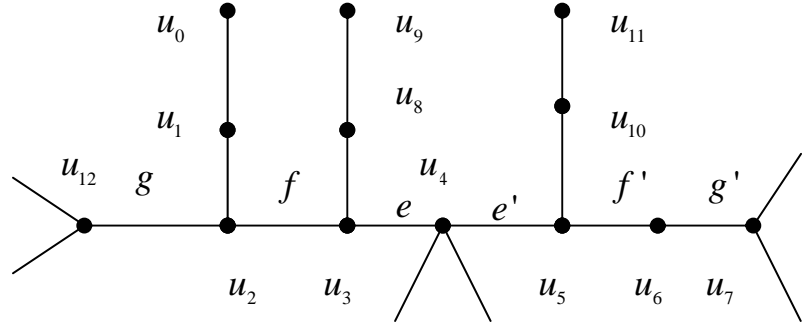

Figure 36

**Proof.** (a) follows from **Lemma 2.46** and **Lemma 2.37**. (b) follows from **Lemma 2.45** and **Lemma 2.37**.

(c) Since $e \in \pi(G), e' \in \pi(G)$, we have
$$\beta(G) = 4 + \beta(G(1,g)) + \beta(G(2,f')) + \beta(G(2,e) \setminus (\{u_3\} \cup V(G(2,e')))),$$
$$\beta(G) = 5 + \beta(G(1,g) - u_2) + \beta(G(2,g')) + \beta(G(1,e') \setminus (\{u_5\} \cup V(G(1,e)))),$$
therefore
$$\beta(G(1,g)) + \beta(G(2,f')) = \beta(G(1,g) - u_2) + \beta(G(2,g')) + 1.$$

(d) **Lemma 2.45** and **Lemma 2.37** imply that it suffices to show that there is $F \in S(G, L(G))$ such that $e' \notin F$.

Using **Lemma 2.48**, **Lemma 2.42**, **Corollary 2.14**, **Lemma 2.49**, we get
$$\Lambda(e) = 3 + L(G(1,g)) + L(G(2,f')) + L(G(2,e) \setminus (\{u_{10}, u_{11}\} \cup V(G(2,f')))) \geq$$
$$\geq L(G(1,g) - u_2) + 3 + L(G(2,g')) + L(G(1,e') \setminus (\{u_8, u_9\} \cup V(G(1,f)))) = \Lambda(e'),$$
therefore, due to **Lemma 2.38**, there is $F \in S(G, L(G))$ such that $e' \notin F$.

(e) **Lemma 2.46** and **Lemma 2.37** imply that it suffices to show that there is $F \in S(G, L(G))$ such that $e \notin F$.

Using **Lemma 2.48**, **Lemma 2.42**, **Corollary 2.14**, **Lemma 2.49**, we get
$$\Lambda(e') = L(G(1,g) - u_2) + 3 + L(G(2,g')) + L(G(1,e') \setminus (\{u_8, u_9\} \cup V(G(1,f)))) \geq$$
$$\geq 3 + L(G(1,g)) + L(G(2,f')) + L(G(2,e) \setminus (\{u_{10}, u_{11}\} \cup V(G(2,f')))) = \Lambda(e),$$
therefore, due to **Lemma 2.38**, there is $F \in S(G, L(G))$ such that $e \notin F$.

(f) **Lemma 2.45** and **Lemma 2.37** imply that it suffices to show that there is $F \in S(G, L(G))$ such that $e' \notin F$.

Using **Lemma 2.49**, **Lemma 2.42, Lemma 2.40**, **Corollary 2.14, Lemma 2.48** and the inequality $L(G(2, f')) \geq 1 + L(G(2, g'))$, we get
$$\Lambda(e) = 3 + L(G(1, g)) + L(G(2, f')) + L(G(2,e) \setminus (\{u_{10}, u_{11}\} \cup V(G(2, f')))) \geq$$
$$\geq L(G(1, g)) + 4 + L(G(2, g')) + L(G(1, e') \setminus (\{u_5\} \cup V(G(1, e)))) = \Lambda(e'),$$
therefore, due to **Lemma 2.38**, there is $F \in S(G, L(G))$ such that $e' \notin F$.

(g) **Lemma 2.45** and **Lemma 2.37** imply that it suffices to show that there is $F \in S(G, L(G))$ such that $e' \notin F$.

Using **Lemma 2.48**, **Lemma 2.42, Corollary 2.14, Lemma 2.49**, we get
$$\Lambda(e) = 3 + L(G(1, g)) + L(G(2, f')) + L(G(2,e) \setminus (\{u_{10}, u_{11}\} \cup V(G(2, f')))) \geq$$
$$\geq L(G(1, g) - u_2) + 3 + L(G(2, g')) + L(G(1, e') \setminus (\{u_8, u_9\} \cup V(G(1, f)))) = \Lambda(e'),$$
therefore, due to **Lemma 2.38**, there is $F \in S(G, L(G))$ such that $e' \notin F$.

(h) **Lemma 2.46** and **Lemma 2.37** imply that it suffices to show that there is $F \in S(G, L(G))$ such that $e \notin F$.

Using **Lemma 2.49**, **Lemma 2.42, Corollary 2.14, Lemma 2.48**, we get
$$\Lambda(e') = L(G(1, g) - u_2) + 3 + L(G(2, g')) + L(G(1, e') \setminus (\{u_8, u_9\} \cup V(G(1, f)))) \geq$$
$$\geq 3 + L(G(1, g)) + L(G(2, f')) + L(G(2,e) \setminus (\{u_{10}, u_{11}\} \cup V(G(2, f')))) = \Lambda(e),$$
therefore, due to **Lemma 2.38**, there is $F \in S(G, L(G))$ such that $e \notin F$. Proof of **Lemma 2.54** is complete.

**Lemma 2.55.** Let $G$ be a tree, $U = \{u_0, ..., u_{14}\} \subseteq V(G)$, $d_G(u_0) = d_G(u_9) = d_G(u_{11}) = d_G(u_8) = 1$, $d_G(u_1) = d_G(u_{10}) = d_G(u_{12}) = d_G(u_7) = 2$, $d_G(u_2) = d_G(u_3) = d_G(u_5) = d_G(u_6) = 3$, $(u_{i-1}, u_i) \in E(G)$ for $i = 1, ..., 8, 10, 12$, $(u_2, u_{13}) \in E(G)$, $(u_3, u_{10}) \in E(G)$, $(u_5, u_{12}) \in E(G)$, $(u_6, u_{14}) \in E(G)$. Set: $g = (u_{13}, u_2)$, $f = (u_2, u_3)$, $e = (u_3, u_4)$, $e' = (u_4, u_5)$, $f' = (u_5, u_6)$, $g' = (u_6, u_{14})$, and let $u_0 \in V(G(2, g))$, $u_0 \in V(G(1, f))$, $u_0 \in V(G(1, e))$, $u_0 \in V(G(1, e'))$, $u_0 \in V(G(1, f'))$, $u_0 \in V(G(1, g'))$ (figure 37). Then

(a) if $e \notin \pi(G)$, then if $F_i \in S(G(i, f), L(G(i, f)))$ and $f \in F_i$, for $i = 1, 2$, then $F_1 \cup F_2 \in S(G, L(G))$;

(b) if $e' \notin \pi(G)$, then if $F_i' \in S(G(i, f'), L(G(i, f')))$ and $f' \in F_i'$, for $i = 1, 2$, then $F_1' \cup F_2' \in S(G, L(G))$;

(c) if $e \in \pi(G)$, $e' \in \pi(G)$ then
$$\beta(G(1, g)) + \beta(G(2, g') - u_6) = \beta(G(1, g) - u_2) + \beta(G(2, g'));$$

(d) if $e \in \pi(G)$, $e' \in \pi(G)$, $\beta(G(1, g)) = \beta(G(1, g) - u_2)$ and $L(G(1, g) - u_2) + L(G(2, g')) \leq$
$\leq L(G(1, g)) + L(G(2, g') - u_6)$, then if $F_i' \in S(G(i, f'), L(G(i, f')))$ and $f' \in F_i'$, for $i = 1, 2$, then $F_1' \cup F_2' \in S(G, L(G))$;

(e) if $e \in \pi(G)$, $e' \in \pi(G)$, $\beta(G(1, g)) = \beta(G(1, g) - u_2)$ and $L(G(1, g)) + L(G(2, g') - u_6) \leq$
$\leq L(G(1, g) - u_2) + L(G(2, g'))$, then if $F_i \in S(G(i, f), L(G(i, f)))$ and $f \in F_i$, for $i = 1, 2$, then $F_1 \cup F_2 \in S(G, L(G))$;

(f) if $e \in \pi(G)$, $e' \in \pi(G)$, $\beta(G(1,g)) = \beta(G(1,g) - u_2) + 1$, $L(G(1,g)) = L(G(1,g) - u_2)$, $L(G(2,g')) = L(G(2,g') - u_6)$, then if $F_i \in S(G(i,g), L(G(i,g)))$ and $g \in F_i$, for $i = 1, 2$, then $F_1 \cup F_2 \in S(G, L(G))$;

(g) if $e \in \pi(G)$, $e' \in \pi(G)$, $\beta(G(1,g)) = \beta(G(1,g) - u_2) + 1$, $L(G(1,g)) = L(G(1,g) - u_2)$, $L(G(2,g')) \leq L(G(2,g') - u_6) - 1$, then if $F_i' \in S(G(i,f'), L(G(i,f')))$ and $f' \in F_i'$, for $i = 1, 2$, then $F_1' \cup F_2' \in S(G, L(G))$;

(h) if $e \in \pi(G)$, $e' \in \pi(G)$, $\beta(G(1,g)) = \beta(G(1,g) - u_2) + 1$, $L(G(1,g)) \leq L(G(1,g) - u_2) - 1$, $L(G(2,g')) = L(G(2,g') - u_6)$, then if $F_i \in S(G(i,f), L(G(i,f)))$ and $f \in F_i$, for $i = 1, 2$, then $F_1 \cup F_2 \in S(G, L(G))$;

(i) if $e \in \pi(G)$, $e' \in \pi(G)$, $\beta(G(1,g)) = \beta(G(1,g) - u_2) + 1$, $L(G(1,g)) \leq L(G(1,g) - u_2) - 1$, $L(G(2,g')) \leq L(G(2,g') - u_6) - 1$, $L(G(1,g)) + L(G(2,g') - u_6) \geq L(G(1,g) - u_2) + L(G(2,g'))$, then if $F_i' \in S(G(i,f'), L(G(i,f')))$ and $f' \in F_i'$, for $i = 1, 2$, then $F_1' \cup F_2' \in S(G, L(G))$;

(j) if $e \in \pi(G)$, $e' \in \pi(G)$, $\beta(G(1,g)) = \beta(G(1,g) - u_2) + 1$, $L(G(1,g)) \leq L(G(1,g) - u_2) - 1$, $L(G(2,g')) \leq L(G(2,g') - u_6) - 1$, $L(G(1,g)) + L(G(2,g') - u_6) \leq L(G(1,g) - u_2) + L(G(2,g'))$, then if $F_i \in S(G(i,f), L(G(i,f)))$ and $f \in F_i$, for $i = 1, 2$, then $F_1 \cup F_2 \in S(G, L(G))$.

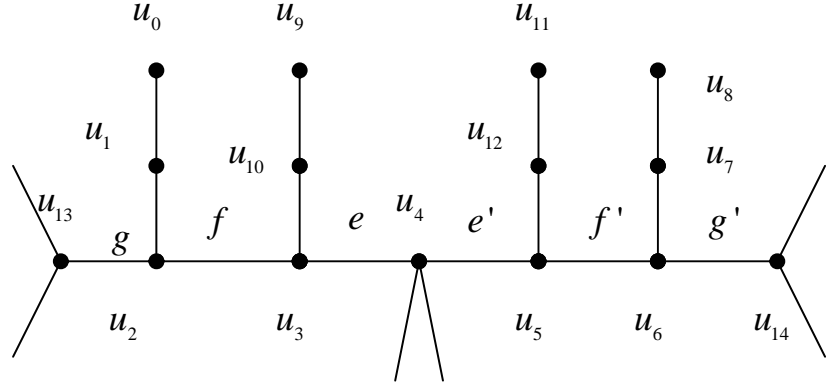

Figure 37

**Proof.** (a) and (b) follow from **Lemma 2.46** and **Lemma 2.37**.

(c) Since $e \in \pi(G)$, $e' \in \pi(G)$, we have

$$\beta(G) = 6 + \beta(G(1,g)) + \beta(G(2,g') - u_6) + \beta(G(2,e) \setminus (\{u_3\} \cup V(G(2,e')))),$$
$$\beta(G) = 6 + \beta(G(1,g) - u_2) + \beta(G(2,g')) + \beta(G(1,e') \setminus (\{u_5\} \cup V(G(1,e)))),$$

therefore

$$\beta(G(1,g)) + \beta(G(2,g') - u_6) = \beta(G(1,g) - u_2) + \beta(G(2,g')).$$

(d) **Lemma 2.46** and **Lemma 2.37** imply that it suffices to show that there is $F \in S(G, L(G))$ such that $e' \notin F$.

Using **Lemma 2.49**, **Lemma 2.42, Corollary 2.14,** we get

$$\Lambda(e) = 4 + L(G(1,g)) + L(G(2,g^{'}) - u_6) + L(G(2,e) \setminus (\{u_{11}, u_{12}\} \cup V(G(2,f^{'})))) \geq$$
$$\geq L(G(1,g) - u_2) + 4 + L(G(2,g^{'})) + L(G(1,e^{'}) \setminus (\{u_9, u_{10}\} \cup V(G(1,f)))) = \Lambda(e^{'}),$$
therefore, due to **Lemma 2.38**, there is $F \in S(G, L(G))$ such that $e^{'} \notin F$.

(e) **Lemma 2.46** and **Lemma 2.37** imply that it suffices to show that there is $F \in S(G, L(G))$ such that $e \notin F$.

Using **Lemma 2.49**, **Lemma 2.42**, **Corollary 2.14,** we get
$$\Lambda(e^{'}) = L(G(1,g) - u_2) + 4 + L(G(2,g^{'})) + L(G(1,e^{'}) \setminus (\{u_9, u_{10}\} \cup V(G(1,f)))) \geq$$
$$\geq 4 + L(G(1,g)) + L(G(2,g^{'}) - u_6) + L(G(2,e) \setminus (\{u_{11}, u_{12}\} \cup V(G(2,f^{'})))) = \Lambda(e),$$
therefore, due to **Lemma 2.38**, there is $F \in S(G, L(G))$ such that $e \notin F$.

(f) **Lemma 2.46** and **Lemma 2.37** imply that it suffices to show that there is $F \in S(G, L(G))$ such that $e \notin F$.

Using **Lemma 2.49**, **Lemma 2.42**, **Corollary 2.14**, **Lemma 2.41,** we get
$$\Lambda(e^{'}) = 5 + L(G(1,g)) + L(G(2,g^{'})) + L(G(1,e^{'}) \setminus (\{u_5\} \cup V(G(1,e)))) = \Lambda(e),$$
therefore, due to **Lemma 2.38**, there is $F \in S(G, L(G))$ such that $e \notin F$.

(g) **Lemma 2.46** and **Lemma 2.37** imply that it suffices to show that there is $F \in S(G, L(G))$ such that $e^{'} \notin F$.

Using **Lemma 2.49**, **Lemma 2.42**, **Corollary 2.14**, **Lemma 2.40**, **Lemma 2.41,** we get
$$\Lambda(e) = 4 + L(G(1,g)) + L(G(2,g^{'}) - u_6) + L(G(2,e) \setminus (\{u_{11}, u_{12}\} \cup V(G(2,f^{'})))) \geq$$
$$\geq L(G(1,g)) + 5 + L(G(2,g^{'})) + L(G(1,e^{'}) \setminus (\{u_5\} \cup V(G(1,e)))) = \Lambda(e^{'}),$$
therefore, due to **Lemma 2.38**, there is $F \in S(G, L(G))$ such that $e^{'} \notin F$.

(h) **Lemma 2.46** and **Lemma 2.37** imply that it suffices to show that there is $F \in S(G, L(G))$ such that $e \notin F$.

Using **Lemma 2.49**, **Lemma 2.42**, **Corollary 2.14**, **Lemma 2.40**, **Lemma 2.41,** we get
$$\Lambda(e^{'}) = L(G(1,g) - u_2) + 4 + L(G(2,g^{'})) + L(G(1,e^{'}) \setminus (\{u_9, u_{10}\} \cup V(G(1,f)))) \geq$$
$$\geq 5 + L(G(1,g)) + L(G(2,g^{'})) + L(G(2,e) \setminus (\{u_3\} \cup V(G(2,e^{'})))) = \Lambda(e),$$
therefore, due to **Lemma 2.38**, there is $F \in S(G, L(G))$ such that $e \notin F$.

(i) **Lemma 2.46** and **Lemma 2.37** imply that it suffices to show that there is $F \in S(G, L(G))$ such that $e^{'} \notin F$.

Using **Lemma 2.49**, **Lemma 2.42**, **Corollary 2.14,** we get
$$\Lambda(e) = 4 + L(G(1,g)) + L(G(2,g^{'}) - u_6) + L(G(2,e) \setminus (\{u_{11}, u_{12}\} \cup V(G(2,f^{'})))) \geq$$
$$\geq L(G(1,g) - u_2) + 4 + L(G(2,g^{'})) + L(G(1,e^{'}) \setminus (\{u_9, u_{10}\} \cup V(G(1,f)))) = \Lambda(e^{'}),$$
therefore, due to **Lemma 2.38**, there is $F \in S(G, L(G))$ such that $e^{'} \notin F$.

(j) **Lemma 2.46** and **Lemma 2.37** imply that it suffices to show that there is $F \in S(G, L(G))$ such that $e \notin F$.

Using **Lemma 2.49**, **Lemma 2.42**, **Corollary 2.14,** we get
$$\Lambda(e^{'}) = L(G(1,g) - u_2) + 4 + L(G(2,g^{'})) + L(G(1,e^{'}) \setminus (\{u_9, u_{10}\} \cup V(G(1,f)))) \geq$$
$$\geq 4 + L(G(1,g)) + L(G(2,g^{'}) - u_6) + L(G(2,e) \setminus (\{u_{11}, u_{12}\} \cup V(G(2,f^{'})))) = \Lambda(e),$$
therefore, due to **Lemma 2.38**, there is $F \in S(G, L(G))$ such that $e \notin F$. Proof of **Lemma 2.55** is complete.

## §3. Elementary trees

For a graph $G$ set:
$$\tilde{V}(G) \equiv \{v \in V(G) / d_G(v) \geq 3 \text{ and } \exists w_v, u_v \in V(G) \text{ such that } d_G(w_v) = 1,$$
$$d_G(u_v) = 2, (w_v, u_v) \in E(G), (u_v, v) \in E(G)\}.$$

For each $\tilde{v} \in \tilde{V}(G)$ choose $w_{\tilde{v}}, u_{\tilde{v}} \in V(G)$ such that
$$d_G(w_{\tilde{v}}) = 1, d_G(u_{\tilde{v}}) = 2, (w_{\tilde{v}}, u_{\tilde{v}}) \in E(G), (u_{\tilde{v}}, \tilde{v}) \in E(G),$$

and assume:
$$P(\tilde{v}) = \{w_{\tilde{v}}, u_{\tilde{v}}\}.$$

Set:
$$G^* \equiv G \setminus \bigcup_{\tilde{v} \in \tilde{V}(G)} P(\tilde{v}).$$

It is not hard to see that the following properties hold:

**Property 3.1.** For every graph $G$ the graph $G^*$ is defined uniquely (up to isomorphism of graphs).

**Property 3.2.** If $G$ is a tree then $G^*$ is also tree.

**Property 3.3.** Let $x \in V(G^*)$. $d_{G^*}(x) = 1$ if and only if $d_G(x) = 1$.

**Definition 3.1.** A tree $G$ is elementary if $G^*$ is a simple path.

It is not hard to see that for an elementary tree $G$ there are $n \in Z^+$, $I \subseteq \{1,...,n-1\}$ and $v_0 \in V(G),..., v_n \in V(G)$, $u_i \in V(G)$, $w_i \in V(G)$ ($i \in I$) such that

(a) $V(G) = \{v_0,...,v_n\} \cup \bigcup_{i \in I} \{w_i, u_i\}$;

(b) $E(G) = \{(v_{i-1}, v_i) / i = 1,...,n\} \cup \bigcup_{i \in I} \{(w_i, u_i), (v_i, u_i)\}$;

(c) $|\tilde{V}(G)| = |I|$.

**Lemma 3.1.** Let $G$ be an elementary tree satisfying the condition $\beta(G) < \dfrac{|V(G)|}{2}$. Then
$$|M(G)| = \frac{n}{2} + 1 + |\{v / d_G(v) = 3 \text{ and the distance between } v \text{ and } v_0 \text{ is even}\}|.$$

**Proof.** Since $G$ does not contain a perfect matching we imply that $n$ is even. Assume that $n = 2k$. Note that
$$|V(G)| = 2|I| + n + 1 = 2(k + |I|) + 1,$$
$$\beta(G) = k + |I|,$$

hence $|V(G)| = 2\beta(G) + 1$. This implies that for each $F \in M(G)$ there is exactly one $x_F \in V(G)$ satisfying the condition $\psi(x_F) \cap F = \varnothing$. It is not hard to see that the vertex $x_F \in V(G)$ uniquely determines the original maximum matching $F$, hence the following equality is true:
$$|M(G)| = |\{x \in V(G) / \exists F \in M(G) \text{ such that } \psi(x) \cap F = \varnothing\}|.$$

On the other hand, it is not hard to see that
$$\{x \in V(G) / \exists F \in M(G) \text{ such that } \psi(x) \cap F = \varnothing\} = \{v_0, v_2,..., v_{2k} = v_n\} \cup \{w_i / i \in I \text{ and } i \text{ is even}\}.$$

Proof of **Lemma 3.1** is complete.

An immediate corollary of the **Lemma 3.1** is the following

**Theorem 3.1.** Let $G$ be an elementary tree. Then the inequality

$$|M(G)| \le \frac{|V(G)|+1}{2}$$

is true.

Note that **Theorem 3.1** implies that using the standard brute-force search we can design an algorithm
(a) which for an elementary tree $G$ returns $F \in S'(G,l(G))$ and the number $l = l(G) = \beta(G \setminus F)$ in time $O(|V(G)| \cdot \log_2 |V(G)|)$;
(b) which for an elementary tree $G$ returns the number $L = L(G) = \beta(G \setminus F)$ and $F \in S(G, L(G))$ that contains the prescribed edge $e \in \overline{\theta}(G)$ in time $O(|V(G)| \cdot \log_2 |V(G)|)$ (**Corollary 2.9**).

## §4. Two polynomial algorithms for constructing special maximum matchings in trees

In this section we will describe two polynomial algorithms which solve the **Problems 1.1** and **1.2** for trees. Here we will assume that we are aware of two algorithms with complexity $O(|V(G)| \cdot \log_2 |V(G)|)$ that solve these problems for elementary trees (§3).

**Algorithm 4.1** *MinMax*
**Input:** $G$-tree and $\Gamma \in P'(G)$
**Output:** $F \in S'(G,l(G))$ with $\Gamma \subseteq F$

**Step 0. If** ($|\Gamma| = \beta(G)$) **then return** $\Gamma$;

**Step 1. If** ($G$ - is an elementary tree) **then** using the algorithm from §3 construct $F \in S'(G,l(G))$ in $G$ and **return** $F$;

**Step 2.** If $G$ satisfies the conditions of the **Lemma 2.23**, **Lemma 2.21, Lemma 2.20**, **Lemma 2.24 - Lemma 2.32**, then using the algorithm from §3 construct $F \in S'(G,l(G))$ according to these lemmata.

**Step 3. return** $F$;

**Theorem 4.1.** For every input $(G,\Gamma)$, where $G$ is a tree and $\Gamma \in P'(G)$, **Algorithm 4.1** returns $F \in S'(G,l(G))$ satisfying the condition $\Gamma \subseteq F$ in time $O(|V(G)|^3)$.

**Proof.** First of all let us show that $F \in S'(G,l(G))$ and $\Gamma \subseteq F$. The proof will be done by induction on $|E(G)|$.

The case $|E(G)| = 1$ is trivial. Assume the statement to be true for trees $G'$ satisfying the condition $|E(G')| \le t$, and let us suppose that $|E(G)| = t+1$, $t \ge 1$. **Corollary 2.4** implies that without loss of

generality we may assume $G$ to satisfy the inequality $\eta(G) < \beta(G)$. We may also assume $G$ to be non-elementary.

Note that all we need at this moment is to show that every non-elementary tree $G$, for which $\eta(G) < \beta(G)$ and which does not satisfy the conditions of **Lemma 2.23**, **Lemma 2.21**, satisfies at least one of the conditions of **Lemma 2.20**, **Lemma 2.24 - Lemma 2.32**.

For tree $G$ consider the tree $G^*$. Since $G$ is non-elementary we have $\left|V_{G^*}(0)\right| \geq 3$, hence the set $\overline{V}$ defined as $\overline{V} \equiv \{v \in V(G^*) \mid d_{G^*}(v) \geq 3\}$ is not empty. Take $\overline{v} \in \overline{V}$ such that $k_{G^*}(\overline{v}) \to \min$. Clearly, $d_{G^*}(\overline{v}) \geq 3$ and $k_{G^*}(\overline{v}) \geq 1$. The choice of the vertex $\overline{v} \in \overline{V}$ implies that there are $U_1, U_2, ..., U_r \subseteq V(G^*)$ ($r \geq 2$) such that

1. $U_i \cap U_j = \emptyset$ $1 \leq i < j \leq r$;
2. for each $i$, $1 \leq i \leq r$ and $w \in U_i$, $k_{G^*}(w) < k_{G^*}(\overline{v})$, $|U_i| \geq 1$;
3. $\max\limits_{1 \leq i \leq r} |U_i| = k_{G^*}(\overline{v})$;
4. for $i = 1, ..., r$ the graph $<U_i \cup \{\overline{v}\}>_{G^*}$ is a simple path (figure 38).

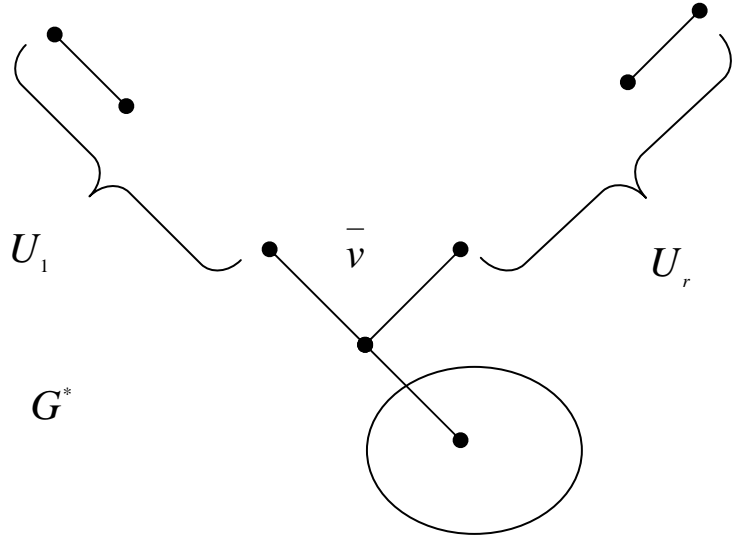

Figure 38

Note that there is no $i, 1 \leq i \leq r$ such that $|U_i| = 2$ (otherwise, we would have $d_{G^*}(\overline{v}) \neq d_G(\overline{v})$, hence $G$ satisfies the condition of **Lemma 2.23**). To complete the proof by induction, let us consider the following possible cases:

(a) there are $i_1, i_2$ $1 \leq i_1 < i_2 \leq r$, such that for each $u \in U_{i_1} \cup U_{i_2}$ $d_{G^*}(u) = d_G(u)$.

Let us consider the case $\left|U_{i_1}\right| = 1$.

If $\left|U_{i_2}\right| = 1$ then $G$ satisfies the condition of **Lemma 2.20**;

If $\left|U_{i_2}\right| = 3$ then $G$ satisfies the condition of **Lemma 2.24**;

If $\left|U_{i_2}\right| = 4$ then $G$ satisfies the condition of **Lemma 2.25**;

If $|U_{i_2}| \geq 5$ then $G$ satisfies the condition of **Lemma 2.26**.

We does not consider the case of $|U_{i_2}| = 2$, since we assumed that $G$ does not satisfy the condition of **Lemma 2.21**.

Now, let us consider the case $|U_{i_1}| = 3$.

If $|U_{i_2}| = 3$ then $G$ satisfies the condition of **Lemma 2.27**;

If $|U_{i_2}| = 4$ then $G$ satisfies the condition of **Lemma 2.28**;

If $|U_{i_2}| \geq 5$ then $G$ satisfies the condition of **Lemma 2.30**.

Let us consider the case $|U_{i_1}| = 4$.

If $|U_{i_2}| = 4$ then $G$ satisfies the condition of **Lemma 2.29**;

If $|U_{i_2}| \geq 5$ then $G$ satisfies the condition of **Lemma 2.31**.

Finally, let us consider the case $|U_{i_1}| \geq 5$ and $|U_{i_2}| \geq 5$. Note that in this case $G$ satisfies the condition of **Lemma 2.32**.

(b) $G$ does not satisfy (a), but there is $i_0, 1 \leq i_0 \leq r$, such that for each $u \in U_{i_0}$ $d_{G^*}(u) = d_G(u)$, and for every $i \in \{1,...,r\} \setminus \{i_0\}$ there is $u_i \in U_i$ such that $d_{G^*}(u_i) \neq d_G(u_i)$.

Let us consider the case $|U_{i_0}| = 1$. Note that in this case $G$ satisfies at least one of the conditions of **Lemma 2.24- Lemma 2.26**.

Now, let us consider the case $|U_{i_0}| = 3$. Note that in this case $G$ satisfies at least one of the conditions of **Lemma 2.27, Lemma 2.28, Lemma 2.30**.

Let us consider the case $|U_{i_0}| = 4$. Note that in this case $G$ satisfies at least one of the conditions of **Lemma 2.28, Lemma 2.29, Lemma 2.31**.

Finally, let us consider the case $|U_{i_0}| \geq 5$. Note that in this case $G$ satisfies at least one of the conditions of **Lemma 2.30-Lemma 2.32**.

(c) $G$ does not satisfy (a),(b), but for every $i$, $1 \leq i \leq r$ there is $u_i \in U_i$ such that $d_{G^*}(u_i) \neq d_G(u_i)$. It is not hard to see that $G$ satisfies at least one of the conditions of **Lemma 2.27-Lemma 2.32**.

Now let us show that **Algorithm 4.1** runs in time $O(|V(G)|^3)$. We will assume that the tree $G$ is presented to the input of **Algorithm 4.1** by its matrix of adjacency.

Note that if $G$ is a non-elementary tree and $\eta(G) < \beta(G)$, then **Algorithm 4.1** calls itself for some subtrees of $G$. Moreover, the number of calls is the number of connected components of the forest $G \setminus U$, if $G$ satisfies the condition of **Lemma 2.20** or **Lemma 2.21**, where $U$ is the subset of $V(G)$ mentioned in the formulation of the lemmata, and is one, otherwise. Let us show that the total number of calls of **Algorithm 4.1** for the input tree $G$ (counting the original call, too) does not exceed $|V(G)|$. Clearly, the number of calls of **Algorithm 4.1** is one when $|V(G)| = 1$. Assume that the statement is true for $|V(G)| \leq t$ and let us consider the case $|V(G)| = t + 1$. The following two cases are possible:

**Case 1**: $G$ satisfies the condition of **Lemma 2.20** or **Lemma 2.21**. Then, by the induction hypothesis, the number of calls of **Algorithm 4.1** does not exceed the number $1+|V(G_1)|+...+|V(G_r)|$ (counting the original call, too), where $G_1,...,G_r$ are the connected components of $G \setminus U$. Note that $|U| \geq 3$, hence the total number of calls of **Algorithm 4.1** does not exceed $|V(G)|$ since $|V(G)|=|U|+|V(G_1)|+...+|V(G_r)|$.

**Case 2**: $G$ does not satisfy the condition of **Lemma 2.20** or **Lemma 2.21**. Then, as we mentioned above, **Algorithm 4.1** calls itself for some subtree $G'$ of $G$. Note that $|V(G)| \geq |V(G')|+2$, hence, due to induction hypothesis, the total number of calls of **Algorithm 4.1** does not exceed $|V(G')|+1 \leq |V(G)|$ (counting the original call, too).

Finally, let us estimate the number of operations for one call of **Algorithm 4.1** when the input graph is some subtree $G'$ of the original input tree $G$ (of course, in the beginning we have $G'=G$).

It is not hard to see that in time $O(|V(G')|^2)$ we can construct:

1. the tree $G'^*$;
2. the sets $V_{G^*}(0), V_{G^*}(1),...,V_{G^*}(r_{G^*})$;
3. the array **data**, which for every vertex $v \in V(G')$ contains the degree of the vertex $v$, the vertices which are adjacent to $v$ and the number $k_{G^*}(v)$ where $v \in V_{G^*}(k_{G^*}(v))$.

Having done this, the **Step 0** can be realized in time $O(|V(G')|)$. The **Step 1** will take no more than $O(|V(G')| \cdot \log_2 |V(G')|)$ (§3). Since in time $O(|V(G')|^2)$ we can decide which is the **Lemma** from the **Step 2** condition of which $G'$ satisfies, and every operation presented in the reduction that is provided by this **Lemma** can be performed in time $O(|V(G')| \cdot \log_2 |V(G')|)$ (§3), we imply that the **Step 2** of **Algorithm 4.1** can be realized in time $O(|V(G')|^2)$. Hence the total time for one call of **Algorithm 4.1** is $O(|V(G')|^2) \leq O(|V(G)|^2)$. Since the number of calls of **Algorithm 4.1** does not exceed $|V(G)|$, we imply that **Algorithm 4.1** runs in time $O(|V(G)|^3)$. Proof of **Theorem 4.1** is complete.

**Algorithm 4.2** *MaxMax*
**Input:** $G$-tree and $e \in \bar{\theta}(G)$
**Output:** $F \in S(G,L(G))$ with $e \in F$

**Step 1. If** ($G$ - is an elementary tree) **then** using the algorithm from §3 construct $F \in S(G,L(G))$ with $e \in F$ and **return** $F$;

**Step 2.** If $G$ satisfies the conditions of the **Corollary 2.14, Lemma 2.41- Lemma 2.43**, **Lemma 2.47- Lemma 2.55**, then using the algorithm from §3 construct $F \in S(G,L(G))$ according to these lemmata.

**Step 3.**
    **If** ($e \notin F$) **then**
     **begin**

**Substep 3.1.** Take $H \in M(G \setminus F)$ such that $e \in H$;

**Substep 3.2.** Consider the maximal alternating path $u_0, e_0, v_0, f_0, v_1, f_1, u_2, ..., u_t, e_t, v_t, f_t, u_{t+1}$ of the tree $G$, satisfying the conditions: $e_i = (u_i, v_i)$, $f_i = (v_i, u_{i+1})$ for $i = 0, ..., t$, $\{e_0, ..., e_t\} \subseteq H$, $\{f_0, ..., f_t\} \subseteq F$, $e_0 = e$. Assume:
$$F := F \setminus \{f_0, ..., f_t\} \cup \{e_0, ..., e_t\};$$

   **end;**
**return** $F$;

**Theorem 4.2.** For every input $(G, e)$, where $G$ is a tree and $e \in \overline{\theta}(G)$, **Algorithm 4.2** returns $F \in S(G, L(G))$ satisfying the condition $e \in F$ in time $O(|V(G)|^3)$.

**Proof.** First of all let us show that $F \in S(G, L(G))$ and $e \in F$. Note that **Corollary 2.9** implies that it suffices to show that $F \in S(G, L(G))$. The proof will be done by induction on $|E(G)|$.

The case $|E(G)| = 1$ is trivial. Assume the statement to be true for trees $G'$ satisfying the condition $|E(G')| \leq t$, and let us suppose that $|E(G)| = t+1$, $t \geq 1$. We may also assume $G$ to be non-elementary.

Note that all we need at this moment is to show that every non-elementary tree $G$ satisfies at least one of the conditions of **Corollary 2.14, Lemma 2.41- Lemma 2.43, Lemma 2.47- Lemma 2.55**.

For the tree $G$ consider the tree $G^*$. Since $G$ is non-elementary we have $|V_{G^*}(0)| \geq 3$, hence the set $\overline{V}$ defined as $\overline{V} \equiv \{v \in V(G^*) | d_{G^*}(v) \geq 3\}$ is not empty. Take $\overline{v} \in \overline{V}$ such that $k_{G^*}(\overline{v}) \to \min$. Clearly, $d_{G^*}(\overline{v}) \geq 3$ and $k_{G^*}(\overline{v}) \geq 1$. The choice of the vertex $\overline{v} \in \overline{V}$ implies that there are $U_1, U_2, ..., U_r \subseteq V(G^*)$ ($r \geq 2$) such that
1. $U_i \cap U_j = \emptyset$  $1 \leq i < j \leq r$;
2. for each $i$, $1 \leq i \leq r$ and $w \in U_i$  $k_{G^*}(w) < k_{G^*}(\overline{v})$, $|U_i| \geq 1$;
3. $\max\limits_{1 \leq i \leq r} |U_i| = k_{G^*}(\overline{v})$;
4. for $i = 1, ..., r$ the graph $<U_i \cup \{\overline{v}\}>_{G^*}$ is a simple path (figure 38).

To complete the proof by induction, let us consider the following possible cases:

(a) there are $i_1, i_2$  $1 \leq i_1 < i_2 \leq r$, such that for each $u \in U_{i_1} \cup U_{i_2}$  $d_{G^*}(u) = d_G(u)$.

Note that if $\max\{|U_{i_1}|, |U_{i_2}|\} \geq 3$, then $G$ satisfies the condition of **Corollary 2.14**, therefore we may assume that $\max\{|U_{i_1}|, |U_{i_2}|\} \leq 2$. On the other hand, if $\min\{|U_{i_1}|, |U_{i_2}|\} = 1$ then $G$ satisfies one of the conditions of **Lemma 2.41, Lemma 2.42**, thus we may also assume that $\min\{|U_{i_1}|, |U_{i_2}|\} \geq 2$. It is not hard to see that $|U_{i_1}| = |U_{i_2}| = 2$. This implies that $d_{G^*}(\overline{v}) \neq d_G(\overline{v})$, hence $G$ satisfies the conditions of **Lemma 2.43**,

(b) $G$ does not satisfy (a), but there is $i_0$, $1 \leq i_0 \leq r$, such that for each $u \in U_{i_0}$  $d_{G^*}(u) = d_G(u)$, and for every $i \in \{1, ..., r\} \setminus \{i_0\}$ there is $u_i \in U_i$ such that $d_{G^*}(u_i) \neq d_G(u_i)$.

Note that if $|U_{i_0}| \geq 3$, then $G$ satisfies the condition of **Corollary 2.14**, and if $|U_{i_0}| = 2$, then $d_{G^*}(\bar{v}) \neq d_G(\bar{v})$, hence $G$ satisfies the conditions of **Lemma 2.43**, therefore we may assume that $|U_{i_0}| = 1$. It is not hard to see that $G$ satisfies at least one of the conditions of **Lemma 2.47- Lemma 2.49**.

(c) $G$ does not satisfy (a),(b), but for every $i$, $1 \leq i \leq r$ there is $u_i \in U_i$ such that $d_{G^*}(u_i) \neq d_G(u_i)$. It is not hard to see that $G$ satisfies at least one of the conditions of **Lemma 2.50- Lemma 2.55**.

Now let us show that **Algorithm 4.2** runs in time $O(|V(G)|^3)$. We will assume that $G$ is presented to the input of **Algorithm 4.2** by its matrix of adjacency.

Note that if $G$ is a non-elementary tree, then **Algorithm 4.2** calls itself for some subtrees of $G$. Moreover, the number of calls is the number of connected components of the forest $G \setminus U$, if $G$ satisfies either of conditions of **Lemma 2.41**, **Lemma 2.42**, where $U$ is the subset of $V(G)$ mentioned in these lemmata, and is one, otherwise. Let us show that the total number of calls of the **Algorithm 4.2** for the input tree $G$ (counting the original call, too) does not exceed $|V(G)|$. Clearly, the number of calls of **Algorithm 4.2** is one when $|V(G)| = 1$. Assume that the statement is true for $|V(G)| \leq t$ and let us consider the case $|V(G)| = t+1$. The following two cases are possible:

**Case 1**: $G$ satisfies at least one of the conditions of **Lemma 2.41** or **Lemma 2.42**. Then, by the induction hypothesis, the number of calls of **Algorithm 4.2** does not exceed the number $1 + |V(G_1)| + ... + |V(G_r)|$ (counting the original call, too), where $G_1,...,G_r$ are the connected components of $G \setminus U$. Note that $|U| \geq 3$, hence the total number of calls of **Algorithm 4.2** does not exceed $|V(G)|$ since $|V(G)| = |U| + |V(G_1)| + ... + |V(G_r)|$.

**Case 2**: $G$ satisfies none of the conditions of **Lemma 2.41** and **Lemma 2.42**. Then, as we mentioned above, **Algorithm 4.2** calls itself for some subtree $G'$ of $G$. Note that $|V(G)| \geq |V(G')| + 2$, hence, due to induction hypothesis, the total number of calls of **Algorithm 4.2** does not exceed $|V(G')| + 1 \leq |V(G)|$ (counting the original call, too).

Finally, let us estimate the number of operations for one call of **Algorithm 4.2** when the input graph is some subtree $G'$ of the original input tree $G$ (of course, in the beginning we have $G' = G$).

It is not hard to see that in time $O(|V(G')|^2)$ we can construct:

1. the tree $G'^*$;
2. the sets $V_{G^*}(0), V_{G^*}(1), ..., V_{G^*}(r_{G^*})$;
3. the array **data**, which for every vertex $v \in V(G')$ contains the degree of the vertex $v$, the vertices which are adjacent to $v$ and the number $k_{G^*}(v)$ where $v \in V_{G^*}(k_{G^*}(v))$.

Having done this, the **Step 1** will take no more than $O(|V(G')| \cdot \log_2 |V(G')|)$ (§3). Since in time $O(|V(G')|^2)$ we can decide which is the **Lemma** from the **Step 2** condition of which $G'$ satisfies, and every operation presented in the reduction that is provided by this **Lemma** can be performed in time $O(|V(G')| \cdot \log_2 |V(G')|)$ (§3), we imply that the **Step 2** of **Algorithm 4.2** can be realized in time

$O(|V(G')|^2)$. Finally, the **Step 3** can be realized in time $O(|V(G')|)$. Hence the total time for one call of **Algorithm 4.2** is $O(|V(G')|^2) \leq O(|V(G)|^2)$. Since the number of calls of **Algorithm 4.2** does not exceed $|V(G)|$, we imply that **Algorithm 4.2** runs in time $O(|V(G)|^3)$. Proof of **Theorem 4.2** is complete.

**Acknowledgement**: We would like to thank Anush Tserunyan for providing us with an exact formula for the number of maximum matchings in elementary trees.